\documentclass[aps,pra,twocolumn,longbibliography,superscriptaddress]{revtex4-1}
\usepackage{epsfig}
\usepackage{epstopdf}
\usepackage{amsmath}
\usepackage{amsfonts}
\usepackage{amssymb}
\usepackage{hyperref}
\usepackage{bm}
\usepackage{makecell}
\usepackage{rotating}
\usepackage{hyperref}
\usepackage{multirow}

\usepackage{graphicx}
\usepackage{dcolumn}
\usepackage{bm}
\usepackage{color}

\usepackage{tikz,xcolor,hyperref}

\definecolor{lime}{HTML}{A6CE39}
\DeclareRobustCommand{\orcidicon}{%
	\begin{tikzpicture}
	\draw[lime, fill=lime] (0,0)
	circle [radius=0.16]
	node[white] {{\fontfamily{qag}\selectfont \tiny ID}};
	\draw[white, fill=white] (-0.0625,0.095)
	circle [radius=0.007];
	\end{tikzpicture}
	\hspace{-2mm}
}

\foreach \x in {A, ..., Z}{%
	\expandafter\xdef\csname orcid\x\endcsname{\noexpand\href{https://orcid.org/\csname orcidauthor\x\endcsname}{\noexpand\orcidicon}}
}

\begin{document}

\title{CdTe and HgTe doped with V, Cr, and Mn -- \\ prospects for the quantum anomalous Hall effect}

\author{Giuseppe Cuono\orcidD}
\affiliation{International Research Centre MagTop, Institute of Physics, Polish Academy of Sciences,
Aleja Lotnik\'ow 32/46, PL-02668 Warsaw, Poland}

\author{Carmine Autieri\orcidA}
\email{autieri@magtop.ifpan.edu.pl}
\affiliation{International Research Centre MagTop, Institute of Physics, Polish Academy of Sciences,
Aleja Lotnik\'ow 32/46, PL-02668 Warsaw, Poland}

\author{Tomasz Dietl\orcidB}
\affiliation{International Research Centre MagTop, Institute of Physics, Polish Academy of Sciences,
Aleja Lotnik\'ow 32/46, PL-02668 Warsaw, Poland}

\date{\today}
\begin{abstract}
Using first principle calculations we examine properties of (Cd,V)Te, (Cd,Cr)Te, (Hg,V)Te, and (Hg,Cr)Te relevant to the quantum anomalous Hall effect (QAHE), such as the position of V- and Cr-derived energy levels and the exchange interactions between magnetic ions. We consider CdTe and HgTe, containing 12.5\% of cation-substitutional V or Cr ions in comparison to the well-known case of (Cd,Mn)Te and (Hg,Mn)Te, and examine their suitability for fabrication of ferromagnetic barriers or ferromagnetic topological quantum wells, respectively. To account for the strong correlation of transition metal $d$ electrons we employ hybrid functionals with different mixing parameters $a_{\text{HSE}}$ focusing on $a_{\text{HSE}}=0.32$, which better reproduces the experimental band gaps in HgTe and Hg$_{0.875}$Mn$_{0.125}$Te.  We find that Cr, like Mn, acts as an isoelectronic dopant but V can be an in-gap donor in CdTe and a resonant donor in HgTe, similar to the case of Fe in HgSe. From a magnetic point of view, the Cr-doped HgTe is ferromagnetic within the general gradient approximation (GGA) but becomes AFM within hybrid functionals. However, (Hg,V)Te is a ferromagnet within both exchange-correlation functionals in a stark contrast to (Hg,Mn)Te for which robust antiferromagnetic coupling is found theoretically and experimentally. Furthermore, we establish that the Jahn-Teller effect is relevant for the magnetism in the case of Cr-doping. Considering lower defect concentrations in HgTe-based quantum wells compared to (Bi,Sb)$_3$Te$_2$ layers, our results imply that HgTe quantum wells or (Cd,Hg)Te barriers containing either V or Cr show advantages over (Bi,Sb,Cr,V)$_3$Te$_2$-based QAHE systems but whether (i) ferromagnetic coupling will dominate in the Cr case and (ii) V will not introduce too many electrons to the quantum well is to be checked experimentally.
\end{abstract}


\maketitle

\section{Introduction}

The theoretical prediction \cite{Yu:2010_S} and the experimental discovery of the quantum
anomalous Hall effect (QAHE) in dilute ferromagnetic semiconductor (Bi,Sb,Cr)$_2$Te$_3$ \cite{Chang:2013_S} and other
systems \cite{Chang:2023_RMP} have triggered research on the prospects of
dissipationless and spin-polarized carrier channels for energy-efficient and
decoherence-free electronic and spintronic classical and quantum devices \cite{Ke:2018_ARCMP,Tokura:2019_NRP}.
Simultaneously, the application potential of the QAHE for
resistance \cite{Fox:2018_PRB,Goetz:2018_APL,Okazaki:2022_NP} and current \cite{Rodenbach:2023_arXiv} standards operating
in the absence of an external magnetic field has been demonstrated.
It has, however, become clear that relatively large native
defect concentrations in (Bi,Sb)$_2$Te$_3$ and related systems, typically
above $10^{19}$\,cm$^{-3}$, and the associated in-gap impurity-band charge transport,
limit the standards' operation to below 100\,mK and 1\,$\mu$A \cite{Fox:2018_PRB,Goetz:2018_APL,Okazaki:2022_NP,Rodenbach:2023_arXiv,Fijalkowski:2021_NC}.
Particularly relevant for the present work is the observation of the QAHE in (Bi,Sb)$_2$Te$_3$  layers sandwiched
between ferromagnetic (Zn,Cr)Te barriers \cite{Watanabe:2019_APL}.

It is, therefore, interesting to consider the metrology prospects of
HgTe and related systems, in which the native defect concentration is at the
$10^{16}$\,cm$^{-3}$ level \cite{Dietl:2023_PRL,Dietl:2023_PRB}. It was found that at the quantum well (QW) thickness corresponding to the topological phase transition, the quantum Hall (QH) plateau $R_{xy} = -h/e^2$ appears in weak magnetic fields
and persists in a broad magnetic range of the magnetic fields \cite{Konig2007,Yahniuk:2021_arXiv},
the observation relevant to the QHE metrology \cite{Yahniuk:2021_arXiv}. The
effect is particularly spectacular in the case of (Cd,Hg)Te/(Hg,Mn)Te QW, where
the broad plateau begins at 50\,mT \cite{Shamim:2020_SA}. Those surprising observations were explained
by an energetic overlap of the acceptor impurity band with the hole portion of the QW Dirac cone \cite{Dietl:2023_PRL,Dietl:2023_PRB}. The Coulomb gap, charge ordering, and the formation of bound magnetic polarons (in the Mn-doped samples) are the essential ingredients of the model \cite{Dietl:2023_PRL,Dietl:2023_PRB}.
In the case of the studied
samples \cite{Shamim:2020_SA}, QHE dominates over a possible QAHE \cite{Liu:2008_PRLb},  as the QAHE requires also the presence of a magnetic field -- ferromagnetic coupling between Mn spins in II-VI compounds appears only if the hole density exceeds $10^{20}$\,cm$^{-3}$ \cite{Dietl:1997_PRB,Ferrand:2000_JCG}, whereas intrinsic Mn-Mn interactions are antiferromagnetic \cite{Mycielski:1984_SSC}.

In the present and companion paper \cite{Sliwa:2023_arXiv}, we address the question of which cation-substitutional transition-metal (TM) impurities in barriers or QWs could lead to the QAHE in HgTe-based systems.
The conditions to be considered experimentally and theoretically include:
\begin{enumerate}
 \item Sufficiently high -- in a couple of percent range -- solubility limits of particular dopants at the cation-substitutional positions and without aggregation. Importantly, the equilibrium limits can often be overcome by appropriate growth protocols \cite{Dietl:2015_RMP} and, for instance, incorporation of Cr to ZnTe by low-temperature molecular beam epitaxy appears successful \cite{Watanabe:2019_APL}.
 \item Ferromagnetic coupling between localized spins. We anticipate that Mn and Co dopants can be excluded, as antiferromagnetic superexchange dominates for these ions in II-VI compounds, also in the case of Hg$_{1-x}$Mn$_x$Te \cite{Mycielski:1984_SSC}. Similarly, Fe impurities do not appear appropriate, as they exhibit Van Vleck's paramagnetism in HgTe \cite{Serre:2006_Pr}. Guided by the early theoretical prediction on the ferromagnetic superexchange in Cr-doped II-VI compounds \cite{Blinowski:1996_PRB} as well by the observation of ferromagnetism at low temperatures in Zn$_{1-x}$Cr$_x$Te \cite{Watanabe:2019_APL}, we consider the Cr and V case.
 \item Sufficiently high Curie temperature $T_{\text{C}}$ at $x$ small enough in the case of QW doping to prevent a transition to the topologically trivial phase. In the case of Hg$_{1-x}$Mn$_x$Te, the inverted band structure disappears at $x_c \simeq 7$\% \cite{Sawicki:1983_Pr} but QW confinement shifts $x_c$ to lower values \cite{Shamim:2020_SA}. However, higher TM doping is possible in the case of barriers.
 \item Isoelectronic character of TM impurities, as charge doping by magnetic ions will hamper shifting of the Fermi level to the QW gap. The internal reference rule \cite{Langer:1988_PRB}, together with the known valence band offsets \cite{Mathieu:1988_PRB,Dietl:1988_PRB,Becker:2000_PRB} and the positions of V levels in CdTe \cite{Selber:1999_SST} and Cr levels in CdTe \cite{Cieplak:1975_pssb} and ZnTe \cite{Kuroda:2007_NM} suggest that the relevant TM$^{2+/3+}$ donor level resides in the HgTe conduction and valence band, respectively. This indicates that V in either (Cd,Hg)Te barriers or HgTe QWs might act as an electron dopant, while Cr as an isoelectronic impurity.
  \item Formation of a single chiral edge channel in the presence of spin-polarized magnetic ions. According to the pioneering theoretical analysis \cite{Liu:2008_PRLb}, the QAHE shows up in magnetically-doped HgTe quantum wells if $p =-\alpha/\beta \gtrsim 0.25$, where $\alpha>0$ and $\beta<0$ are the $s$-$d$ and $p$-$d$ exchange integrals, respectively. As discussed in the companion paper \cite{Sliwa:2023_arXiv}, this condition can be relaxed by tilting the magnetization vector away from the growth direction, as in such a magnetization orientation, spin-orbit interaction diminishes spin-splitting of heavy-hole-like subbands and reduces an effective $|\beta|$ value.
 \item Single-ion magnetic anisotropy. Except for Mn$^{2+}$ for which orbital momentum is zero, magnetic ions such as V$^{2+}$ and Cr$^{2+}$ exhibit sizable single-ion anisotropy enlarged by the Jahn-Teller distortion. Accordingly, properties of HgTe QWs with those dopants are expected to be more sensitive to epitaxial strain than (Hg,Mn)Te QWs.
 \end{enumerate}

Our {\em ab initio} approach to the issues listed above is built on the previous extensive first-principles investigations of non-topological semiconductors \cite{Sato:2010_RMP,Dietl:2015_RMP} and topological Bi-Sb chalcogenides doped with transition metals \cite{Zhang:2012_PRL,Vergniory:2014_PRB, Bouaziz:2018_PRB}, as well as on results obtained recently for a wide family of Eu compounds, such EuCd$_2$As$_2$ and EuCd$_2$Bi$_2$ \cite{Cuono23Eu}. 
Those studies have demonstrated that strong correlations on localized magnetic ions and narrow band gaps specific to topological systems require a careful selection and testing of the density-functional theory (DFT) implementation.

In this paper, we exploit a range of {\em ab initio} methods to asses three aspects of HgTe doped with 12.5\% of cation-substitutional V and Cr ions.
First, the concentration of V and Cr opens the band gap in (Hg,TM)Te.  As mentioned, in the case of Hg$_{1-x}$Mn$_x$Te, the transition between the topological and non-topological phase occurs for $x_c =0.07$ but we
find that $x_c$ can be larger for V and Cr.

Second, the positions of states brought about by V and Cr ions in CdTe and HgTe in comparison to the better-known case of Mn doping. Our results indicate that V impurities, in agreement with experimental results, act as mid-gap donors in CdTe but are close to the bottom of the conduction band in HgTe, so their isoelectronic character has to be checked experimentally. In contrast, Cr impurities do not provide carriers, as the relevant donor state is close to the valence band maximum in CdTe and deeper in the valence band of HgTe. However, compared to the Mn case, Cr donor level is much closer to the Fermi energy resulting in a relatively high magnitude of the $p$-$d$ exchange integral $|\beta|$ compared to the (Hg,Mn)Te value.

Third, the magnitudes and signs of coupling between magnetic ions in those systems. We conclude that ferromagnetism dominates in the case of V ions, whereas there is a competition between ferromagnetic and antiferromagnetic coupling in the case of neighbor Cr spins.

\begin{figure}[tb]
\includegraphics[width=0.3\textwidth,angle=270]{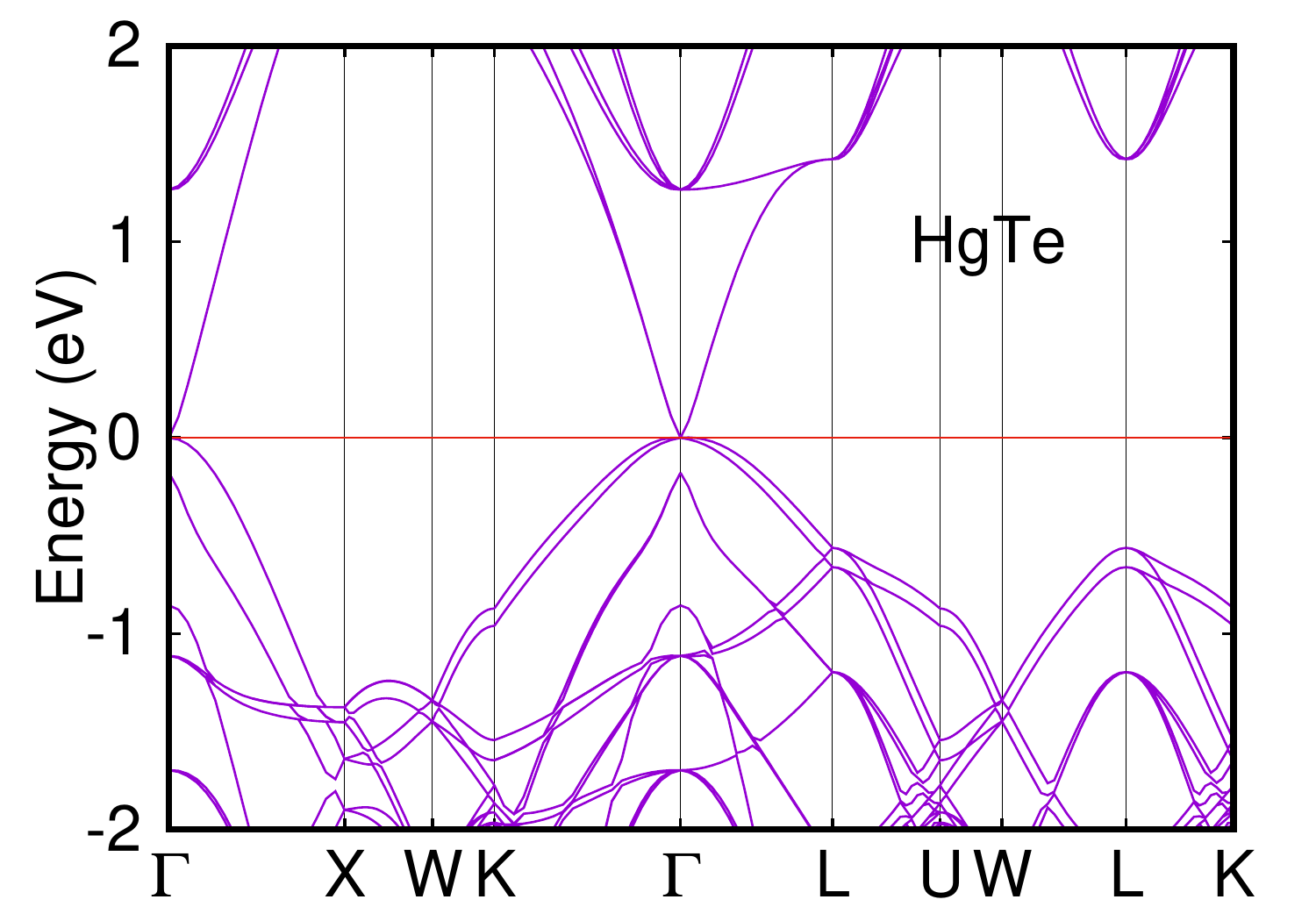}
\caption{HSE band structure for  HgTe;  the value of $a_{\text{HSE}}=0.32$ has been used.
$\Gamma_6$, $\Gamma_7$ and $\Gamma_8$ are the irreducible representations of the states at the $\Gamma$ point of the Brillouin zone taking into account the spin-orbit interaction. $\Gamma_6$ (s = 1/2) represents states that transform under the point group operations like $s$ orbitals; $\Gamma_7$ (j = 1/2) and $\Gamma_8$ (j = 3/2) represent states that transform as $p$-orbitals. The band inversion is determined by the energy difference E$_{\Gamma_8}$ $-$ E$_{\Gamma_6}$ \cite{Islam22,PhysRevB.107.125102}. The Fermi level is set to zero.}
\label{HgTe_HSE}
\end{figure}

\begin{figure*}[t!]
\centering
\includegraphics[width=0.23\textwidth,angle=270]{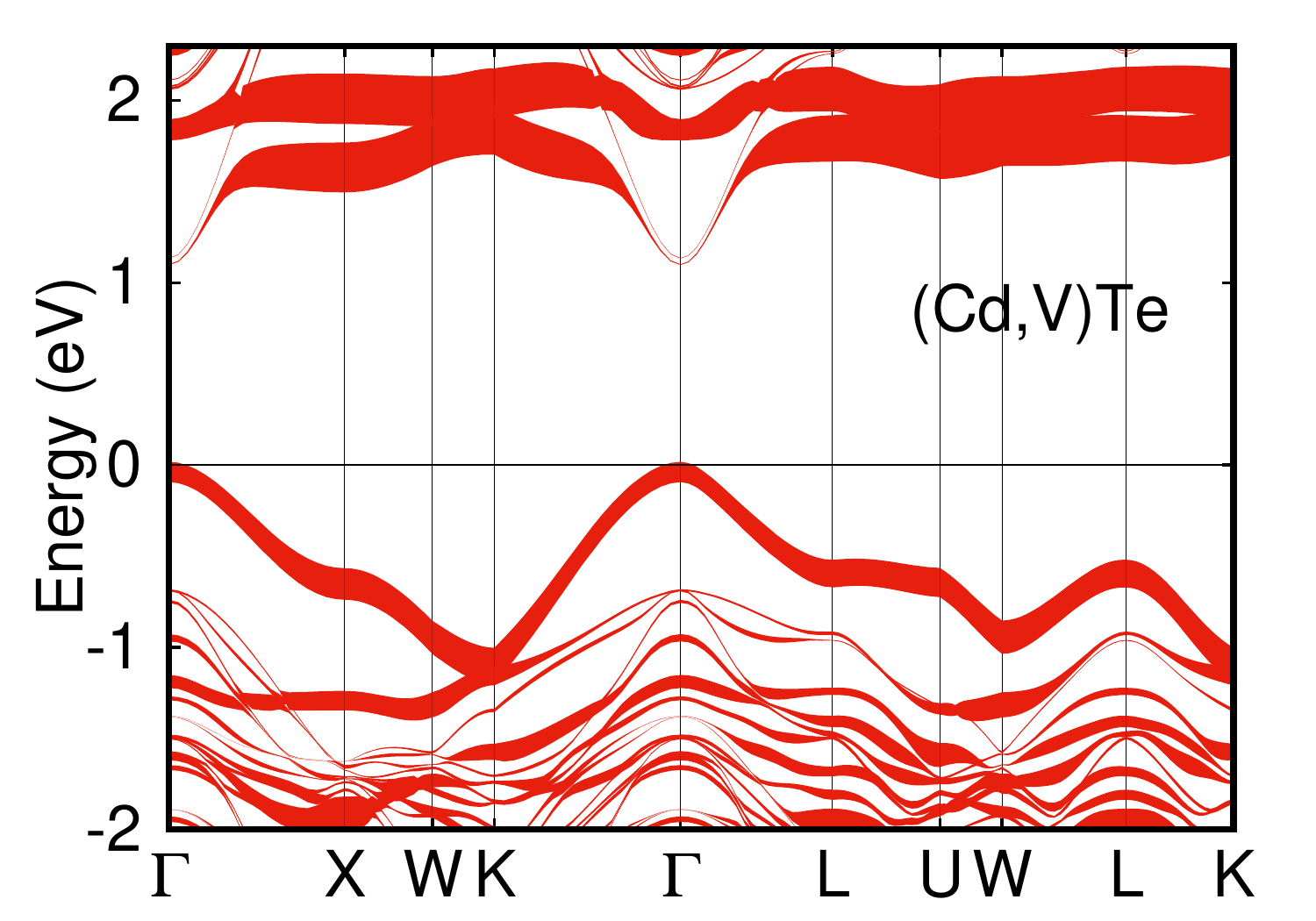}
\includegraphics[width=0.23\textwidth,angle=270]{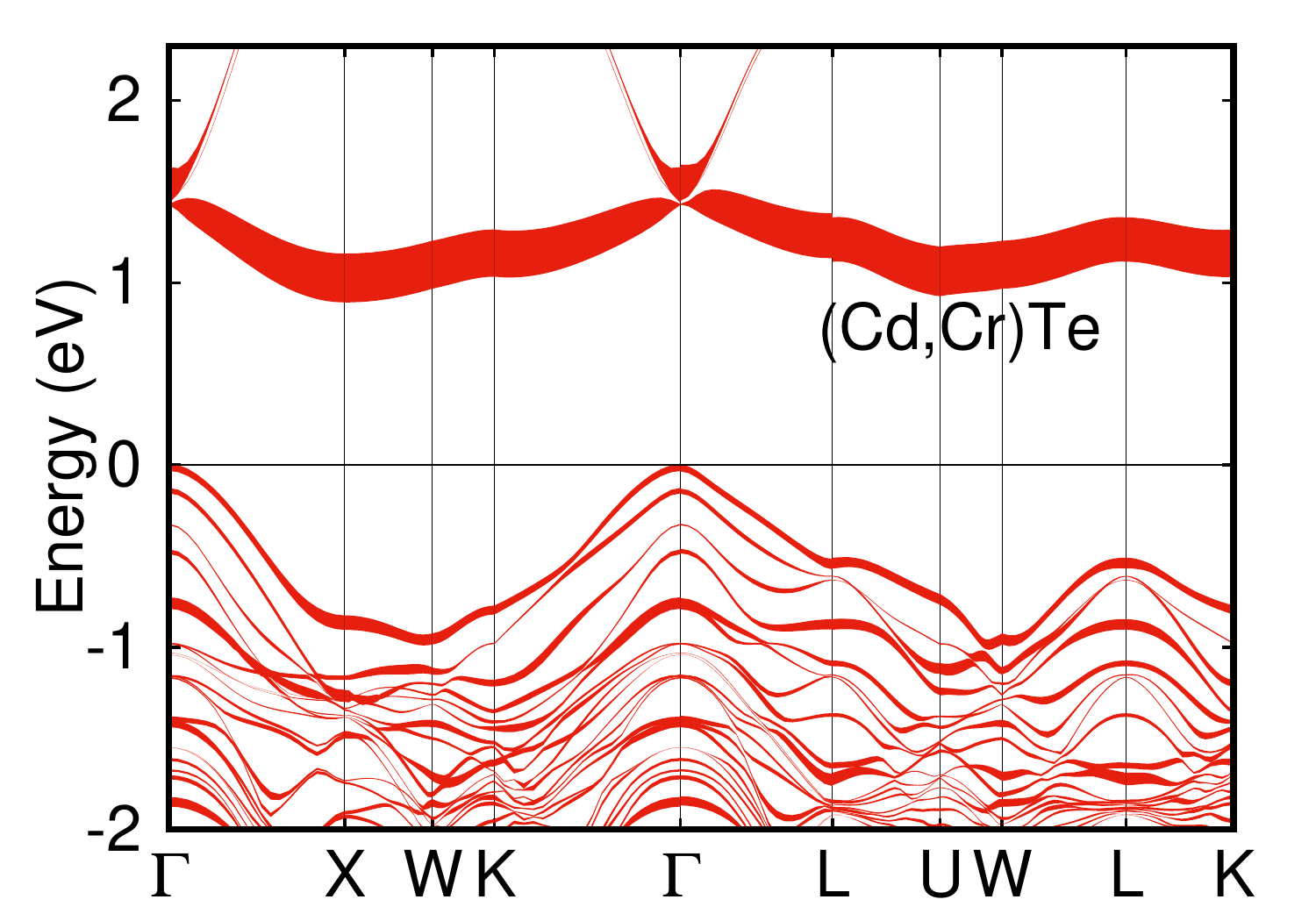}
\includegraphics[width=0.23\textwidth,angle=270]{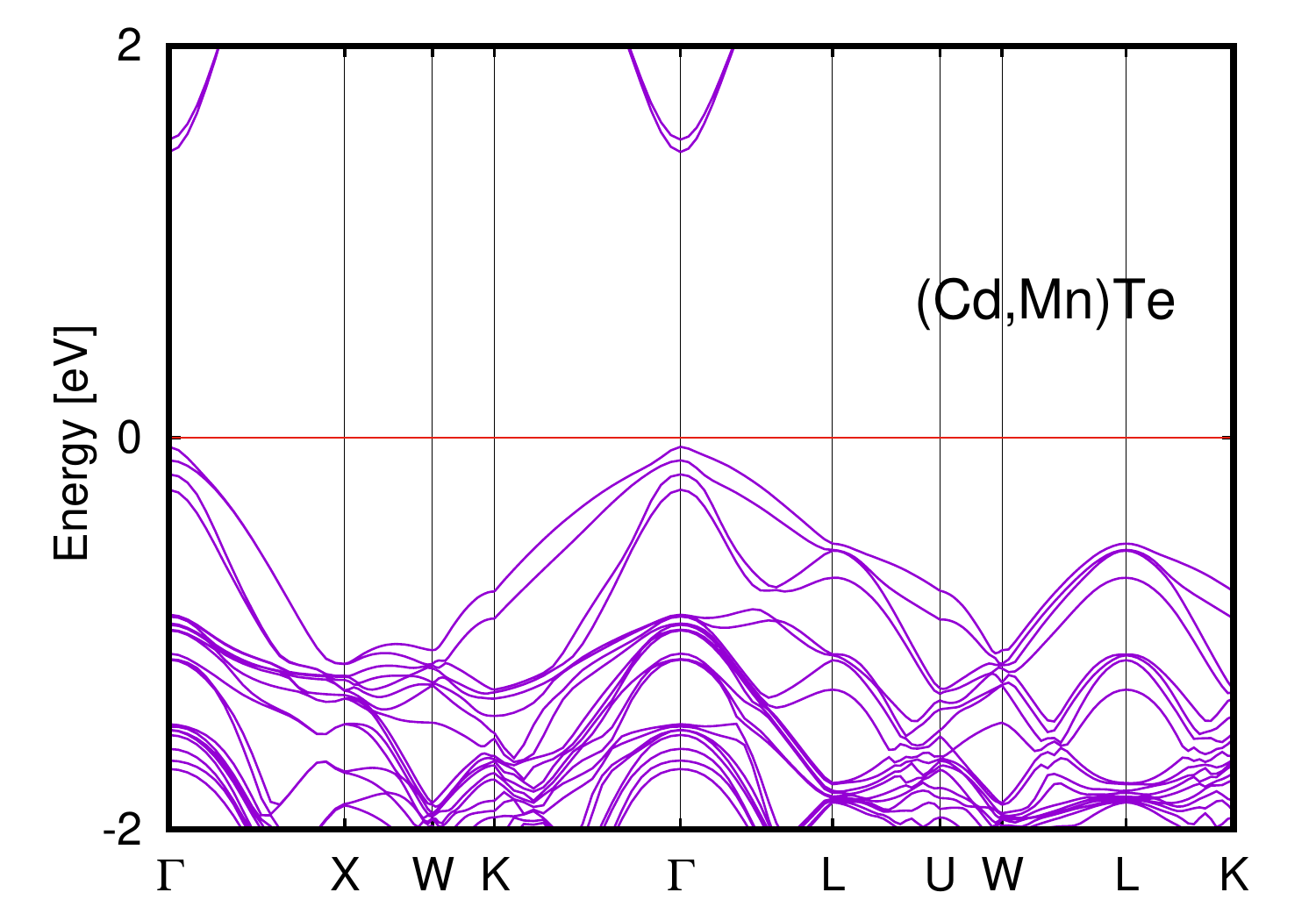}
\caption{HSE band structure for  Cd$_{0.875}$V$_{0.125}$Te and Cd$_{0.875}$Cr$_{0.125}$Te  in
comparison to Cd$_{0.875}$Mn$_{0.125}$Te, where the Mn donor- and acceptor-like $d$ states are relatively far from
the Fermi energy, below the top of the valence band and above the bottom of the conduction band, respectively.  The line thickness describes
a relative contribution of the V and Cr $d$ states. The Fermi level is set at zero energy
and appears to be pinned by the in-gap V donor-like $d$ states. Spins of magnetic ions are ordered ferromagnetically along the [001] direction, which results in a $k$-dependent exchange spin-splittings of bands.
The value of $a_{\text{HSE}}=0.32$ has been used. }
\label{CdTe_HSE_fat}
\end{figure*}

\begin{figure*}[t!]
\centering
\includegraphics[width=0.23\textwidth,angle=270]{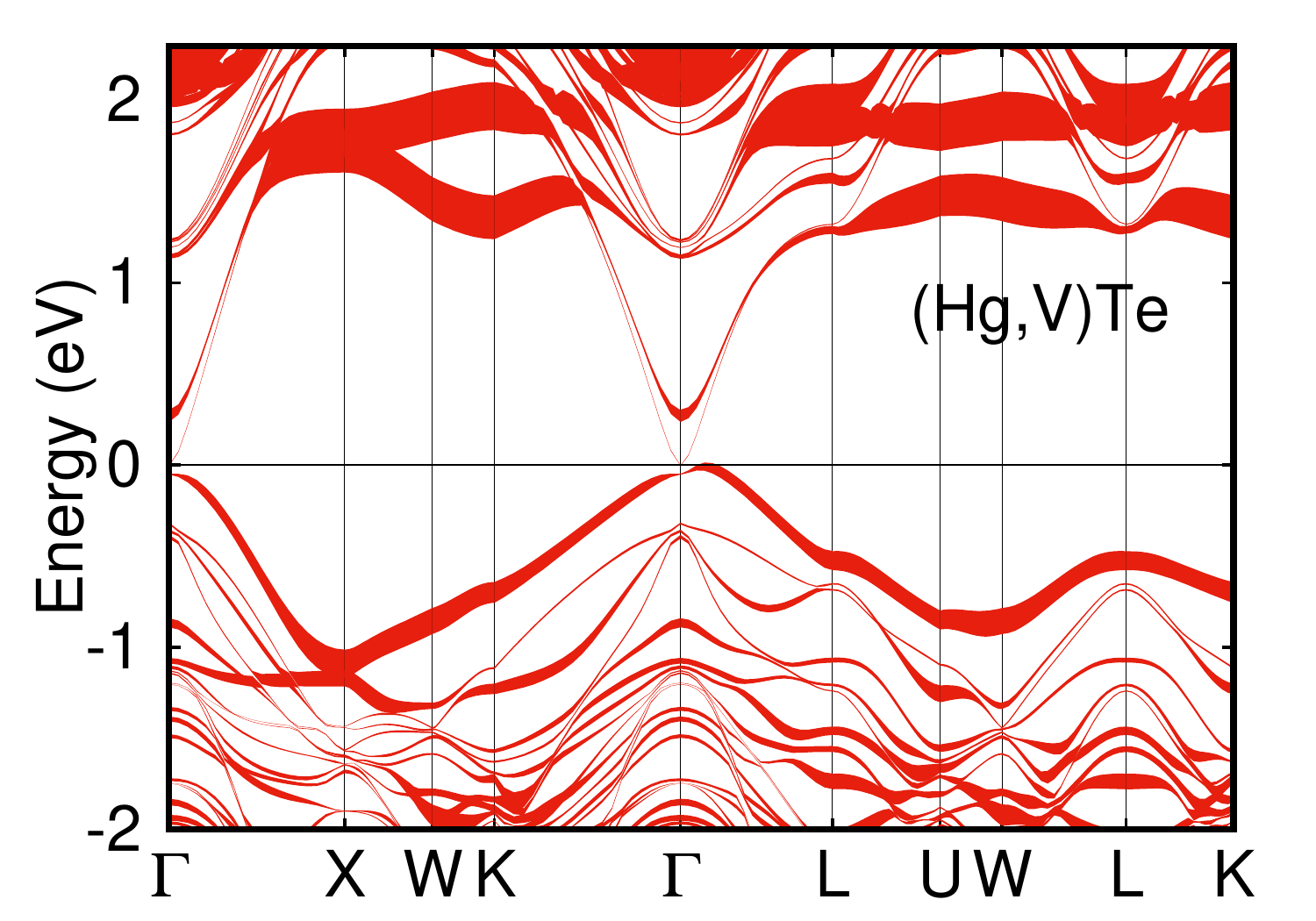}
\includegraphics[width=0.23\textwidth,angle=270]{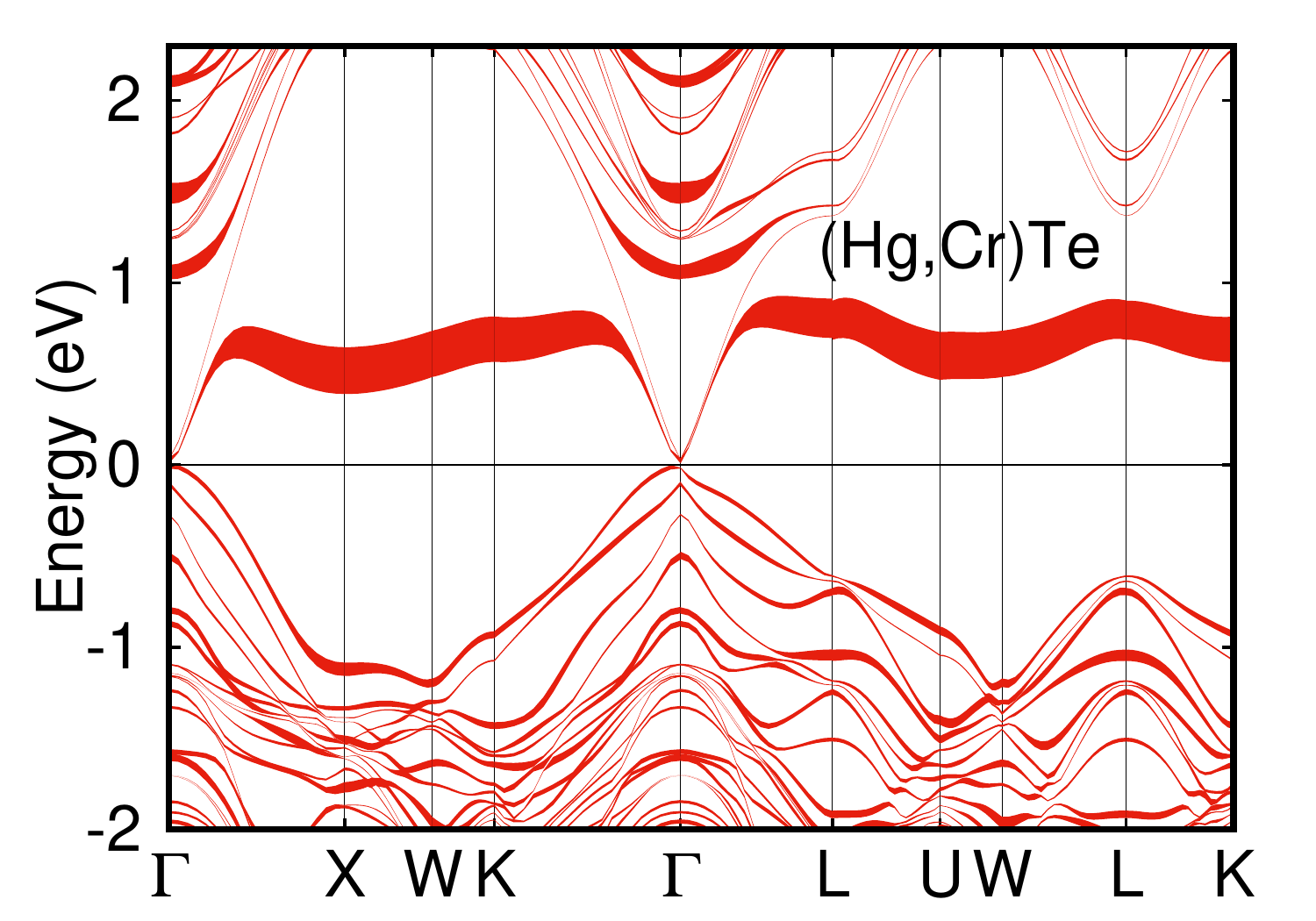}
\includegraphics[width=0.23\textwidth,angle=270]{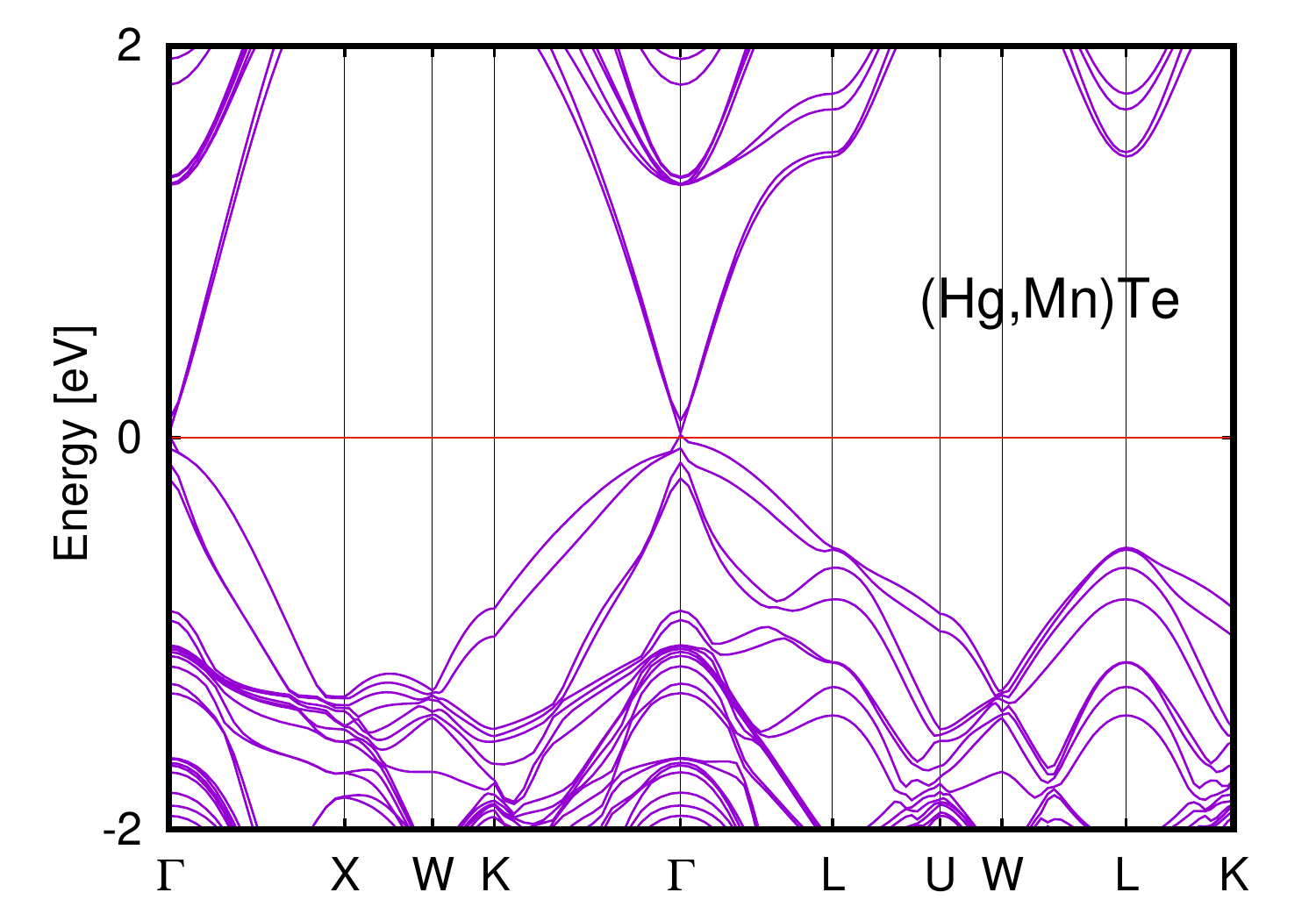}
\caption{HSE band structure for Hg$_{0.875}$V$_{0.125}$Te and Hg$_{0.875}$Cr$_{0.125}$Te in comparison to
Hg$_{0.875}$Mn$_{0.125}$Te, where the Mn donor- and acceptor-like $d$ states are relatively far from the Fermi level, below the top of the valence band and above the bottom of the conduction band, respectively.
The line thickness describes
a relative contribution of the V and Cr $d$ states.  The Fermi level is set at zero energy and appears to be pinned by the V donor-like $d$ states. Spins of magnetic ions are ordered ferromagnetically along the [001] direction, which results in a $k$-dependent exchange spin-splittings of bands. The value of $a_{\text{HSE}}=0.32$ has been used.}
\label{HgTe_HSE_fat}
\end{figure*}

\section{Computational details}

In order to describe properly the band gap and strong correlation of electrons in the TM $d$ bands, we have used the  Heyd-Scuseria-Ernzerhof 2006 (HSE06) hybrid functional\cite{Paier06}. The mixing parameter is often set to around 0.20-0.30, and the value $a_{\text{HSE}}=0.25$ was used for Co-doped  CdTe \cite{Dou21}. Other II-IV semiconductors have been studied with $a_{\text{HSE}}$ up to 0.36 \cite{Lyons2017,Kavanagh2021}. We have carried out computations for $a_{\text{HSE}}= 0.25$, 0.32, and 0.5 with spin-orbit coupling (SOC) taken into account. Guided by our results presented in Sec.~\ref{sec:bands} and Appendix A, we focus on data for $a_{\text{HSE}}=0.32$, which reproduces with accuracy of 0.1 eV experimental band gaps of CdTe, HgTe, Cd$_{0.875}$Mn$_{0.125}$Te, and  Hg$_{0.875}$Mn$_{0.125}$Te.
We have performed the band structure calculations using a plane-wave energy cutoff of 400\,eV and an 8$\times$8$\times$8 $k$-points grid centered in $\Gamma$ with 512 $k$-points in the Brillouin zone,
adopting the experimental lattice parameters,  $a_0 = 6.46152$\,{\AA} for HgTe and 6.4815\,{\AA} for CdTe  \cite{Skauli01} and considering cation-substitutional transition metal content of 12.5$\%$ for all the compounds investigated. We put the spin polarization along the [001] direction. The band structure of undoped HgTe obtained with $a_{\text{HSE}}=0.32$ is shown in Fig.~\ref{HgTe_HSE}. For comparison, we have also performed computations within the GGA +$U$ approach. The obtained results
are most similar to those obtained for $a_{\text{HSE}}=0.25$.

Since V and Cr atoms are distributed periodically in supercells, the V- and Cr-derived states form bands. In reality, for randomly distributed magnetic ions, Anderson-Mott localization will result, at least at low magnetic ion concentrations, in strongly localized band gap levels or resonant states, which have a donor or acceptor character, and correspond to $d^n/d^{n-1}$ or $d^n/d^{n+1}$ states, respectively, where $n = 3, 4, 5$ for V, Cr, and Mn, respectively. Furthermore, for the band structure determination, we arrange TM spins ferromagnetically along the [001] direction. Such a TM spins' configuration leads to $sp$-$d$ exchange splitting of bands, which is $\bm{k}$-dependent due to the interplay of the $sp$-$d$ exchange with $kp$ and spin-orbit interactions\cite{Autieri21}.

\begin{figure}[t!]
\centering
\includegraphics[width=4.7cm,angle=270]{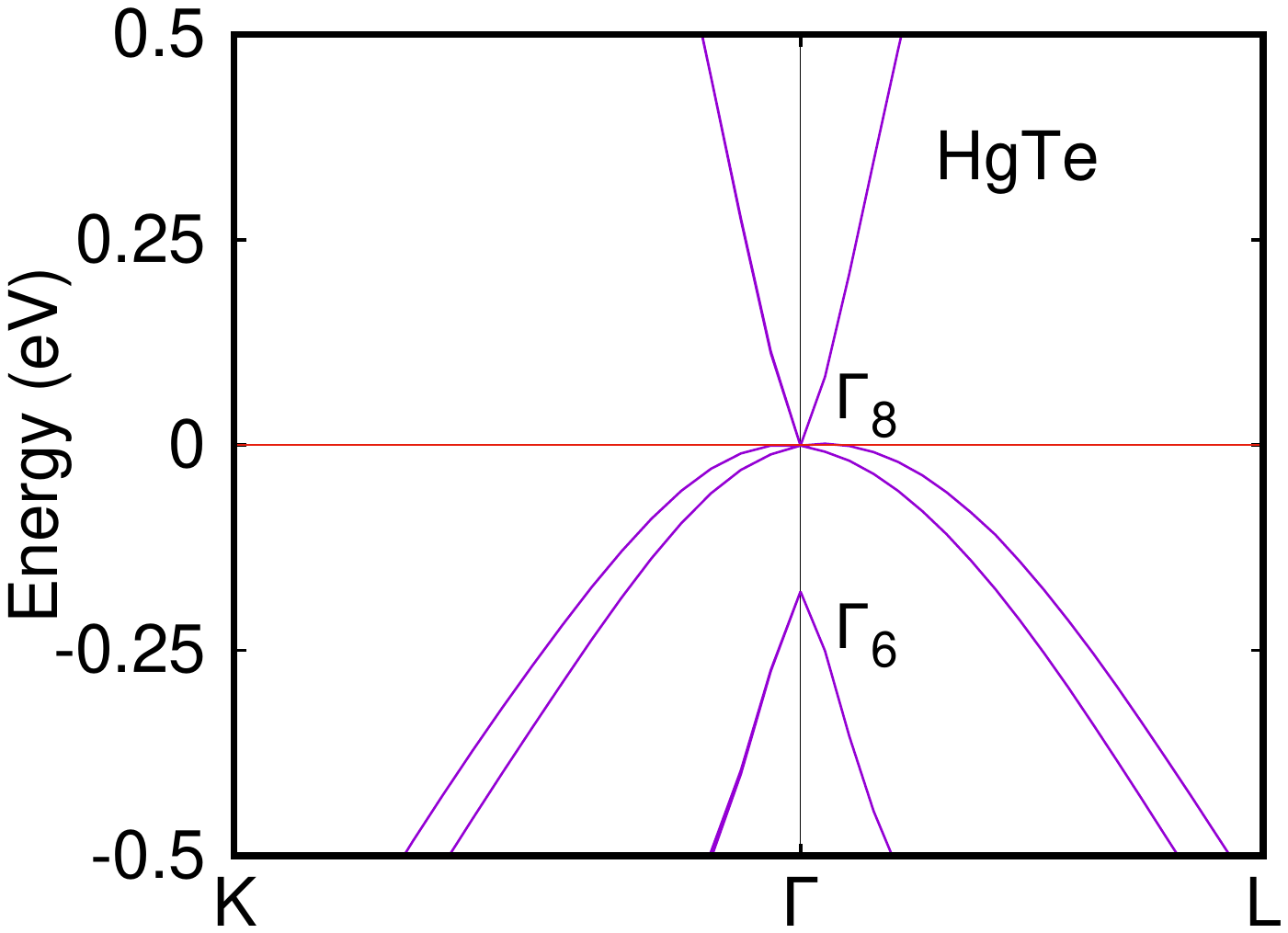}
\includegraphics[width=4.7cm,angle=270]{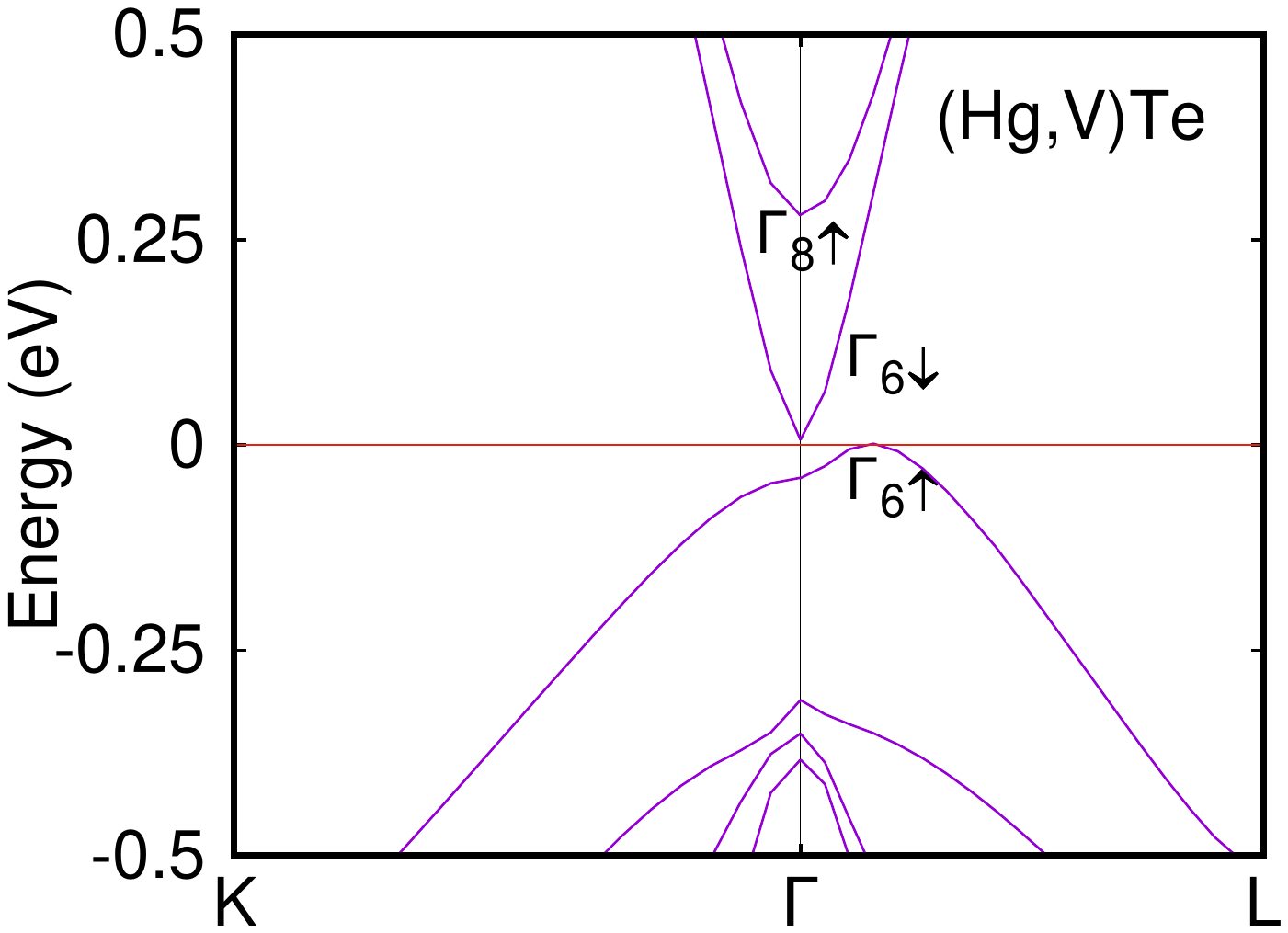}
\includegraphics[width=4.7cm,angle=270]{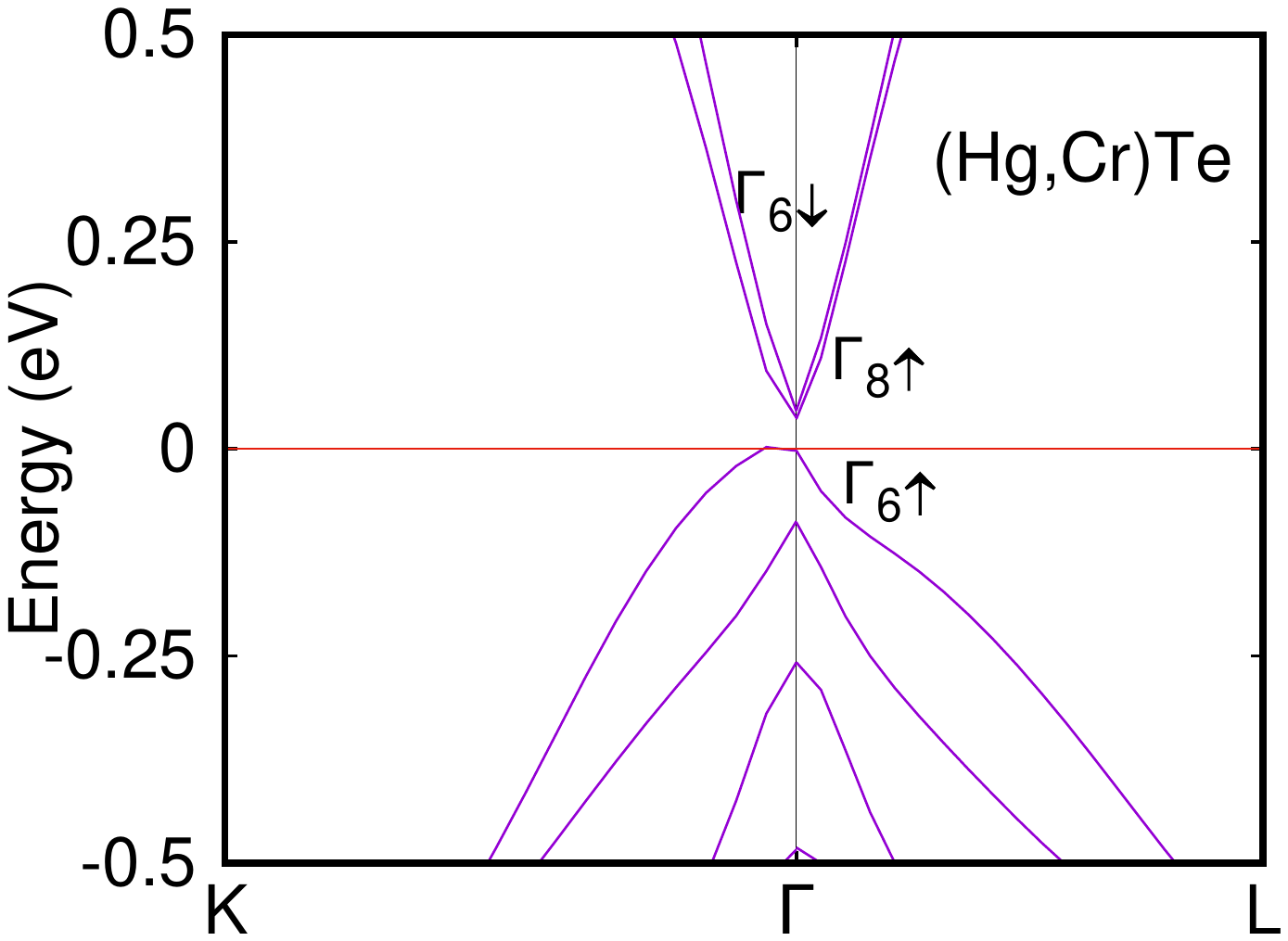}
\includegraphics[width=4.7cm,angle=270]{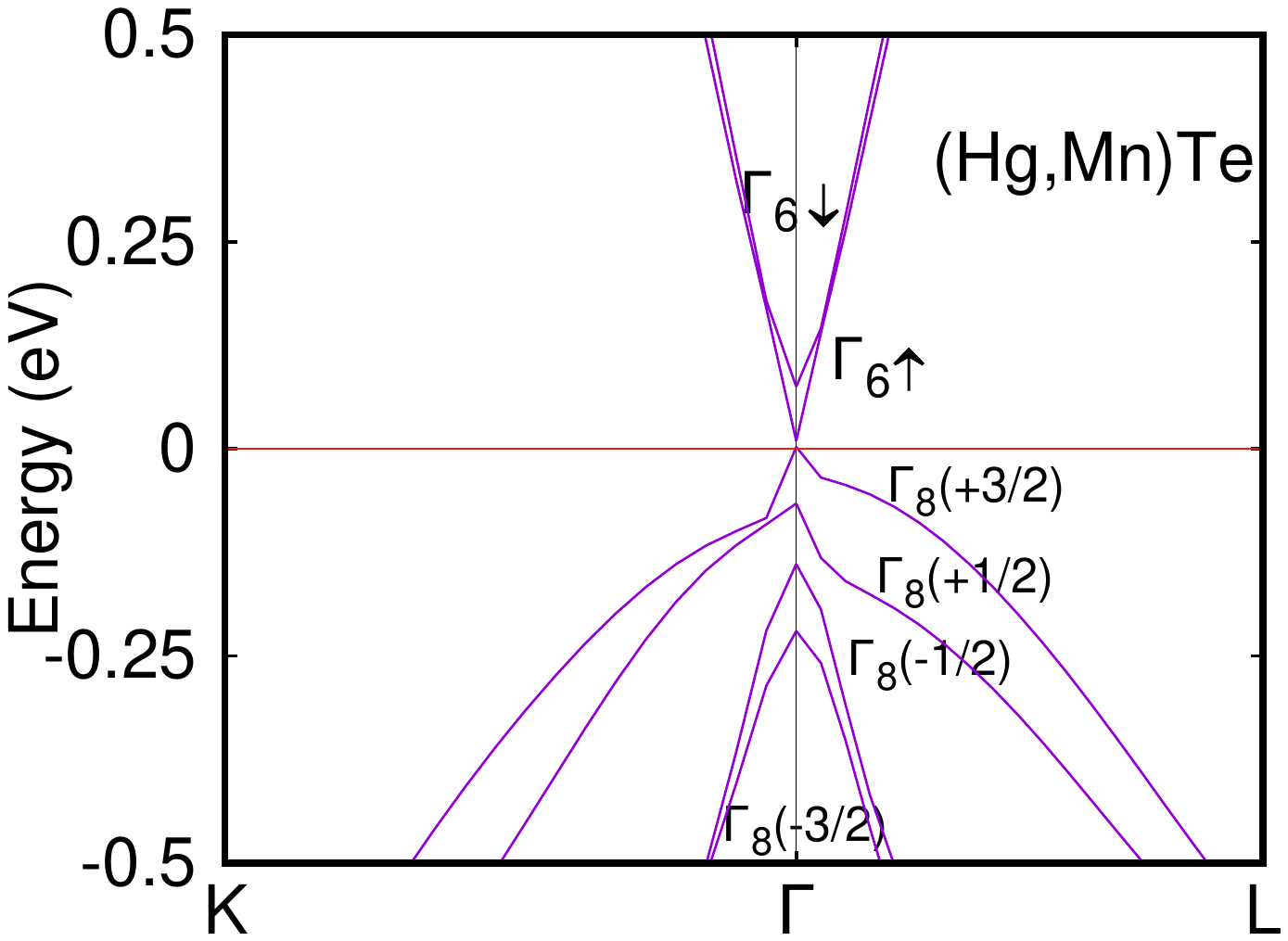}
\caption{Zoom of the HSE band structure along the lines K-$\Gamma$-L for  HgTe, and for Hg$_{0.875}$V$_{0.125}$Te, Hg$_{0.875}$Cr$_{0.125}$Te and Hg$_{0.875}$Mn$_{0.125}$Te, with a ferromagnetic spin arrangement along the [001] direction; the value of a$_{HSE}$ = 0.32 has been used. Labels show two $\Gamma_6$ (s$_z$ = $\pm$1/2) and four $\Gamma_8$ multiplets (j$_z$ =$\pm$3/2, $\pm$1/2), generated by sp-d exchange coupling of carrier bands to the V, Cr, and Mn localized spins.  The Fermi level is set at zero energy.}
\label{Zoom_HgTe}
\end{figure} 


\section{Band structures without distortions and prospects for QAHE}
\label{sec:bands}
We start by showing
electronic properties of the CdTe and HgTe doped with V and Cr without taking into account the distortions produced by the structural relaxation and assuming ferromagnetic ordering of TM spins.
The band structures obtained within the HSE approach with $a_{\text{HSE}}=0.32$ for doped CdTe and HgTe with SOC interaction included and for ferromagnetic and periodic spin arrangement in comparison to the Mn-doped  CdTe and HgTe are shown in Figs.~\ref{CdTe_HSE_fat} and ~\ref{HgTe_HSE_fat}, respectively.
In contrast to the Mn case, where $d$ states are high in the conduction band and deep in the valence band, the states derived from V and Cr $d$ levels reside close to the bottom of the conduction band and the top of the valence band. In accord with experimental observations for CdTe:V \cite{Selber:1999_SST}, the uppermost occupied $V$-derived donor band in Cd$_{0.875}$V$_{0.125}$Te lies in the middle of the band gap and the lowest
unoccupied V-derived acceptor band near the bottom of the conduction band. In the case of Cd$_{0.875}$Cr$_{0.125}$Te, the occupied Cr-derived $d$ donor bands are strongly hybridized with the host valence band, whereas the Cr unoccupied acceptor band
is near the bottom of the conduction band, again in accord with experimental data for CdTe:Cr \cite{Cieplak:1975_pssb}. In Fig. \ref{Zoom_HgTe}, a zoom of the bands close to the Fermi level is reported for HgTe and for HgTe doped with V, Cr, and Mn assuming a ferromagnetic ordering of the TM spins. The exchange splitting of the $\Gamma$ bands into six components corresponding to $\Gamma_6$ (s$_z$ = $\pm$ 1/2) and  $\Gamma_8$ (j$_z$ =$\pm$ 3/2, $\pm$ 1/2) levels,  is clearly seen. Interestingly, the subband arrangement and the magnitude of splitting differ for particular magnetic ions. In particular, the splitting of the $\Gamma_8$ band is seen to be larger in (Hg,V)Te and (Hg,Cr)Te compared to the (Hg,Mn)Te. We assign this difference to the proximity of the V and Cr $d$ levels to the top of the valence band, which enhances the $p$-$d$ hybridization and the $p$-$d$ Schrieffer-Wolff kinetic exchange, as discussed in detail in the companion paper \cite{Sliwa:2023_arXiv}.   At the same time, the $s$-$d$ splitting of the $\Gamma_6$ band is smaller and roughly independent of the TM ion, as could be expected for the potential exchange.
Another relevant information concerns the relative position of the $\Gamma_6$ and $\Gamma_8$ bands. In the case of HgTe, the band structure is inverted, i. e., the $\Gamma_6$ band is below the $\Gamma_8$ bands, which accounts for topological properties of HgTe and its quantum structures. According to results in Fig. \ref{Zoom_HgTe}, Cr and Mn doping shifts the $\Gamma_6$ subbands up with respect to the $\Gamma_8$ subbands, the fact already known experimentally for (Hg,Mn)Te \cite{Furdyna:1988_JAP} and (Hg,Mn)Te QWs \cite{Shamim:2020_SA},  meaning that QWs with Cr content below 10$\%$ can show topological properties. Furthermore, $k$-dependent exchange splitting of bands leads to an overlap of the valence and conduction band states for certain $k$ directions in TM-doped HgTe, which points to outstanding carrier transport properties in and near the topological regime in those dilute magnetic systems. Prospects, in the light of our results, of HgTe doped with TM ions for studies and applications of the quantum anomalous Hall effect are discussed in Sec. VI.

A close proximity of the $d$ states to the top of the valence band results in a large magnitude of the $p$-$d$ exchange integral $\beta$ and, therefore, in a larger splitting of the topmost valence band subbands near the $\Gamma$ point of the Brillouin zone in the Cr case compared to Mn doping, the effect clearly visible in Fig.~\ref{CdTe_HSE_fat}. In Fig. \ref{Zoom_HgTe} a zoom of the bands of HgTe doped with V and Cr and close to the Fermi level is reported. The splitting of the $\Gamma$ bands is shown. There is a band inversion between the $\Gamma_{6}$ and the $\Gamma_{8}$ bands of the spin-up subsector. The topological invariant, which is the Chern number in this case, is extremely challenging to compute within the DFT approach for a strongly-correlated system with a supercell due to the doping. However, the detailed calculation of the Chern number $\mathcal{C}$ was performed for the same system within the tight-binding model in the companion paper\cite{Sliwa:2023_arXiv} reporting the expected value of $\mathcal{C}$=1. The energy difference between $\Gamma_{6}\downarrow$ and $\Gamma_{6}\uparrow$ is 46.6 meV for Hg$_{0.875}$V$_{0.125}$Te and 47.9 meV for Hg$_{0.875}$Cr$_{0.125}$Te. In accord with experimental results for Hg$_{1-x}$Mn$_{x}$Te, the band gap at the $\Gamma$ point in a paramagnetic phase, i.e., the distance between the centers of two split conduction subbands and four valence subbands implies normal band ordering with a gap of 0.2\,eV. A similar band gap is implied by our results for Hg$_{0.875}$Cr$_{0.125}$Te meaning that QWs with Cr concentrations below 10\% can show the QAHE.  Furthermore, $\bm{k}$-dependent exchange splitting of bands leads to an overlap of the valence and conduction band states for certain $\bm{k}$ directions in TM-doped HgTe, which points to the presence of outstanding carrier transport properties in and near the topological regime in those dilute magnetic systems.

\begin{table*} 
    \caption{Bond lengths at 12.5\% concentration of magnetic doping. We report the bond connecting the magnetic atoms M (V or Cr) and Te atoms, angles, polyhedral volume (PV), bond angle variance (BAV), distortion index (DI) of the structures with the distortions produced by the Jahn-Teller effect and energy difference between the undistorted $E_{u}$ and distorted structure $E_{d}$. The bonds M-Te$_{3}$ and M-Te$_{4}$ have equal lengths. $\alpha$ and $\beta$ are equivalent, if we exchange them the total energy will be the same.}
     \vspace{3pt}
     \label{JahnTeller}
    \begin{tabular}{c|c|c|c|c}
    \hline
     & Cd$_{1-x}$V$_x$Te & Cd$_{1-x}$Cr$_x$Te & Hg$_{1-x}$V$_x$Te & Hg$_{1-x}$Cr$_x$Te \\
     \hline
     Bond length ({\AA}) & & &  & \\
     \hline
     M-Te$_{1}$ &2.8019&	2.7622&	2.8028&	2.7282\\
     M-Te$_{2}$ &2.8040&	2.7520&	2.8057&	2.7266\\
     M-Te$_{3}$ &2.7733&	2.7600&	2.7589&	2.7271\\
     \hline
     Angles (deg)	& & & &\\	
    \hline		
     $\alpha$&	102.80&	112.60&	101.13&	109.73\\
    $\beta$&	115.72&	107.64&	117.07&	109.25\\
     \hline
     PV(\AA$^3$)&	11.053&	10.739&	10.932&	10.409\\
    \hline
    BAV (deg$^2$)&	43.054&	2.794&	69.191&	0.025\\
     DI (bond length)&	0.0040&	0.0017&	0.0062&	0.0002\\
     $E_u-E_d$ (meV)&	34.696&	4.400&	45.05&	20.094\\
   \hline
      \end{tabular}
\end{table*}

Previous extensive studies of dilute magnetic materials \cite{Bonanni:2021_HB}, together with the results for Hg$_{0.815}$V$_{0.125}$Te presented in Fig.~\ref{HgTe_HSE_fat} for $a_{\text{HSE}}=0.32$ and also in Appendix A  for $a_{\text{HSE}}=0.25$ and 0.5, indicate that four different scenarios relevant to the QAHE are possible:
 \begin{enumerate}
 \item for certain V concentrations: (i) the band structure remains inverted and (ii) V donor states are in the valence and, therefore, V acts as an isoelectronic impurity, similarly to the case of Mn and, presumably Cr, in HgTe;
 \item V forms a resonant state in the conduction band, similarly to the case of Fe in HgSe \cite{Mycielski:1988_JAP} and Sc in CdSe \cite{Glod:1994_PRB};
 \item V acts as an electron dopant but does not give rise to the presence of resonant states;
 \item substantial hybridization between V $d$ orbitals and band states leads to unusual band ordering and minigaps in the vicinity of the Fermi energy.
 \end{enumerate}
In the case (1), (Hg,V)Te quantum wells can show the QAHE, as--according to our results presented in Sec.~\ref{sec:magnetism}--there is a ferromagnetic interaction between V spins in HgTe. The QAHE could be observed as long the QW is topological, i.e., neither a reduction in the QW thickness nor V doping makes the band ordering topologically trivial. By contrast, within scenarios (2) and (3), it will be difficult to shift the Fermi level from the conduction band to the topological gap for V concentrations sufficiently high to result in a ferromagnetic ground state. Finally, in the fourth case, presumably approximately described by  {\em ab initio} results for Hg$_{0.815}$V$_{0.125}$Te with ferromagnetic and periodic spin arrangement (Fig.~\ref{HgTe_HSE_fat}), hybridization between host and dopant states leads to band reconstruction and hybridization gaps. It is unclear on whether the QAHE is possible under these conditions. Finally, we note that V in the (Cd,Hg,V)Te/HgTe quantum wells may act as a modulation electron dopant.

\begin{figure}[ht]
	\centering
\includegraphics[width=\columnwidth, angle=0]{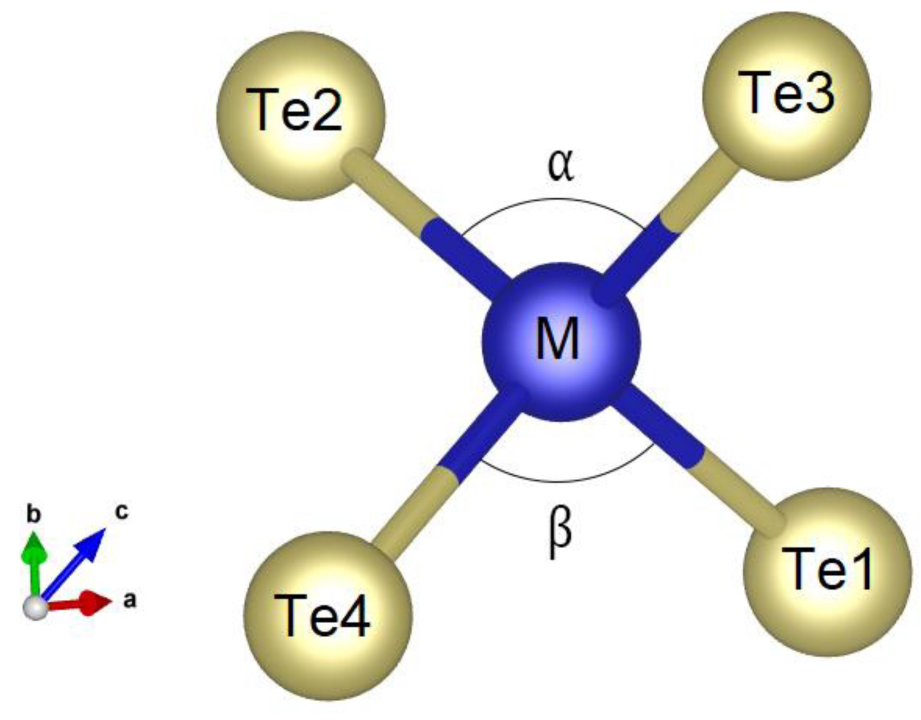}
	\caption{Angles $\alpha$ and $\beta$ that change with the Jahn-Teller distortion. In brown  Te atoms, in blue the central TM atom (Cr or V).
	}
	\label{JT}
\end{figure}

Another situation occurs in (Hg,Cr)Te QWs or (Cd,Hg,Cr)Te barriers, where the Cr donor-like $d$ states reside in the valence band but closer to the $\Gamma_8$ point compared to the Mn case. Accordingly, we expect Cr to be an isoelectronic dopant and to generate larger, compared to Mn doping, exchange splitting of $\Gamma_8$ subbands. These large splittings are seen in Fig.~\ref{HgTe_HSE_fat} which presents computational results for the ferromagnetic arrangement of TM spins along the [001] direction. These results substantiate also the tight-binding model of the (Hg,Cr)Te band structure employed to evaluate the sign and magnitude of exchange interactions between pairs of Cr impurities in HgTe \cite{Sliwa:2023_arXiv}. As tight-binding presented in the companion paper\cite{Sliwa:2023_arXiv} and {\em ab initio} results presented in Sec.~\ref{sec:magnetism} point to a competition between antiferromagnetic and ferromagnetic interactions, the observation of the QAHE may require the application of a magnetic field in order to polarize Cr spins.

\section{Local distortions produced by the Jahn-Teller effect}

While the long-range cooperative Jahn-Teller (JT) effect in homogeneous bulk systems can be described with a finite number of phonon modes as in molecules, the description of the JT effect in solids with JT-active defects involves an infinite number of phonon modes. A particularly tricky case is represented by the case of JT-active impurities in tetrahedra environments like CdTe and HgTe which are analyzed in this paper. 
Recently, Cruz et al.\cite{APSMarch} presented a recipe to tackle this problem in the case of topologically trivial CdSe which is similar to CdTe, while we are not aware of attempting to solve the same problem in topologically non-trivial systems like HgTe.
Therefore, the lattice distortions produced by the JT effect in zinc-blende II-IV semiconductors are not straightforward and the effect can be relevant. For instance, the impurities in zinc-blende ZnSe$_x$S$_{1-x}$ were shown to produce octahedral rotations and changes of bond-angles \cite{PhysRevB.105.184201}.
The JT lifted the degeneracy of the t$_{2g}$ and e$_g$ manifold for both spins in both Cr and V doped systems, this was established through the check of the electronic levels at the $\Gamma$ point. The lifting of the degeneracy of the manifold is necessary to observe an insulating phase.
The main effect is a change in the Te-M-Te bond angle. In regular tetrahedra, we have four angles equal to 109.5 degrees. Among these four angles, two remain almost unchanged in the case of doping, one of these angles becomes smaller and another becomes larger.
In Fig.~\ref{JT} we show the crystal structure and the angles that change with the JT distortion.
The bond angle variance (BAV) is much larger for the V-doping than for the Cr-doping. While the e$_g$ manifold for the majority spin is full, the V-doping consists of 1 electron in the t$_{2g}$ manifold, while the Cr-doping consists of 2 electrons in the t$_{2g}$ manifold. We assume that the BAV is larger for the case of a single electron in the manifold as happens for the d$^4$ in an octahedral crystal field.


When we dope the systems with Cr, the angle $\alpha$ becomes larger and $\beta$ becomes smaller, the opposite happens when we dope with V.
This situation is different from what happens in other systems  \cite{Forte18,Ni20}, where the two angles $\alpha$ and $\beta$ become larger or smaller together.
The JT allows the breaking of the degeneracy of the energy levels compatible with the cubic symmetry of the system.
Two larger angles or two smaller angles will be more efficient in producing the JT splitting, and they will create a local tetragonal atom arrangement.
In Table~\ref{JahnTeller}, we describe the local distortions in different systems and the gain in the total energy resulting from the JT distortion.
These JT distortions help in stabilizing the localized character of states derived from TM $d$ levels.

\section{Exchange couplings between transition metal spins}
\label{sec:magnetism}
In this Section, we present computational results on exchange couplings between TM spins employing $a_{\text{HSE}}=0.32$.
In Table~\ref{HYBRIDJ}, we report energy differences between antiferromagnetic (AFM) and ferromagnetic (FM) configurations of TM spins in CdTe and HgTe doped with  V, Cr, and Mn without taking the Jahn-Teller distortion into account.
As seen, FM couplings prevail for the V case, whereas the interaction is AFM for Cr- and Mn-doped compounds. Once the Jahn-Teller distortion is taken into account, FM couplings show up in the Cr-compounds, too, as shown in Table~\ref{HYBRIDJT}. 
Regarding the V-doped systems, the ground state is still FM if we include the Jahn-Teller distortions, as shown in Table~\ref{HYBRIDJT_V}. 
A comparison of data for $a_{\text{HSE}}=0.32$ presented here to results obtained for $a_{\text{HSE}}=0.25$ and 0.50 summarized in Appendix, demonstrate a strong sensitivity of the coupling strength to the employed theoretical framework. This sensitivity confirms the presence of a delicate balance between FM and AFM contributions to the TM coupling in the case of early transition metals in II-VI compounds \cite{Sliwa:2023_arXiv}.

\begin{figure*}[t!]
\centering
\includegraphics[width=0.23\textwidth,angle=270]{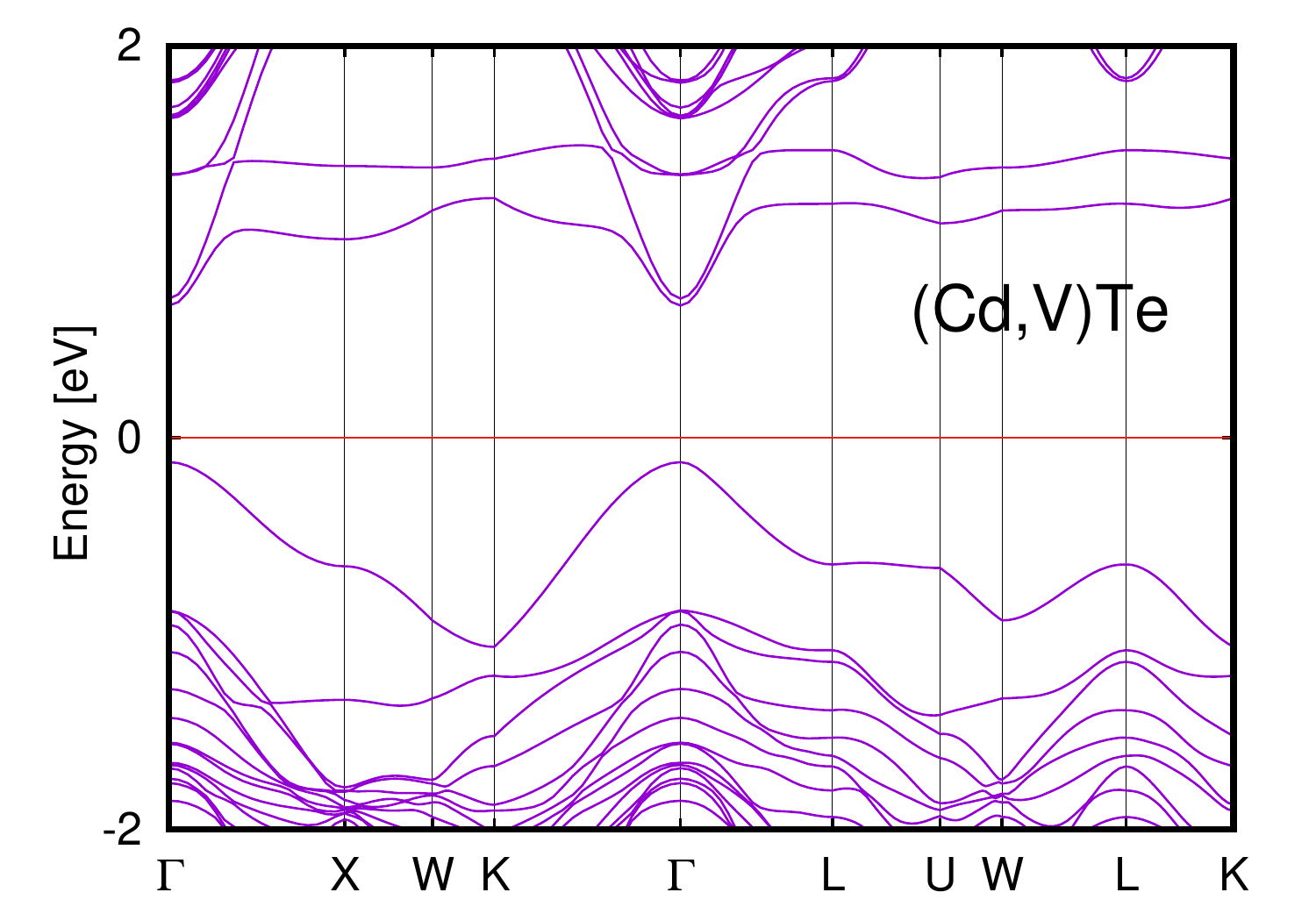}
\includegraphics[width=0.23\textwidth,angle=270]{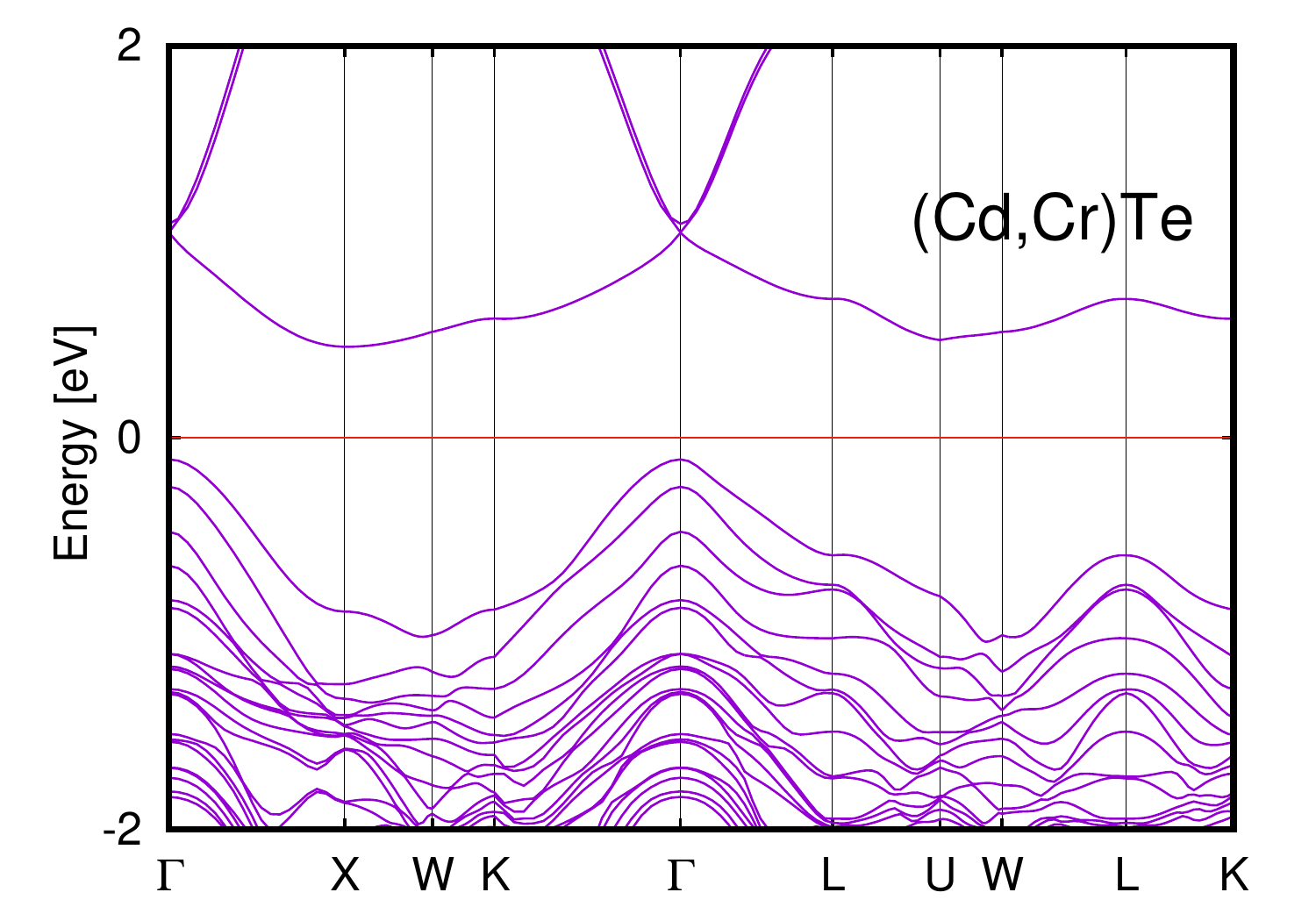}
\includegraphics[width=0.23\textwidth,angle=270]{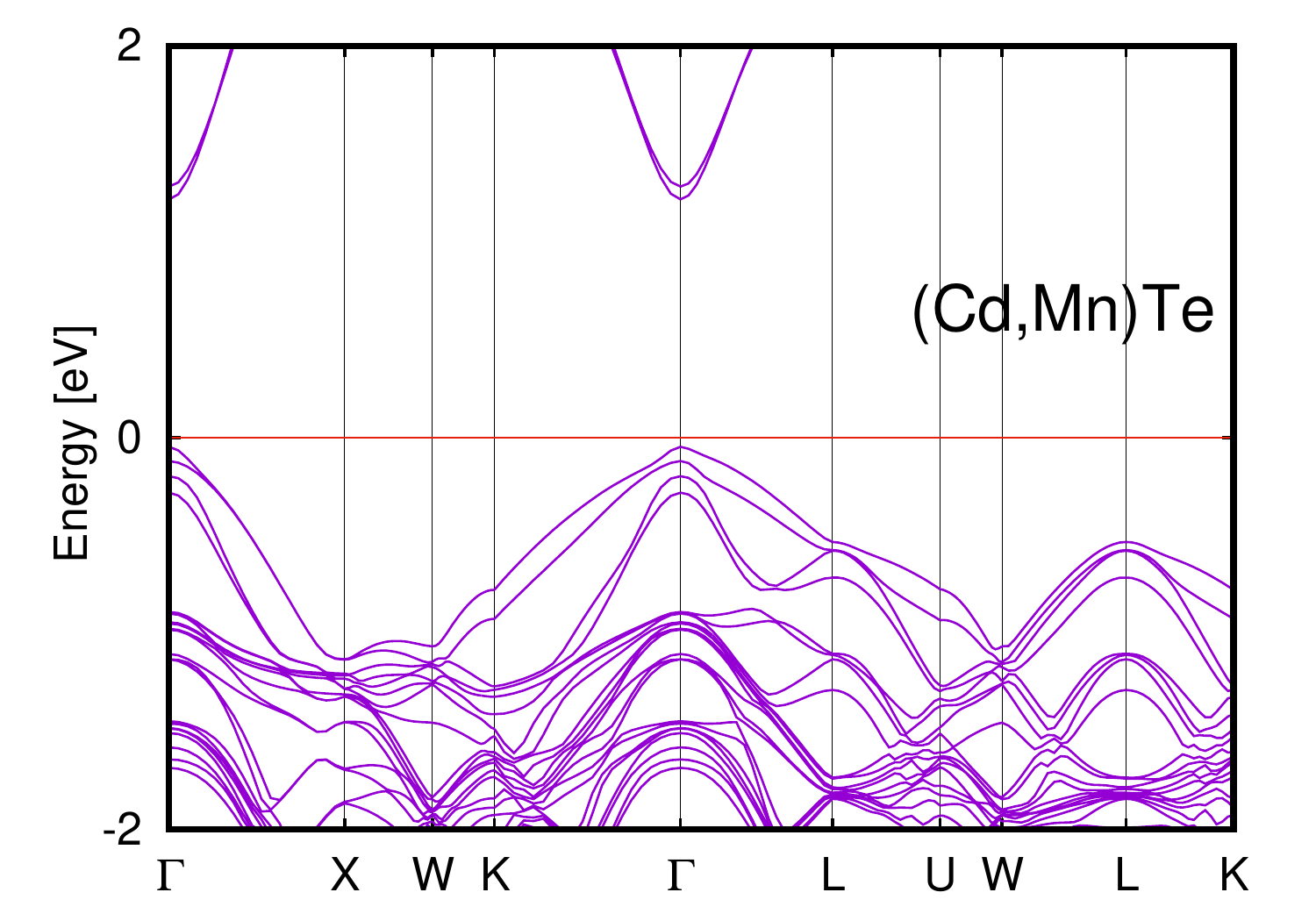}
\caption{HSE band structure for Cd$_{0.875}$V$_{0.125}$Te, Cd$_{0.875}$Cr$_{0.125}$Te, and Cd$_{0.875}$Mn$_{0.125}$Te with ferromagnetic arrangement of TM spins and $a_{\text{HSE}}=0.25$. The Fermi level is set at zero energy.}
\label{CdTe_HSE_025}
\end{figure*} 

\begin{figure*}[t!]
\centering
\includegraphics[width=0.23\textwidth,angle=270]{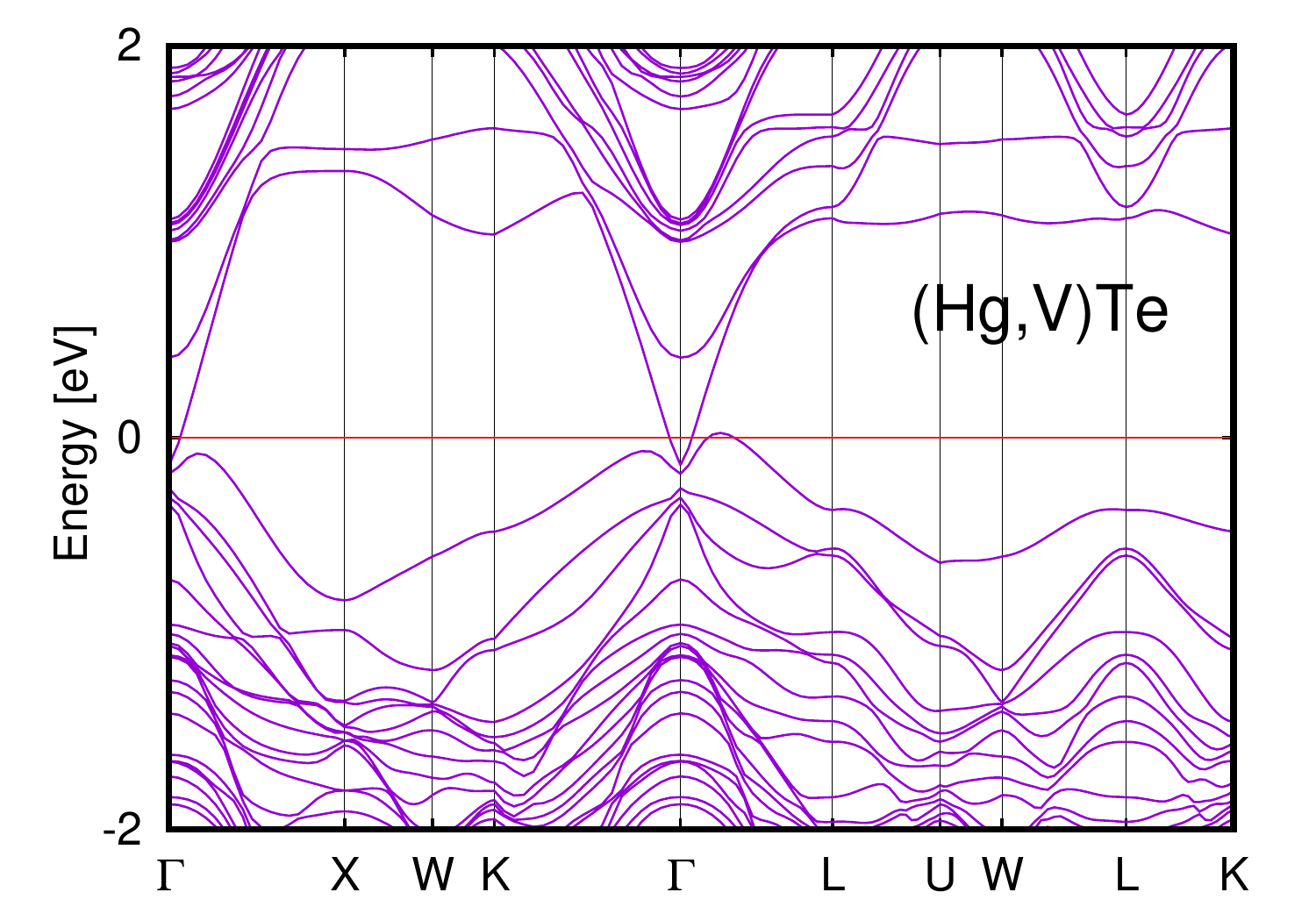}
\includegraphics[width=0.23\textwidth,angle=270]{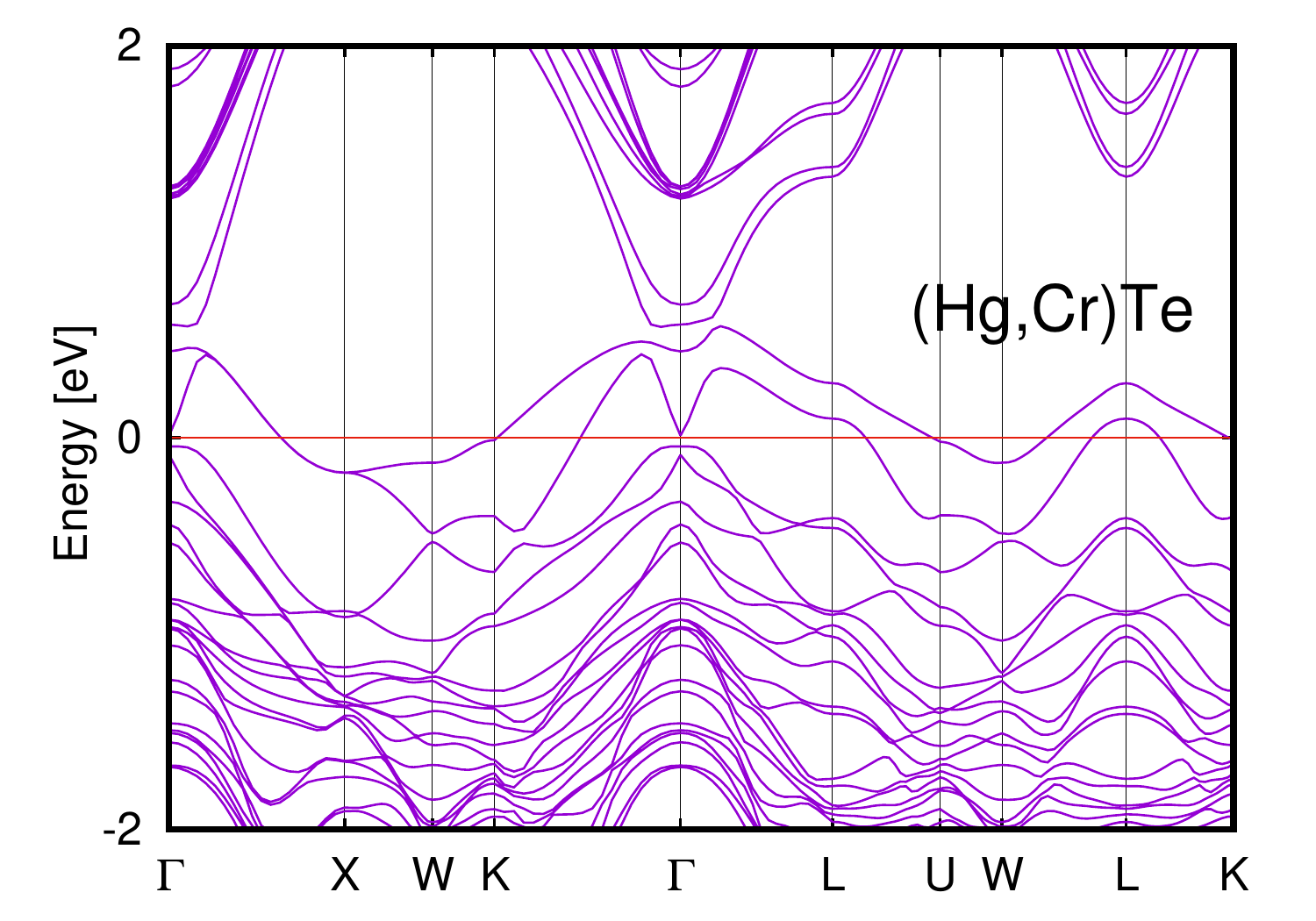}
\includegraphics[width=0.23\textwidth,angle=270]{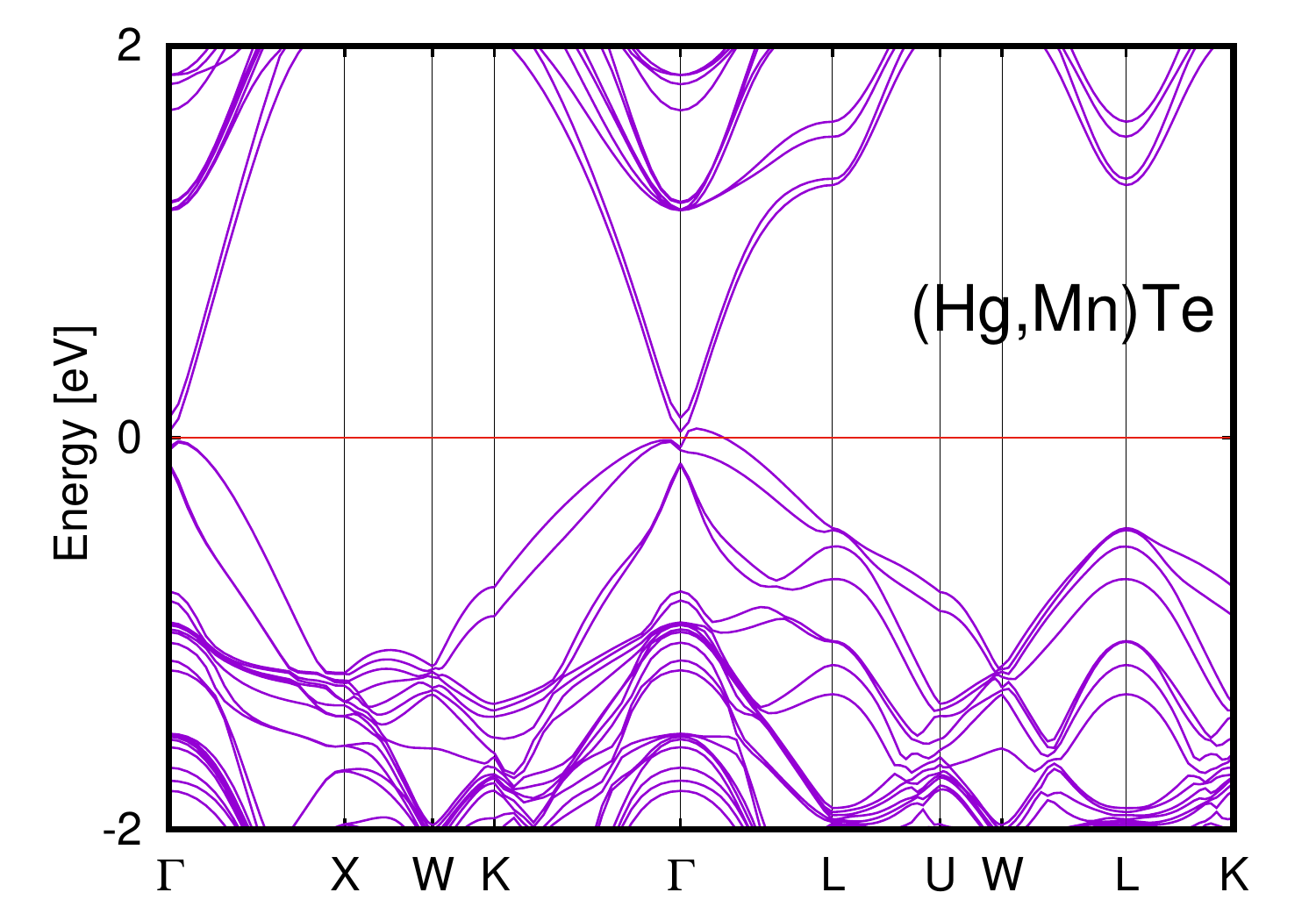}
\caption{HSE band structure for Hg$_{0.875}$V$_{0.125}$Te, Hg$_{0.875}$Cr$_{0.125}$Te, and Hg$_{0.875}$Mn$_{0.125}$Te with ferromagnetic arrangement of TM spins and $a_{\text{HSE}}=0.25$.  The Fermi level is set at zero energy.}
\label{HgTe_HSE_025}
\end{figure*} 

\begin{table}
\begin{center}
    \caption{Energy differences (meV) between AFM and FM configurations for CdTe and HgTe doped with V, Cr, and Mn ions. $\Delta E_1$, $\Delta E_2$ and $\Delta E_3$ refer to the energy differences between the AFM and FM configurations reported in Appendix B; namely $\Delta E_1$ is the energy difference between AFM1 and FM1 and so on. For all the compounds $a_{\text{HSE}}=0.32$ has been used. We highlighted in bold FM couplings.}
  \vspace{3pt}
    \label{HYBRIDJ}
    \begin{tabular}{c|c|c|c}
    \hline
      compound & $\Delta E_1$ & $\Delta E_2$ & $\Delta E_3$ \\
     \hline
      Cd$_{1-x}$V$_x$Te & -22.3   & \textbf{78.3}  & -3.4 \\
    Cd$_{1-x}$Cr$_x$Te & -57.9 & -22.6  & -1.6 \\
     Cd$_{1-x}$Mn$_x$Te & -16.2 & -16.1 & -0.3  \\
     \hline
      Hg$_{1-x}$V$_x$Te & \textbf{35.5}  & \textbf{41.2} & -2.7   \\
     Hg$_{1-x}$Cr$_x$Te & -313.4 & -150.1  & -4.0   \\
        Hg$_{1-x}$Mn$_x$Te & -17.0  & -16.9  & -1.6   \\
        \hline
        \end{tabular}
        \end{center}
\end{table}
%
%

\begin{table}
\begin{center}
    \caption{Energy differences (meV) between AFM and FM configurations of Cr spins obtained without Jahn-Teller (JT) distortion, and with its half and full magnitude for $a_{\text{HSE}} = 0.32$.  We highlighted in bold FM couplings.} %
 \vspace{3pt}
    \label{HYBRIDJT}
    \begin{tabular}{c|c|c|c}
     \hline
    compound   & $\Delta E_1$ & $\Delta E_2$ & $\Delta E_3$  \\
     \hline
     Cd$_{1-x}$Cr$_x$Te NO JT & -57.9 & -22.6 & -1.6  \\
     Cd$_{1-x}$Cr$_x$Te JT/2 & \textbf{10.4} & \textbf{4.8} & -1.6   \\
    Cd$_{1-x}$Cr$_x$Te JT & \textbf{5.9} & \textbf{34.5} & -1.7   \\
     \hline
     Hg$_{1-x}$Cr$_x$Te NO JT  & -313.4 & -150.1 & -4.0  \\
      Hg$_{1-x}$Cr$_x$Te JT/2 & -100.4 & -154.4 & -4.6  \\
      Hg$_{1-x}$Cr$_x$Te JT & -54.8 & -231.0 & -5.6  \\
        \hline
        \end{tabular}
        \end{center}
\end{table}

\begin{table} 
\begin{center}
    \caption{Energy differences (meV) between AFM and FM configurations of V spins obtained without Jahn-Teller (JT) distortion, and with its full magnitude for $a_{\text{HSE}} = 0.32$.  We highlighted in bold FM couplings.} 
 \vspace{3pt}
    \label{HYBRIDJT_V}
    \begin{tabular}{c|c|c|c}
     \hline
    compound   & $\Delta E_1$ & $\Delta E_2$ & $\Delta E_3$  \\
     \hline
     Cd$_{1-x}$V$_x$Te NO JT & -22.3   & \textbf{78.3}  & -3.4 \\
    Cd$_{1-x}$V$_x$Te JT & \textbf{28.3} & \textbf{28.4} & -3.1   \\
     \hline
     Hg$_{1-x}$V$_x$Te NO JT & \textbf{35.5}  & \textbf{41.2} & -2.7   \\
      Hg$_{1-x}$V$_x$Te JT & \textbf{0.2} & \textbf{30.5} & -7.5  \\
        \hline
        \end{tabular}
        \end{center}
\end{table}

\section{Implications to the anomalous Hall effect}

Quantum wells of HgTe constitute a rare example of systems showing the quantum spin Hall effect. Our studies presented here and in the companion paper \cite{Sliwa:2023_arXiv} were designed to address the question about the suitability of HgTe QWs and related systems for the demonstration of the quantum anomalous Hall effect (QAHE) with characteristics opening a door or applications in the quantum metrology. As mentioned in Introduction, this question is timely considering the fact that  Hall resistance quantization in (Bi,Sb,Cr,V) layers is perturbed above 50 mK by hopping conductivity \cite{Kawamura:2017_PRL} involving numerous in-gap states brought about by antisites and other defects \cite{Zhang:2012_PRL}. 
As already noted by Liu et al. \cite{Liu:2008_PRLb}, the appearance of the chiral edge states requires the presence of: (1) localized spin polarization that breaks the time-reversal symmetry and generates giant exchange spin splitting of bands;  (2) a gaped inverted band structure in one spin channel ($p$-like states above $s$-like states), and the gaped normal band arrangement in the other spin channel – to achieve this situation a proper ratio between $s$-$d$ and $p$-$d$ exchange integrals is necessary; (3) the Fermi level in the gap region. 
That pioneering theory of the QAHE considered (Hg,Mn)Te QW as a system in which that phenomenon could be observed \cite{Liu:2008_PRLb}, albeit in the presence of an external magnetic field, as interactions between Mn spins are antiferromagnetic according to our and previous studies \cite{Mycielski:1984_SSC,Sliwa:2021_PRB}. However, recent experimental findings have indicated that the QHE dominates in (Hg,Mn)Te and becomes visible already in 50 mT \cite{Shamim:2020_SA}, the fact linked to the resonant character of acceptor states and the formation of bound magnetic polarons \cite{Dietl:2023_PRL,Dietl:2023_PRB}. Under these conditions, (Hg,Mn)Te looks as a promising QHE resistance standard working in surprisingly low magnetic fields, though its performance above 50 mK is to be verified experimentally.
Our first-principles results for (Hg,V)Te point to the presence of a FM coupling between V spins, an encouraging finding from the QAHE perspective. At the same time, however, our data reveal that the V impurities form a resonant donor state with the conduction band. This would mean that V doping will introduce electrons that would preclude a shift of the Fermi level toward the topological gap by a gate voltage. 
In the case of (Hg,Cr)Te, our {\em ab initio} studies indicate that the Cr donor level is degenerate with the valence band and, therefore, Cr acts as an isoelectronic impurity, similarly to Mn. Accordingly, it should be possible to shift the Fermi level towards the topological gap either by a suitable co-doping with electrically active impurities or by gating. At the same time, our first principles data substantiate two conclusions of the tight-binding computations \cite{Sliwa:2023_arXiv}: 
(i)  The exchange coupling between Cr spins is on the borderline between the FM case of V ions and AFM interactions between Mn ions. This opens a door for manipulation of magnetism by an electric field, strain, or pressure but suggests that the application of an external magnetic field may be necessary to polarize Cr spins. 
(ii) A close proximity of the Cr donor level to the top of the valence band results in a large $p$-$d$ hybridization and, thus, in a large $p$-$d$ exchange interaction between band carriers and localized spins, which we see as large splittings of ${\Gamma}_{7}$ and ${\Gamma}_{8}$ bands in Figs.  \ref{HgTe_HSE_fat} and \ref{Zoom_HgTe}. It has been shown in the companion paper \cite{Sliwa:2023_arXiv}, exploiting the $kp$ methodology developed earlier for HgTe and (Hg,Mn)Te QWs \cite{Novik:2005_PRB,Dietl:2023_PRB},  that a large ratio between the $p$-$d$ and $s$-$d$ exchange integrals (denoted as $\beta$ and $\alpha$ in the literature) precludes the appearance of the QAHE for the magnetization orientation along the growth direction [001] in (Hg,Cr)Te/(Cd,Hg)Te QWs. However, the spin-orbit interaction reduces the $p$-$d$ splitting for a tilted magnetization direction. A direct $kp$ calculation for the [111] magnetization direction, 3$\%$ of Hg-substitutional Cr ions in HgTe QW with a thickness of 6 nm, has demonstrated the presence of the gaped inverted band structure in one spin channel and the normal gaped band arrangement for the other spin direction\cite{Sliwa:2023_arXiv}. We conclude, therefore, that (Hg,Cr)Te QWs are a perspective system for the observation of the QAHE.  Furthermore, the determined magnitude of the topological energy gap of the order of 3 meV \cite{Sliwa:2023_arXiv} shows the range of Fermi level energies for which the Hall resistance will assume the quantized value R$_{xy}$=$\pm$ $h/e^2$ \cite{Liu:2008_PRLb}, corresponding to the Chern number C = $\pm$1. The gate voltage width of the Hall resistance plateau will depend on the density of the in-gap localized states in a given QW \cite{Dietl:2023_PRL}.

\section{Conclusions}

According to theoretical studies summarized in the present and the companion paper \cite{Sliwa:2023_arXiv}, the predicted magnitude and often sign of magnetic coupling between cation-substitutional Cr ions in II-VI compounds depends on the approach employed. In particular, our results show that differing conclusions about the sign of the interaction (FM vs. AFM) come out from the GGA+$U$ approach and the hybrid functional method. This fact reflects a delicate balance between FM superexchange and mostly AFM Bloembergen-Rowland and two-electron terms \cite{Sliwa:2023_arXiv}. Altogether, the data point out that while coupling between Mn ions is AFM, the FM interactions take over in the V case, but Cr in II-VI constitutes a borderline situation – FM and AFM components are of a similar strength. This may mean that the observation of the QAHE may require the application of a magnetic field to polarize Cr spins. At the same time, it might be possible to manipulate with magnetism and, thus, with topological properties by electric field, strain, pressure, and chemical content.

Another relevant question concerns the position of states brought about by magnetic impurities, which determines their electrical activity and the degree of hybridization with the band states. It has been known for a long time \cite{Dietl:2014_RMP}  that,  in II-VI compounds, Mn $d^{4/5}$  donor and $d^{5/6}$  acceptor states reside deep in the valence band and high in the conduction band,  respectively, so that Mn acts as an isoelectronic impurity, the fact reconfirmed by our data. Because the $p$-$d$ interaction between Te $p$ orbitals and Hg $d$ levels, the top of the $\Gamma_8$ valence is high in HgTe, which results in a relatively small value of the $p$-$d$ exchange integral $\beta$ in HgTe compared to other II-VI compounds \cite{Dietl:2014_RMP,Autieri21}.

Furthermore, our results confirm the experimental finding that the V  $d^{2/3}$  and Cr $d^{3/4}$  donor states are the gap of CdTe. However, according to the outcome of our computations, V  $d^{2/3}$ states lie the conduction band of HgTe, meaning that V ions act as resonant donors. If this were the case, it would be hard to shift the Fermi level to the topological gap in (Hg,V)Te QWs. Our results indicated that the Cr  $d^{3/4}$  donor states reside in the HgTe valence band, so that Cr ions constitute isoelectronic impurities. Since, however, those states are relatively close to the top of the valence band, we predict a relatively sizable magnitude of $\beta$ in (Hg,Cr)Te. According to previous \cite{Liu:2008_PRLb} and recent works  \cite{Sliwa:2023_arXiv}, a large ratio of $\beta$ to the $s$-$d$ exchange integral $\alpha$ in HgTe/(Cd,Hg)Te QWs doped with transition metals either in the QW or barriers, results in such an ordering of spin subbands,  which impedes the appearance of the QAHE.  Fortunately, if magnetization is tilted out of the growth direction, the spin-orbit interaction reduces the $p$-$d$ spitting of QW states without affecting the $s$-$d$ exchange. The presence of the QAHE is predicted for the [111] direction of magnetization in the (Hg,Cr)Te QWs \cite{Sliwa:2023_arXiv}.  In any case, a close energetic proximity of $d$ states and the Fermi level opens doors for a new physics in both (Hg,V)Te and (Hg,Cr)Te, not encountered in the case of (Hg,Mn)Te. A pool of other possibilities are offered by engineering superlattices or heterostructures involving TM-doped Hg-based topological systems, some of possibilities being indicated elsewhere \cite{Islam22,PhysRevB.107.125102}.

%
%
%

\begin{figure*}[t!]
\centering
\includegraphics[width=0.23\textwidth,angle=270]{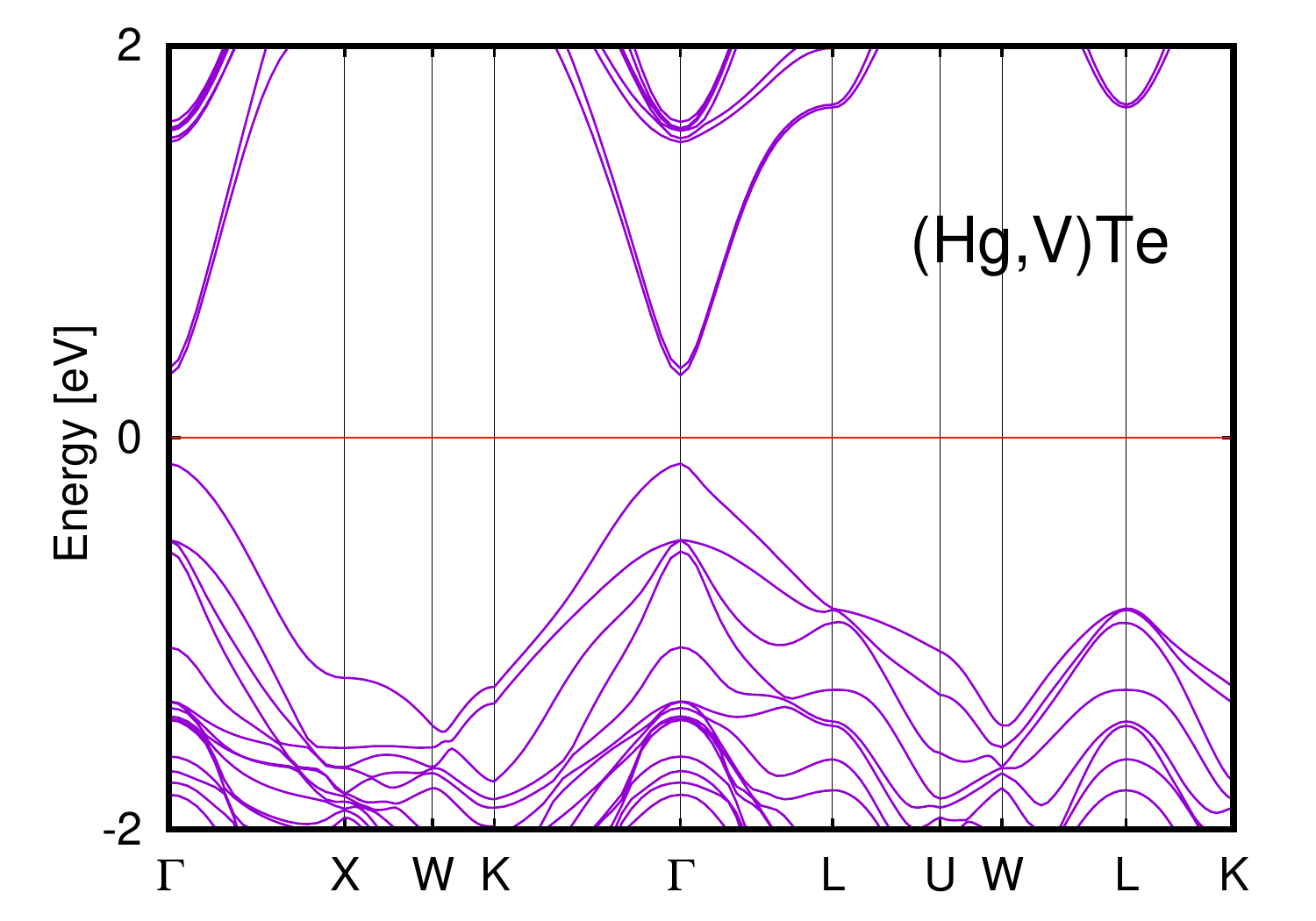}
\includegraphics[width=0.23\textwidth,angle=270]{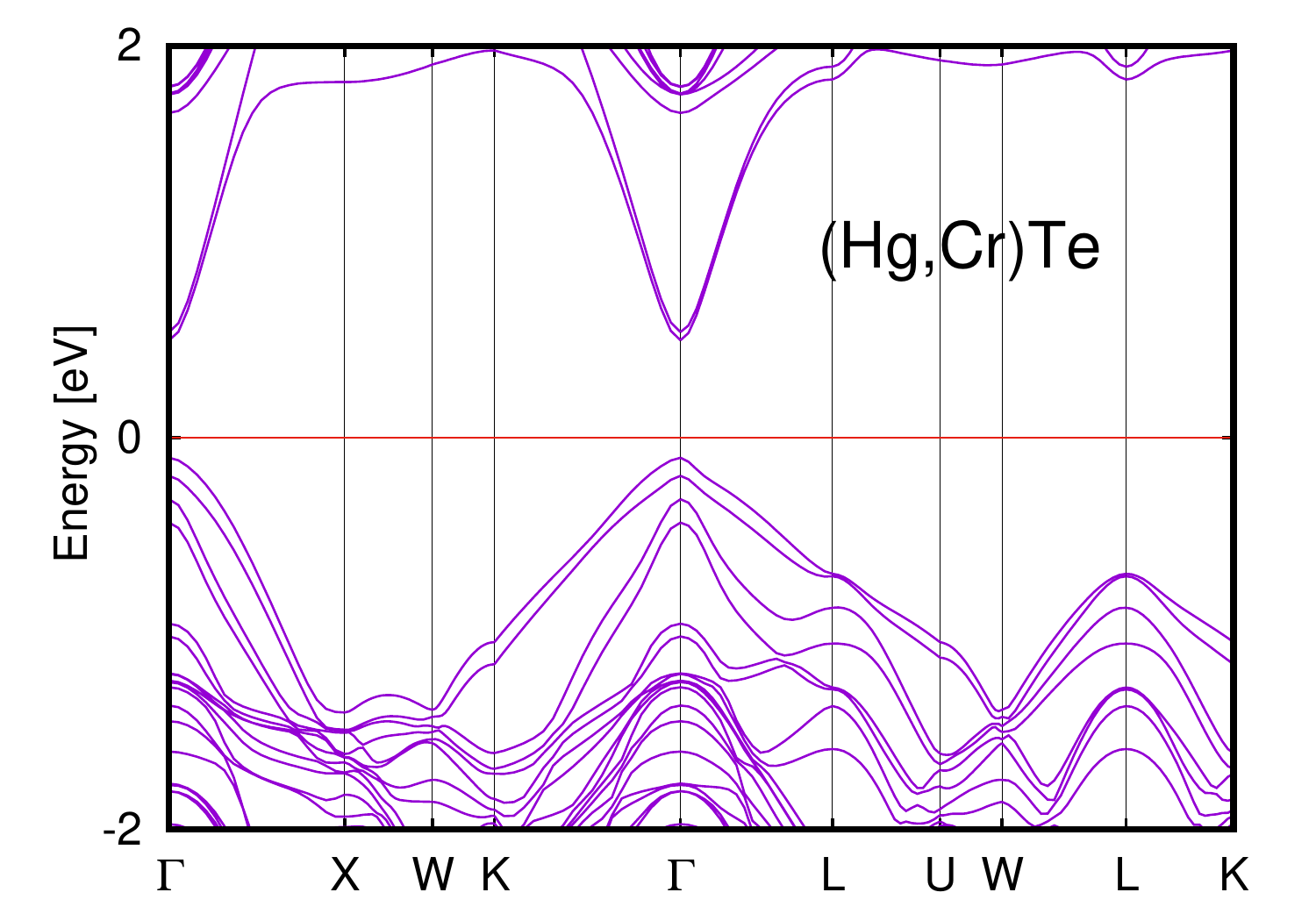}
\includegraphics[width=0.23\textwidth,angle=270]{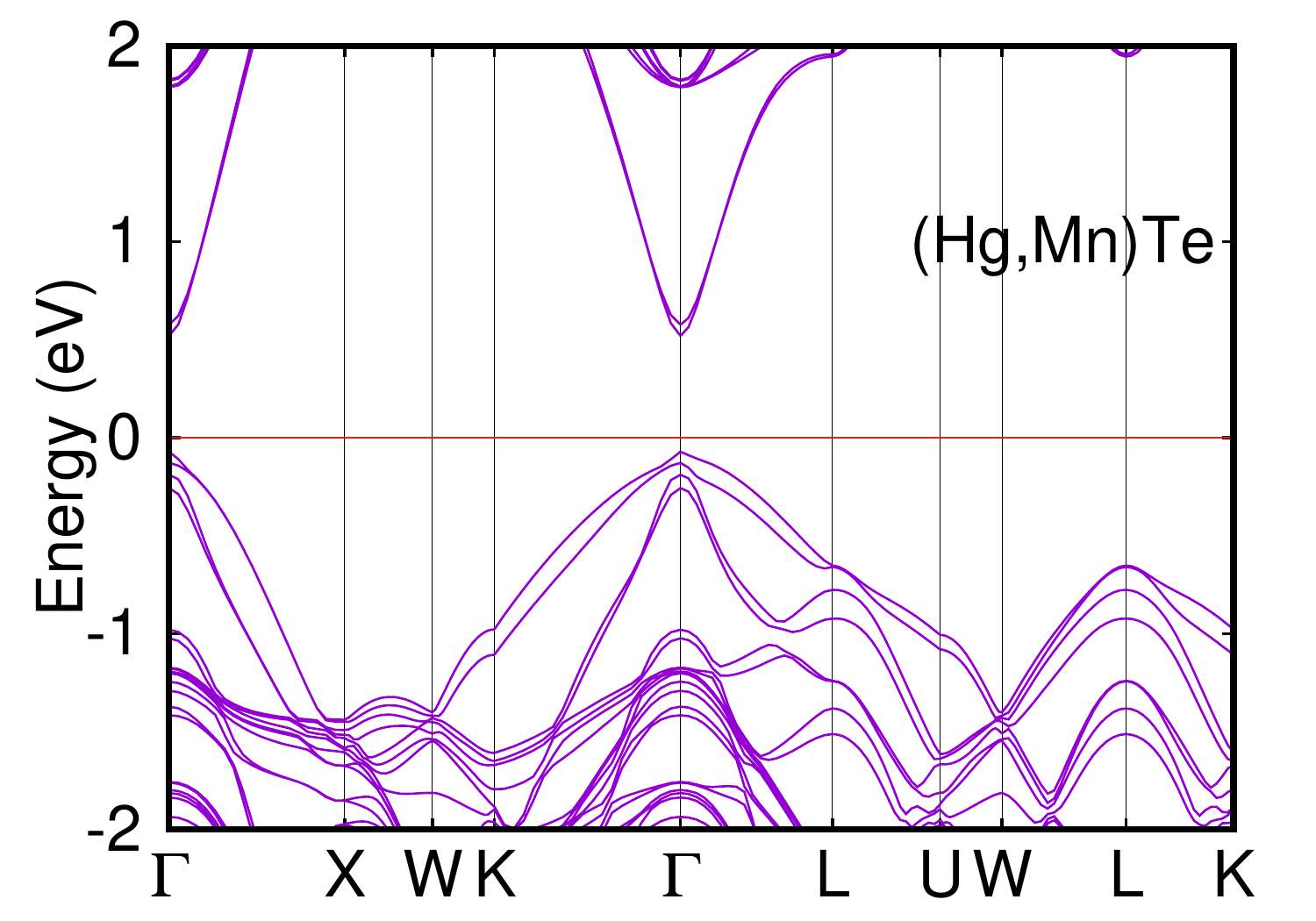}
\caption{HSE band structure for Hg$_{0.875}$V$_{0.125}$Te,  Hg$_{0.875}$Cr$_{0.125}$Te, and Hg$_{0.875}$Mn$_{0.125}$Te with ferromagnetic arrangement of TM spins and $a_{\text{HSE}} =0.5$. The Fermi level is set at zero energy.}
\label{HgTe_HSE_05}
\end{figure*} 

\section*{Acknowledgments}

This research was supported by the Foundation for Polish Science through the IRA Programme co-financed by the EU within SG OP and EFSE (project “MagTop” no. FENG.02.01-IP.05-0028/23), by funds from the state budget allocated by the Minister of Science (Poland) as part of the Polish Metrology II programme project no. PM-II/SP/0012/2024/02. We acknowledge the access to the computing facilities of the Interdisciplinary Center of Modeling at the University of Warsaw, Grant G84-0, GB84-1, and GB84-7. We acknowledge the CINECA award under the ISCRA initiative  IsC85 "TOPMOST" and IsC93 "RATIO" grant, for the availability of high-performance computing resources and support. We acknowledge the access to the computing facilities of the Poznan Supercomputing and Networking Center Grant No.\,609.

\appendix

\section{Transition metal doped compounds with 25\% and 50\% of the exact exchange}

In this Appendix, we present band structures and the magnetic properties for TM-doped CdTe and HgTe obtained employing the hybrid functional with  $a_{\text{HSE}}=0.25$ and $a_{\text{HSE}}=0.50$ of the exact exchange. The band structure results displayed in Figs.~\ref{CdTe_HSE_025}, \ref{HgTe_HSE_025}, and \ref{HgTe_HSE_05} further substantiate the use of $a_{\text{HSE}}=0.32$ in the main body of the text. In particular, in the case of Hg$_{0.875}$Mn$_{0.125}$Te, the experimental value of the band gap  $E_g =E_{\Gamma6} -E_{\Gamma8} =0.2$\,eV \cite{Furdyna:1988_JAP} is close to the computational data for $a_{\text{HSE}}=0.32$ (Fig.~\ref{HgTe_HSE_fat}) but $E_g$ is too small and too large for $a_{\text{HSE}}=0.25$ and 0.5, respectively, as shown in Figs.~\ref{HgTe_HSE_025} and \ref{HgTe_HSE_05}.

\begin{figure*}[t!]
\centering
\includegraphics[width=0.35\textwidth,angle=0]{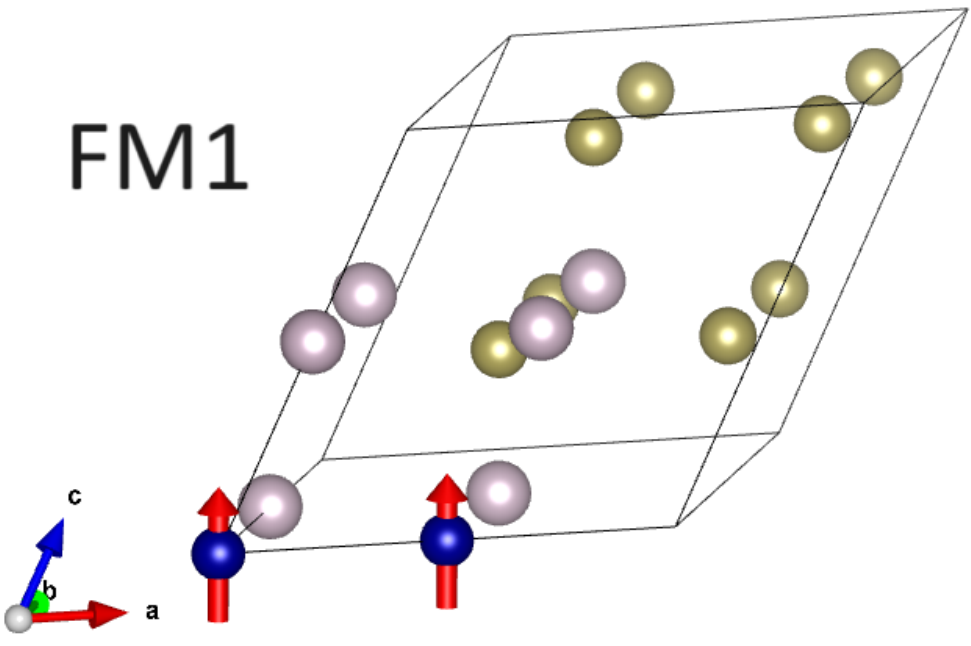}
\includegraphics[width=0.30\textwidth,angle=0]{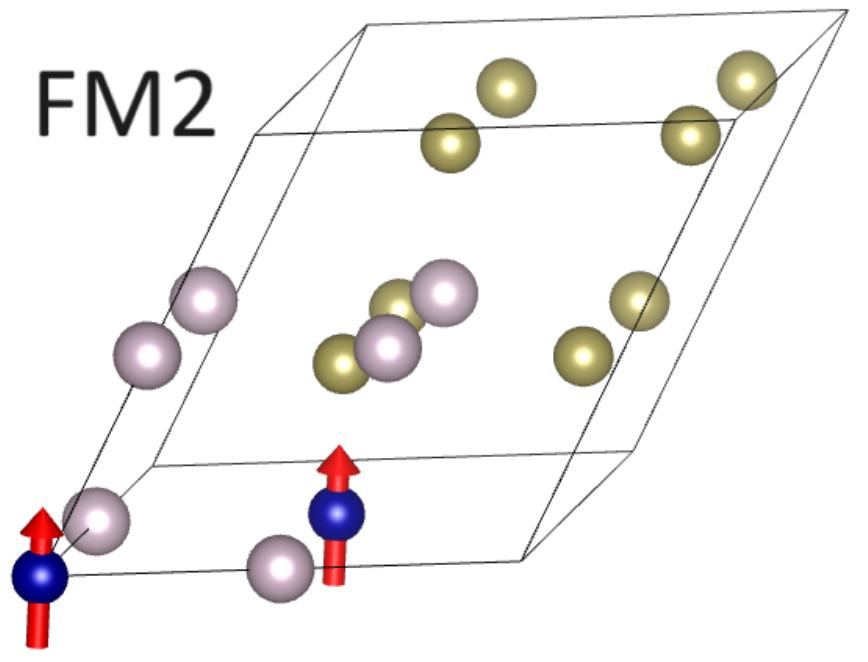}
\includegraphics[width=0.30\textwidth,angle=0]{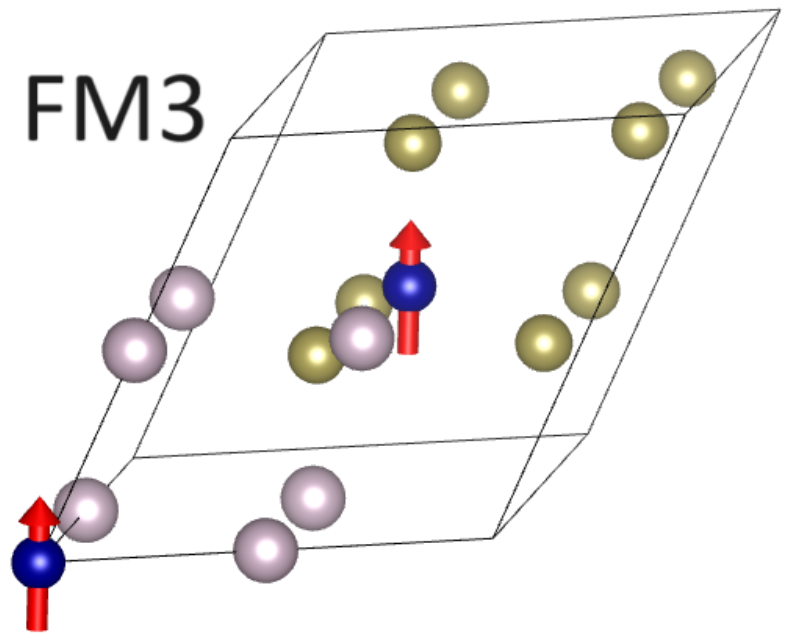}
\caption{Ferromagnetic configurations FM1, FM2 and FM3 investigated in the paper. The red arrows are the directions of the spins. The blue spheres are the magnetic dopants Cr or V, while the white spheres represent Hg or Cd, and the brown spheres are the Te atoms.}
\label{FM}
\end{figure*} 

\begin{figure*}[t!]
\centering
\includegraphics[width=0.30\textwidth,angle=0]{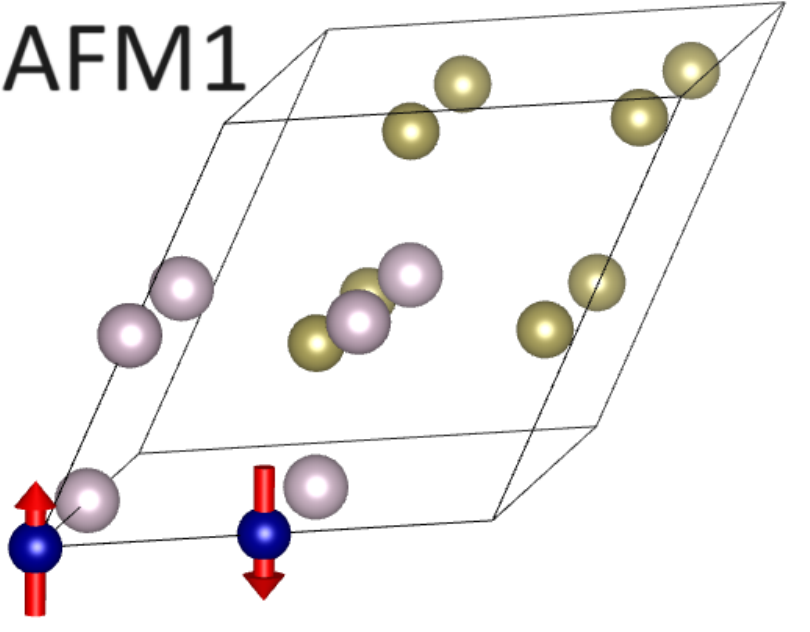}
\includegraphics[width=0.30\textwidth,angle=0]{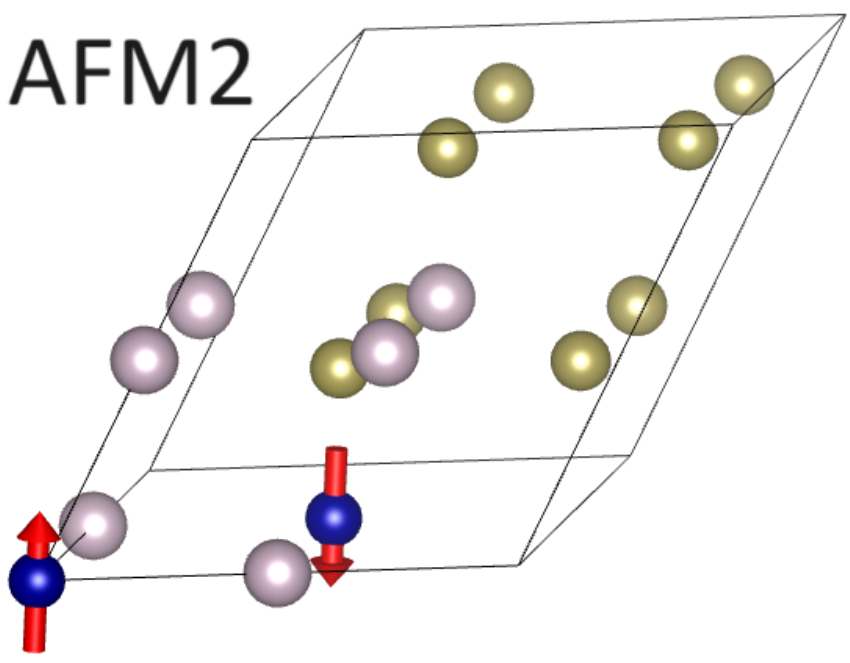}
\includegraphics[width=0.30\textwidth,angle=0]{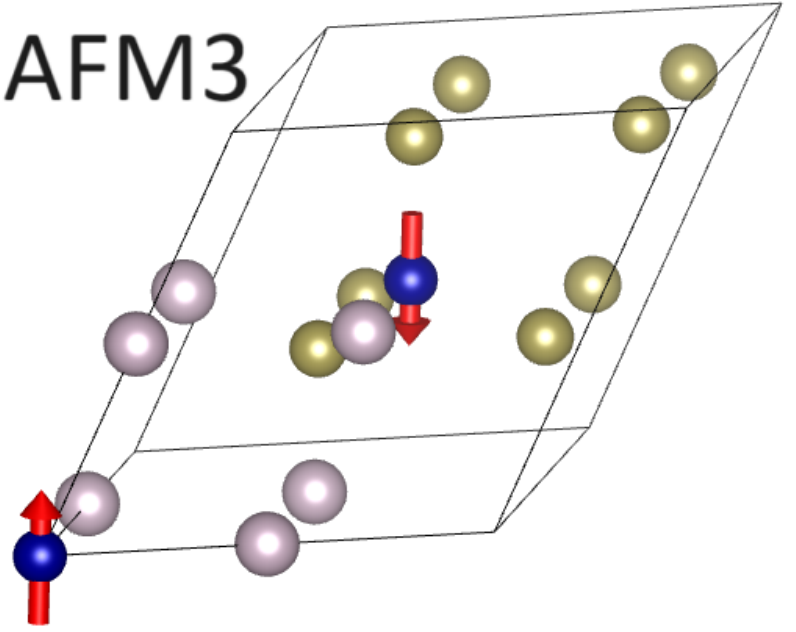}
\caption{Antiferromagnetic configurations AFM1, AFM2 and AFM3 investigated in the paper. The red arrows are the directions of the spins. The blue spheres are the magnetic dopants Cr or V, while the white spheres represent Hg or Cd, and the brown spheres are the Te atoms.}
\label{AFM}
\end{figure*} 

As we can see in Fig.~\ref{HgTe_HSE_025}, for $a_{\text{HSE}} =0.25$   we find a metallic phase in the case of HgTe doped with Cr because acceptor $d$-states of the dopants undergo a shift below the top of the valence band.
The band structures obtained within the GGA+$U$ approach resemble the results obtained at $a_{\text{HSE}} =0.25$ for all compounds and, therefore, we do not display them.

In Tables \ref{HYBRIDJ025} and Table \ref{HYBRIDJ05} we report the values of the spin-spin couplings for the Cr-doped compounds obtained by using $a_{\text{HSE}} =0.25$ and 0.5, respectively. In the latter case, we find that HgTe doped with Cr shows ferromagnetic couplings.

\begin{table}
\begin{center}
\caption{Energy differences (meV) between AFM and FM spin configurations for TM doped CdTe and HgTe. $\Delta E_1$, $\Delta E_2$, and $\Delta E_3$ refer to the energy differences between the AFM and FM configurations reported in Appendix B; namely $\Delta E_1$ is the energy difference between AFM1 and FM1 and so on. For all compounds, $a_{\text{HSE}} =0.25$ is used, except for values with asterisks, which have been obtained with $a_{\text{HSE}} =0.35$, because the calculations with $a_{\text{HSE}} =0.25$ do not converge in those cases. We have highlighted in bold FM couplings.}
\vspace{3pt}
\label{HYBRIDJ025}
\begin{tabular}{ c|c|c|c }
 \hline
      compound & $\Delta E_1$ & $\Delta E_2$ & $\Delta E_3$ \\
    \hline
   Cd$_{1-x}$V$_x$Te & -38.1   & \textbf{22.8}  & 0.6 \\
    Cd$_{1-x}$Cr$_x$Te & -10.5* & -16.8*  & -4.8 \\
    Cd$_{1-x}$Mn$_x$Te & -21.9 & -21.8 & -0.5  \\
    \hline
   Hg$_{1-x}$V$_x$Te & \textbf{47.6}  & \textbf{51.5} & -1.6   \\
    Hg$_{1-x}$Cr$_x$Te & -77.9  & -35.9  & -5.1   \\
    Hg$_{1-x}$Mn$_x$Te & -22.3  & -22.2  & -2.0   \\
    \hline
\end{tabular}
\end{center}
\end{table}

\begin{table}
\begin{center}
    \caption{Energy differences (meV) between AFM and FM spin configurations with respect to the FM state taking the Jahn-Teller distortion into account for Cr-doped compounds. With JT/2 we indicate that we consider an average of the positions of the atoms without and with the JT distortion. We have highlighted in bold FM couplings. The results have been obtained for $a_{\text{HSE}} =0.50$.}
   \vspace{3pt}
    \label{HYBRIDJ05}
    \begin{tabular}{ c|c|c}
     \hline
    compound   & $\Delta E_1$ & $\Delta E_2$  \\
     \hline
     Cd$_{1-x}$Cr$_x$Te NO JT & -218.7 & -160.3  \\
      Cd$_{1-x}$Cr$_x$Te JT/2 & -62.3 & \textbf{24.1}   \\
   \hline
     Hg$_{1-x}$Cr$_x$Te NO JT  & \textbf{74.0} & \textbf{7.8}  \\
      Hg$_{1-x}$Cr$_x$Te JT/2 & \textbf{88.6}  & -22.7  \\
        \hline
        \end{tabular}
        \end{center}
\end{table}

\section{Magnetic configurations}

In this Section, for completeness, we report the graphical representation of the magnetic configurations investigated in the paper. We have used 2$\times$2$\times$2 supercells with two magnetic impurities and we have investigated six different configurations, three in the ferromagnetic case and three in the antiferromagnetic case. The FM configurations are reported in Fig. \ref{FM}, while the AFM configurations are shown in Fig. \ref{AFM}. 
In the unit cell, we have 8 cations and 8 anions.
In the first configuration, the two magnetic dopants replace two cations first-neighbor, in the second configuration the magnetic dopants are second-neighbor, while in the third one, the magnetic dopants are third-neighbors. Since there are also the atoms in the repeated unit cell to be considered, $\Delta{E_i}$ (i=1,2,3) are not exactly proportional to the first-neighbor, second-neighbor and third-neighbor magnetic couplings. However, they give a strong indication on these magnetic couplings.

\section{Results within GGA + U approximation}

In Tab. \ref{GGAJ} we report the values obtained by using GGA + $U$ calculations.
As we can see from Tab. \ref{GGAJ}, in the GGA approximation, in the case of doping with Mn, we have antiferromagnetic couplings for both HgTe and CdTe. In the case of doping with Cr and V, the couplings are ferromagnetic.
The values of the third neighbor energy differences for Hg$_{1-x}$Cr$_x$Te and Cd$_{1-x}$Cr$_x$Te in GGA + U are unphysical.
The antiferromagnetism of the couplings in the case of Mn-doped systems is still present when we use the HSE functional, therefore this property is robust, and also because already within GGA + $U$ we get the experimental insulating configuration for these systems. 
In GGA+$U$ the ferromagnetism is overestimated due to the overestimation of the metallicity. In Tab. \ref{GGAJT} we report the results obtained with GGA + U for the case of doping with Cr with JT distortions included. As in the main body of the paper, with JT/2 we indicate that we are considering the average of the positions of the atoms without and with the Jahn-Teller distortion.

\begin{table}
\begin{center}
    \caption{Energy differences (meV) between the AFM and the FM configurations in GGA+U approximation and calculated with respect to the FM state. U=5 eV on d orbitals of the Cr atoms is used. $\Delta$E$_1$, $\Delta$E$_2$ and $\Delta$E$_3$ refer to the energy differences between the configurations reported in the new Appendix B. We highlighted in bold FM couplings.}
    \label{GGAJ}
    \begin{tabular}{|c|c|c|c|c|}
    \hline
       & $\Delta$E$_1$ & $\Delta$E$_2$ & $\Delta$E$_3$ \\
     \hline
     Cd$_{1-x}$Cr$_x$Te & \textbf{63.5} & \textbf{215.9}  & \textbf{107.6} \\
     \hline
     Cd$_{1-x}$V$_x$Te & \textbf{497.6}   & \textbf{28.2}  & -1.7 \\
     \hline
     Cd$_{1-x}$Mn$_x$Te & -15.7 & -15.6 & -0.7   \\
        \hline
      Hg$_{1-x}$Cr$_x$Te & \textbf{168.3}  & \textbf{178.4}  & -36.1    \\
      \hline
       Hg$_{1-x}$V$_x$Te & \textbf{44.1}  & \textbf{44.2} & -1.8   \\
       \hline
        Hg$_{1-x}$Mn$_x$Te & -13.0  & -12.9  & 0.4   \\
        \hline
        \end{tabular}
        \end{center}
\end{table}

\begin{table}
\begin{center}
    \caption{Energy differences (meV) between the AFM and the FM configurations once we introduce the crystal field for the Cr-doped compounds, obtained in GGA + U approximation and calculated with respect to the FM state. $\Delta$E$_1$, $\Delta$E$_2$ and $\Delta$E$_3$ refer to the first, the second and the third neighbour. With JT/2 we indicate that we are considering the average of the positions of the atoms without and with the crystal field. We highlighted in bold FM couplings.}
    \label{GGAJT}
    \begin{tabular}{|c|c|c|c|c|}
    \hline
       & $\Delta$E$_1$ & $\Delta$E$_2$ & $\Delta$E$_3$ \\
     \hline
      Cd$_{1-x}$Cr$_x$Te NO JT & \textbf{63.5} & \textbf{215.9}  & \textbf{107.6} \\
      \hline
      Hg$_{1-x}$Cr$_x$Te NO JT & \textbf{168.3}  & \textbf{178.4}  & -36.1   \\
      \hline
     Cd$_{1-x}$Cr$_x$Te JT/2 & \textbf{70.3} & \textbf{220.9}  & \textbf{100.9} \\
     \hline
      Hg$_{1-x}$Cr$_x$Te JT/2 & \textbf{171.5}  & \textbf{174.4}  & -32.9   \\
      \hline
     Cd$_{1-x}$Cr$_x$Te JT & \textbf{74.8} & \textbf{140.1}  & \textbf{91.4} \\
     \hline
      Hg$_{1-x}$Cr$_x$Te JT & \textbf{171.6}  & \textbf{170.5}  & -30.4   \\
        \hline
        \end{tabular}
        \end{center}
\end{table}

%

\newpage
\bibliography{HgTe22Dec2023}

\begin{thebibliography}{57}%
\makeatletter
\providecommand \@ifxundefined [1]{%
 \@ifx{#1\undefined}
}%
\providecommand \@ifnum [1]{%
 \ifnum #1\expandafter \@firstoftwo
 \else \expandafter \@secondoftwo
 \fi
}%
\providecommand \@ifx [1]{%
 \ifx #1\expandafter \@firstoftwo
 \else \expandafter \@secondoftwo
 \fi
}%
\providecommand \natexlab [1]{#1}%
\providecommand \enquote  [1]{``#1''}%
\providecommand \bibnamefont  [1]{#1}%
\providecommand \bibfnamefont [1]{#1}%
\providecommand \citenamefont [1]{#1}%
\providecommand \href@noop [0]{\@secondoftwo}%
\providecommand \href [0]{\begingroup \@sanitize@url \@href}%
\providecommand \@href[1]{\@@startlink{#1}\@@href}%
\providecommand \@@href[1]{\endgroup#1\@@endlink}%
\providecommand \@sanitize@url [0]{\catcode `\\12\catcode `\$12\catcode `\&12\catcode `\#12\catcode `\^12\catcode `\_12\catcode `\%12\relax}%
\providecommand \@@startlink[1]{}%
\providecommand \@@endlink[0]{}%
\providecommand \url  [0]{\begingroup\@sanitize@url \@url }%
\providecommand \@url [1]{\endgroup\@href {#1}{\urlprefix }}%
\providecommand \urlprefix  [0]{URL }%
\providecommand \Eprint [0]{\href }%
\providecommand \doibase [0]{http://dx.doi.org/}%
\providecommand \selectlanguage [0]{\@gobble}%
\providecommand \bibinfo  [0]{\@secondoftwo}%
\providecommand \bibfield  [0]{\@secondoftwo}%
\providecommand \translation [1]{[#1]}%
\providecommand \BibitemOpen [0]{}%
\providecommand \bibitemStop [0]{}%
\providecommand \bibitemNoStop [0]{.\EOS\space}%
\providecommand \EOS [0]{\spacefactor3000\relax}%
\providecommand \BibitemShut  [1]{\csname bibitem#1\endcsname}%
\let\auto@bib@innerbib\@empty
\bibitem [{\citenamefont {Yu}\ \emph {et~al.}(2010)\citenamefont {Yu}, \citenamefont {Zhang}, \citenamefont {Zhang}, \citenamefont {Zhang}, \citenamefont {Dai},\ and\ \citenamefont {Fang}}]{Yu:2010_S}%
  \BibitemOpen
  \bibfield  {author} {\bibinfo {author} {\bibfnamefont {Rui}\ \bibnamefont {Yu}}, \bibinfo {author} {\bibfnamefont {Wei}\ \bibnamefont {Zhang}}, \bibinfo {author} {\bibfnamefont {Hai-Jun}\ \bibnamefont {Zhang}}, \bibinfo {author} {\bibfnamefont {Shou-Cheng}\ \bibnamefont {Zhang}}, \bibinfo {author} {\bibfnamefont {Xi}~\bibnamefont {Dai}}, \ and\ \bibinfo {author} {\bibfnamefont {Zhong}\ \bibnamefont {Fang}},\ }\bibfield  {title} {\enquote {\bibinfo {title} {Quantized anomalous {H}all effect in magnetic topological insulators},}\ }\href {\doibase 10.1126/science.1187485} {\bibfield  {journal} {\bibinfo  {journal} {Science}\ }\textbf {\bibinfo {volume} {329}},\ \bibinfo {pages} {61--64} (\bibinfo {year} {2010})}\BibitemShut {NoStop}%
\bibitem [{\citenamefont {Chang}\ \emph {et~al.}(2013)\citenamefont {Chang}, \citenamefont {Zhang}, \citenamefont {Feng}, \citenamefont {Shen}, \citenamefont {Zhang}, \citenamefont {Guo}, \citenamefont {Li}, \citenamefont {Ou}, \citenamefont {Wei}, \citenamefont {Wang}, \citenamefont {Ji}, \citenamefont {Feng}, \citenamefont {Ji}, \citenamefont {Chen}, \citenamefont {Jia}, \citenamefont {Dai}, \citenamefont {Fang}, \citenamefont {Zhang}, \citenamefont {He}, \citenamefont {Wang}, \citenamefont {Lu}, \citenamefont {Ma},\ and\ \citenamefont {Xue}}]{Chang:2013_S}%
  \BibitemOpen
  \bibfield  {author} {\bibinfo {author} {\bibfnamefont {Cui-Zu}\ \bibnamefont {Chang}}, \bibinfo {author} {\bibfnamefont {Jinsong}\ \bibnamefont {Zhang}}, \bibinfo {author} {\bibfnamefont {Xiao}\ \bibnamefont {Feng}}, \bibinfo {author} {\bibfnamefont {Jie}\ \bibnamefont {Shen}}, \bibinfo {author} {\bibfnamefont {Zuocheng}\ \bibnamefont {Zhang}}, \bibinfo {author} {\bibfnamefont {Minghua}\ \bibnamefont {Guo}}, \bibinfo {author} {\bibfnamefont {Kang}\ \bibnamefont {Li}}, \bibinfo {author} {\bibfnamefont {Yunbo}\ \bibnamefont {Ou}}, \bibinfo {author} {\bibfnamefont {Pang}\ \bibnamefont {Wei}}, \bibinfo {author} {\bibfnamefont {Li-Li}\ \bibnamefont {Wang}}, \bibinfo {author} {\bibfnamefont {Zhong-Qing}\ \bibnamefont {Ji}}, \bibinfo {author} {\bibfnamefont {Yang}\ \bibnamefont {Feng}}, \bibinfo {author} {\bibfnamefont {Shuaihua}\ \bibnamefont {Ji}}, \bibinfo {author} {\bibfnamefont {Xi}~\bibnamefont {Chen}}, \bibinfo {author} {\bibfnamefont {Jinfeng}\ \bibnamefont {Jia}}, \bibinfo {author} {\bibfnamefont
  {Xi}~\bibnamefont {Dai}}, \bibinfo {author} {\bibfnamefont {Zhong}\ \bibnamefont {Fang}}, \bibinfo {author} {\bibfnamefont {Shou-Cheng}\ \bibnamefont {Zhang}}, \bibinfo {author} {\bibfnamefont {Ke}~\bibnamefont {He}}, \bibinfo {author} {\bibfnamefont {Yayu}\ \bibnamefont {Wang}}, \bibinfo {author} {\bibfnamefont {Li}~\bibnamefont {Lu}}, \bibinfo {author} {\bibfnamefont {Xu-Cun}\ \bibnamefont {Ma}}, \ and\ \bibinfo {author} {\bibfnamefont {Qi-Kun}\ \bibnamefont {Xue}},\ }\bibfield  {title} {\enquote {\bibinfo {title} {Experimental observation of the quantum anomalous {H}all effect in a magnetic topological insulator},}\ }\href {\doibase 10.1126/science.1234414} {\bibfield  {journal} {\bibinfo  {journal} {Science}\ }\textbf {\bibinfo {volume} {340}},\ \bibinfo {pages} {167--170} (\bibinfo {year} {2013})}\BibitemShut {NoStop}%
\bibitem [{\citenamefont {Chang}\ \emph {et~al.}(2023)\citenamefont {Chang}, \citenamefont {Liu},\ and\ \citenamefont {MacDonald}}]{Chang:2023_RMP}%
  \BibitemOpen
  \bibfield  {author} {\bibinfo {author} {\bibfnamefont {Cui-Zu}\ \bibnamefont {Chang}}, \bibinfo {author} {\bibfnamefont {Chao-Xing}\ \bibnamefont {Liu}}, \ and\ \bibinfo {author} {\bibfnamefont {A.~H.}\ \bibnamefont {MacDonald}},\ }\bibfield  {title} {\enquote {\bibinfo {title} {Quantum anomalous {H}all effect},}\ }\href {\doibase 10.1103/RevModPhys.95.011002} {\bibfield  {journal} {\bibinfo  {journal} {Rev. Mod. Phys.}\ }\textbf {\bibinfo {volume} {95}} (\bibinfo {year} {2023}),\ 10.1103/RevModPhys.95.011002}\BibitemShut {NoStop}%
\bibitem [{\citenamefont {Ke}\ \emph {et~al.}(2018)\citenamefont {Ke}, \citenamefont {Wang},\ and\ \citenamefont {Xue}}]{Ke:2018_ARCMP}%
  \BibitemOpen
  \bibfield  {author} {\bibinfo {author} {\bibfnamefont {He}~\bibnamefont {Ke}}, \bibinfo {author} {\bibfnamefont {Yayu}\ \bibnamefont {Wang}}, \ and\ \bibinfo {author} {\bibfnamefont {Qi-Kun}\ \bibnamefont {Xue}},\ }\bibfield  {title} {\enquote {\bibinfo {title} {Topological materials: quantum anomalous {Hall} system},}\ }\href {\doibase 10.1146/annurev-conmatphys-033117-054144} {\bibfield  {journal} {\bibinfo  {journal} {Annu. Rev. Cond. Mat. Phys.}\ }\textbf {\bibinfo {volume} {9}},\ \bibinfo {pages} {329--344} (\bibinfo {year} {2018})}\BibitemShut {NoStop}%
\bibitem [{\citenamefont {Tokura}\ \emph {et~al.}(2019)\citenamefont {Tokura}, \citenamefont {Yasuda},\ and\ \citenamefont {Tsukazaki}}]{Tokura:2019_NRP}%
  \BibitemOpen
  \bibfield  {author} {\bibinfo {author} {\bibfnamefont {Y.}~\bibnamefont {Tokura}}, \bibinfo {author} {\bibfnamefont {K.}~\bibnamefont {Yasuda}}, \ and\ \bibinfo {author} {\bibfnamefont {A.}~\bibnamefont {Tsukazaki}},\ }\bibfield  {title} {\enquote {\bibinfo {title} {Magnetic topological insulators},}\ }\href {\doibase 10.1038/s42254-018-0011-5} {\bibfield  {journal} {\bibinfo  {journal} {Nat. Rev. Phys.}\ }\textbf {\bibinfo {volume} {1}},\ \bibinfo {pages} {126--143} (\bibinfo {year} {2019})}\BibitemShut {NoStop}%
\bibitem [{\citenamefont {Fox}\ \emph {et~al.}(2018)\citenamefont {Fox}, \citenamefont {Rosen}, \citenamefont {Yang}, \citenamefont {Jones}, \citenamefont {Elmquist}, \citenamefont {Kou}, \citenamefont {Pan}, \citenamefont {Wang},\ and\ \citenamefont {Goldhaber-Gordon}}]{Fox:2018_PRB}%
  \BibitemOpen
  \bibfield  {author} {\bibinfo {author} {\bibfnamefont {E.~J.}\ \bibnamefont {Fox}}, \bibinfo {author} {\bibfnamefont {I.~T.}\ \bibnamefont {Rosen}}, \bibinfo {author} {\bibfnamefont {Yanfei}\ \bibnamefont {Yang}}, \bibinfo {author} {\bibfnamefont {G.~R.}\ \bibnamefont {Jones}}, \bibinfo {author} {\bibfnamefont {R.~E.}\ \bibnamefont {Elmquist}}, \bibinfo {author} {\bibfnamefont {Xufeng}\ \bibnamefont {Kou}}, \bibinfo {author} {\bibfnamefont {Lei}\ \bibnamefont {Pan}}, \bibinfo {author} {\bibfnamefont {Kang~L.}\ \bibnamefont {Wang}}, \ and\ \bibinfo {author} {\bibfnamefont {D.}~\bibnamefont {Goldhaber-Gordon}},\ }\bibfield  {title} {\enquote {\bibinfo {title} {Part-per-million quantization and current-induced breakdown of the quantum anomalous {Hall} effect},}\ }\href {\doibase 10.1103/PhysRevB.98.075145} {\bibfield  {journal} {\bibinfo  {journal} {Phys. Rev. B}\ }\textbf {\bibinfo {volume} {98}},\ \bibinfo {pages} {075145} (\bibinfo {year} {2018})}\BibitemShut {NoStop}%
\bibitem [{\citenamefont {Goetz}\ \emph {et~al.}(2018)\citenamefont {Goetz}, \citenamefont {Fijalkowski}, \citenamefont {Pesel}, \citenamefont {Hartl}, \citenamefont {Schreyeck}, \citenamefont {Winnerlein}, \citenamefont {Grauer}, \citenamefont {Scherer}, \citenamefont {Brunner}, \citenamefont {Gould}, \citenamefont {Ahlers},\ and\ \citenamefont {Molenkamp}}]{Goetz:2018_APL}%
  \BibitemOpen
  \bibfield  {author} {\bibinfo {author} {\bibfnamefont {M.}~\bibnamefont {Goetz}}, \bibinfo {author} {\bibfnamefont {K.~M.}\ \bibnamefont {Fijalkowski}}, \bibinfo {author} {\bibfnamefont {E.}~\bibnamefont {Pesel}}, \bibinfo {author} {\bibfnamefont {M.}~\bibnamefont {Hartl}}, \bibinfo {author} {\bibfnamefont {S.}~\bibnamefont {Schreyeck}}, \bibinfo {author} {\bibfnamefont {M.}~\bibnamefont {Winnerlein}}, \bibinfo {author} {\bibfnamefont {S.}~\bibnamefont {Grauer}}, \bibinfo {author} {\bibfnamefont {H.}~\bibnamefont {Scherer}}, \bibinfo {author} {\bibfnamefont {K.}~\bibnamefont {Brunner}}, \bibinfo {author} {\bibfnamefont {C.}~\bibnamefont {Gould}}, \bibinfo {author} {\bibfnamefont {F.~J.}\ \bibnamefont {Ahlers}}, \ and\ \bibinfo {author} {\bibfnamefont {L.~W.}\ \bibnamefont {Molenkamp}},\ }\bibfield  {title} {\enquote {\bibinfo {title} {Precision measurement of the quantized anomalous {Hall} resistance at zero magnetic field},}\ }\href {\doibase 10.1063/1.5009718} {\bibfield  {journal} {\bibinfo  {journal}
  {Appl. Phys. Lett.}\ }\textbf {\bibinfo {volume} {112}},\ \bibinfo {pages} {072102} (\bibinfo {year} {2018})}\BibitemShut {NoStop}%
\bibitem [{\citenamefont {Yuma~Okazaki}\ \emph {et~al.}(2022)\citenamefont {Yuma~Okazaki}, \citenamefont {Kawamura}, \citenamefont {Yoshimi}, \citenamefont {Nakamura}, \citenamefont {Takada}, \citenamefont {Mogi}, \citenamefont {Takahashi}, \citenamefont {Tsukazaki}, \citenamefont {Kawasaki}, \citenamefont {Tokura},\ and\ \citenamefont {Kaneko}}]{Okazaki:2022_NP}%
  \BibitemOpen
  \bibfield  {author} {\bibinfo {author} {\bibfnamefont {Takehiko~Oe}\ \bibnamefont {Yuma~Okazaki}}, \bibinfo {author} {\bibfnamefont {Minoru}\ \bibnamefont {Kawamura}}, \bibinfo {author} {\bibfnamefont {Ryutaro}\ \bibnamefont {Yoshimi}}, \bibinfo {author} {\bibfnamefont {Shuji}\ \bibnamefont {Nakamura}}, \bibinfo {author} {\bibfnamefont {Shintaro}\ \bibnamefont {Takada}}, \bibinfo {author} {\bibfnamefont {Masataka}\ \bibnamefont {Mogi}}, \bibinfo {author} {\bibfnamefont {Kei~S.}\ \bibnamefont {Takahashi}}, \bibinfo {author} {\bibfnamefont {Atsushi}\ \bibnamefont {Tsukazaki}}, \bibinfo {author} {\bibfnamefont {Masashi}\ \bibnamefont {Kawasaki}}, \bibinfo {author} {\bibfnamefont {Yoshinori}\ \bibnamefont {Tokura}}, \ and\ \bibinfo {author} {\bibfnamefont {Nobu-Hisa}\ \bibnamefont {Kaneko}},\ }\bibfield  {title} {\enquote {\bibinfo {title} {Quantum anomalous {Hall} effect with a permanent magnet defines a quantum resistance standard},}\ }\href {\doibase 10.1038/s41567-021-01424-8} {\bibfield  {journal}
  {\bibinfo  {journal} {Nat. Phys.}\ }\textbf {\bibinfo {volume} {18}},\ \bibinfo {pages} {25--29} (\bibinfo {year} {2022})}\BibitemShut {NoStop}%
\bibitem [{\citenamefont {Rodenbach}\ \emph {et~al.}(2023)\citenamefont {Rodenbach}, \citenamefont {{Ngoc Thanh Mai Tran}}, \citenamefont {Underwood}, \citenamefont {Panna}, \citenamefont {Andersen}, \citenamefont {Barcikowski}, \citenamefont {Payagala}, \citenamefont {{Peng Zhang}}, \citenamefont {{Lixuan Tai}}, \citenamefont {{Kang L. Wang}}, \citenamefont {Elmquist}, \citenamefont {Jarrett}, \citenamefont {Newell}, \citenamefont {Rigosi},\ and\ \citenamefont {Goldhaber-Gordon}}]{Rodenbach:2023_arXiv}%
  \BibitemOpen
  \bibfield  {author} {\bibinfo {author} {\bibfnamefont {K.}~\bibnamefont {Rodenbach}}, \bibinfo {author} {\bibnamefont {{Ngoc Thanh Mai Tran}}}, \bibinfo {author} {\bibfnamefont {J.~M.}\ \bibnamefont {Underwood}}, \bibinfo {author} {\bibfnamefont {A.~R.}\ \bibnamefont {Panna}}, \bibinfo {author} {\bibfnamefont {M.P.}\ \bibnamefont {Andersen}}, \bibinfo {author} {\bibfnamefont {Z.~S.}\ \bibnamefont {Barcikowski}}, \bibinfo {author} {\bibfnamefont {S.~U.}\ \bibnamefont {Payagala}}, \bibinfo {author} {\bibnamefont {{Peng Zhang}}}, \bibinfo {author} {\bibnamefont {{Lixuan Tai}}}, \bibinfo {author} {\bibnamefont {{Kang L. Wang}}}, \bibinfo {author} {\bibfnamefont {R.~E.}\ \bibnamefont {Elmquist}}, \bibinfo {author} {\bibfnamefont {D.~G.}\ \bibnamefont {Jarrett}}, \bibinfo {author} {\bibfnamefont {D.~B.}\ \bibnamefont {Newell}}, \bibinfo {author} {\bibfnamefont {A.~F.}\ \bibnamefont {Rigosi}}, \ and\ \bibinfo {author} {\bibfnamefont {D.}~\bibnamefont {Goldhaber-Gordon}},\ }\bibfield  {title} {\enquote {\bibinfo
  {title} {Realization of the quantum ampere using the quantum anomalous {Hall} and {Josephson} effects},}\ }\href {\doibase 10.48550/arXiv.2308.00200} {\bibfield  {journal} {\bibinfo  {journal} {arXiv:2308.00200}\ } (\bibinfo {year} {2023}),\ 10.48550/arXiv.2308.00200}\BibitemShut {NoStop}%
\bibitem [{\citenamefont {Fijalkowski}\ \emph {et~al.}(2021)\citenamefont {Fijalkowski}, \citenamefont {Liu}, \citenamefont {Mandal}, \citenamefont {Schreyeck}, \citenamefont {Brunner}, \citenamefont {Gould},\ and\ \citenamefont {and}}]{Fijalkowski:2021_NC}%
  \BibitemOpen
  \bibfield  {author} {\bibinfo {author} {\bibfnamefont {Kajetan~M.}\ \bibnamefont {Fijalkowski}}, \bibinfo {author} {\bibfnamefont {Nan}\ \bibnamefont {Liu}}, \bibinfo {author} {\bibfnamefont {Pankaj}\ \bibnamefont {Mandal}}, \bibinfo {author} {\bibfnamefont {Steffen}\ \bibnamefont {Schreyeck}}, \bibinfo {author} {\bibfnamefont {Karl}\ \bibnamefont {Brunner}}, \bibinfo {author} {\bibfnamefont {Charles}\ \bibnamefont {Gould}}, \ and\ \bibinfo {author} {\bibfnamefont {Laurens W.~Molenkamp}\ \bibnamefont {and}},\ }\bibfield  {title} {\enquote {\bibinfo {title} {Quantum anomalous {Hall} edge channels survive up to the {Curie} temperature},}\ }\href {\doibase 10.1038/s41467-021-25912-w} {\bibfield  {journal} {\bibinfo  {journal} {Nat. Commun.}\ }\textbf {\bibinfo {volume} {12}},\ \bibinfo {pages} {5599} (\bibinfo {year} {2021})}\BibitemShut {NoStop}%
\bibitem [{\citenamefont {Watanabe}\ \emph {et~al.}(2019)\citenamefont {Watanabe}, \citenamefont {Yoshimi}, \citenamefont {Kawamura}, \citenamefont {Mogi}, \citenamefont {Tsukazaki}, \citenamefont {Yu}, \citenamefont {Nakajima}, \citenamefont {Takahashi}, \citenamefont {Kawasaki},\ and\ \citenamefont {Tokura}}]{Watanabe:2019_APL}%
  \BibitemOpen
  \bibfield  {author} {\bibinfo {author} {\bibfnamefont {R.}~\bibnamefont {Watanabe}}, \bibinfo {author} {\bibfnamefont {R.}~\bibnamefont {Yoshimi}}, \bibinfo {author} {\bibfnamefont {M.}~\bibnamefont {Kawamura}}, \bibinfo {author} {\bibfnamefont {M.}~\bibnamefont {Mogi}}, \bibinfo {author} {\bibfnamefont {A.}~\bibnamefont {Tsukazaki}}, \bibinfo {author} {\bibfnamefont {X.~Z.}\ \bibnamefont {Yu}}, \bibinfo {author} {\bibfnamefont {K.}~\bibnamefont {Nakajima}}, \bibinfo {author} {\bibfnamefont {K.~S.}\ \bibnamefont {Takahashi}}, \bibinfo {author} {\bibfnamefont {M.}~\bibnamefont {Kawasaki}}, \ and\ \bibinfo {author} {\bibfnamefont {Y.}~\bibnamefont {Tokura}},\ }\bibfield  {title} {\enquote {\bibinfo {title} {Quantum anomalous {H}all effect driven by magnetic proximity coupling in all-telluride based heterostructure},}\ }\href {\doibase 10.1063/1.5111891} {\bibfield  {journal} {\bibinfo  {journal} {Appl. Phys. Lett.}\ }\textbf {\bibinfo {volume} {115}},\ \bibinfo {pages} {102403} (\bibinfo {year}
  {2019})}\BibitemShut {NoStop}%
\bibitem [{\citenamefont {Dietl}(2023{\natexlab{a}})}]{Dietl:2023_PRL}%
  \BibitemOpen
  \bibfield  {author} {\bibinfo {author} {\bibfnamefont {T.}~\bibnamefont {Dietl}},\ }\bibfield  {title} {\enquote {\bibinfo {title} {Effects of charge dopants in quantum spin {Hall} materials},}\ }\href {\doibase 10.1103/PhysRevLett.130.086202} {\bibfield  {journal} {\bibinfo  {journal} {Phys. Rev. Lett.}\ }\textbf {\bibinfo {volume} {130}},\ \bibinfo {pages} {086202} (\bibinfo {year} {2023}{\natexlab{a}})}\BibitemShut {NoStop}%
\bibitem [{\citenamefont {Dietl}(2023{\natexlab{b}})}]{Dietl:2023_PRB}%
  \BibitemOpen
  \bibfield  {author} {\bibinfo {author} {\bibfnamefont {T.}~\bibnamefont {Dietl}},\ }\bibfield  {title} {\enquote {\bibinfo {title} {Quantitative theory of backscattering in topological {HgTe} and {(Hg,Mn)Te} quantum wells: Acceptor states, {Kondo} effect, precessional dephasing, and bound magnetic polaron},}\ }\href {\doibase 10.1103/PhysRevB.107.085421} {\bibfield  {journal} {\bibinfo  {journal} {Phys. Rev. B}\ }\textbf {\bibinfo {volume} {107}},\ \bibinfo {pages} {085421} (\bibinfo {year} {2023}{\natexlab{b}})}\BibitemShut {NoStop}%
\bibitem [{\citenamefont {K{\"o}nig}\ \emph {et~al.}(2007)\citenamefont {K{\"o}nig}, \citenamefont {Wiedmann}, \citenamefont {Br{\"u}ne}, \citenamefont {Roth}, \citenamefont {Buhmann}, \citenamefont {Molenkamp}, \citenamefont {Qi},\ and\ \citenamefont {Zhang}}]{Konig2007}%
  \BibitemOpen
  \bibfield  {author} {\bibinfo {author} {\bibfnamefont {M.}~\bibnamefont {K{\"o}nig}}, \bibinfo {author} {\bibfnamefont {S.}~\bibnamefont {Wiedmann}}, \bibinfo {author} {\bibfnamefont {C.}~\bibnamefont {Br{\"u}ne}}, \bibinfo {author} {\bibfnamefont {A.}~\bibnamefont {Roth}}, \bibinfo {author} {\bibfnamefont {H.}~\bibnamefont {Buhmann}}, \bibinfo {author} {\bibfnamefont {L.~W.}\ \bibnamefont {Molenkamp}}, \bibinfo {author} {\bibfnamefont {Xiao-Liang}\ \bibnamefont {Qi}}, \ and\ \bibinfo {author} {\bibfnamefont {Shou-Cheng}\ \bibnamefont {Zhang}},\ }\bibfield  {title} {\enquote {\bibinfo {title} {Quantum spin {Hall} insulator state in {HgTe} quantum wells},}\ }\href {\doibase 10.1126/science.1148047} {\bibfield  {journal} {\bibinfo  {journal} {Science}\ }\textbf {\bibinfo {volume} {318}},\ \bibinfo {pages} {766--770} (\bibinfo {year} {2007})}\BibitemShut {NoStop}%
\bibitem [{\citenamefont {Yahniuk}\ \emph {et~al.}(2021)\citenamefont {Yahniuk}, \citenamefont {Kazakov}, \citenamefont {Jouault}, \citenamefont {Krishtopenko}, \citenamefont {Kret}, \citenamefont {Grabecki}, \citenamefont {Cywi\'nski}, \citenamefont {Mikhailov}, \citenamefont {Dvoretskii}, \citenamefont {Przybytek}, \citenamefont {Gavrilenko}, \citenamefont {Teppe}, \citenamefont {Dietl},\ and\ \citenamefont {Knap}}]{Yahniuk:2021_arXiv}%
  \BibitemOpen
  \bibfield  {author} {\bibinfo {author} {\bibfnamefont {I.}~\bibnamefont {Yahniuk}}, \bibinfo {author} {\bibfnamefont {A.}~\bibnamefont {Kazakov}}, \bibinfo {author} {\bibfnamefont {B.}~\bibnamefont {Jouault}}, \bibinfo {author} {\bibfnamefont {S.~S.}\ \bibnamefont {Krishtopenko}}, \bibinfo {author} {\bibfnamefont {S.}~\bibnamefont {Kret}}, \bibinfo {author} {\bibfnamefont {G.}~\bibnamefont {Grabecki}}, \bibinfo {author} {\bibfnamefont {G.}~\bibnamefont {Cywi\'nski}}, \bibinfo {author} {\bibfnamefont {N.~N.}\ \bibnamefont {Mikhailov}}, \bibinfo {author} {\bibfnamefont {S.~A.}\ \bibnamefont {Dvoretskii}}, \bibinfo {author} {\bibfnamefont {J.}~\bibnamefont {Przybytek}}, \bibinfo {author} {\bibfnamefont {V.~I.}\ \bibnamefont {Gavrilenko}}, \bibinfo {author} {\bibfnamefont {F.}~\bibnamefont {Teppe}}, \bibinfo {author} {\bibfnamefont {T.}~\bibnamefont {Dietl}}, \ and\ \bibinfo {author} {\bibfnamefont {W.}~\bibnamefont {Knap}},\ }\bibfield  {title} {\enquote {\bibinfo {title} {{HgTe} quantum wells for {QHE}
  metrology under soft cryomagnetic conditions: permanent magnets and liquid {$^4$He} temperatures},}\ }\href {\doibase 10.48550/arXiv.2111.07581} {\bibfield  {journal} {\bibinfo  {journal} {arXiv e-prints}\ } (\bibinfo {year} {2021}),\ 10.48550/arXiv.2111.07581},\ \Eprint {http://arxiv.org/abs/2111.07581} {arXiv:2111.07581} \BibitemShut {NoStop}%
\bibitem [{\citenamefont {Shamim}\ \emph {et~al.}(2020)\citenamefont {Shamim}, \citenamefont {Beugeling}, \citenamefont {B\"ottcher}, \citenamefont {Shekhar}, \citenamefont {Budewitz}, \citenamefont {Leubner}, \citenamefont {Lunczer}, \citenamefont {Hankiewicz}, \citenamefont {Buhmann},\ and\ \citenamefont {Molenkamp}}]{Shamim:2020_SA}%
  \BibitemOpen
  \bibfield  {author} {\bibinfo {author} {\bibfnamefont {S.}~\bibnamefont {Shamim}}, \bibinfo {author} {\bibfnamefont {W.}~\bibnamefont {Beugeling}}, \bibinfo {author} {\bibfnamefont {J.}~\bibnamefont {B\"ottcher}}, \bibinfo {author} {\bibfnamefont {P.}~\bibnamefont {Shekhar}}, \bibinfo {author} {\bibfnamefont {A.}~\bibnamefont {Budewitz}}, \bibinfo {author} {\bibfnamefont {P.}~\bibnamefont {Leubner}}, \bibinfo {author} {\bibfnamefont {L.}~\bibnamefont {Lunczer}}, \bibinfo {author} {\bibfnamefont {E.~M.}\ \bibnamefont {Hankiewicz}}, \bibinfo {author} {\bibfnamefont {H.}~\bibnamefont {Buhmann}}, \ and\ \bibinfo {author} {\bibfnamefont {L.~W.}\ \bibnamefont {Molenkamp}},\ }\bibfield  {title} {\enquote {\bibinfo {title} {Emergent quantum {Hall} effects below 50 {mT} in a two-dimensional topological insulator},}\ }\href {\doibase 10.1126/sciadv.aba4625} {\bibfield  {journal} {\bibinfo  {journal} {Adv. Sci.}\ }\textbf {\bibinfo {volume} {6}},\ \bibinfo {pages} {eaba4625} (\bibinfo {year} {2020})}\BibitemShut
  {NoStop}%
\bibitem [{\citenamefont {Liu}\ \emph {et~al.}(2008)\citenamefont {Liu}, \citenamefont {Qi}, \citenamefont {Dai}, \citenamefont {Fang},\ and\ \citenamefont {Zhang}}]{Liu:2008_PRLb}%
  \BibitemOpen
  \bibfield  {author} {\bibinfo {author} {\bibfnamefont {Chao-Xing}\ \bibnamefont {Liu}}, \bibinfo {author} {\bibfnamefont {Xiao-Liang}\ \bibnamefont {Qi}}, \bibinfo {author} {\bibfnamefont {Xi}~\bibnamefont {Dai}}, \bibinfo {author} {\bibfnamefont {Zhong}\ \bibnamefont {Fang}}, \ and\ \bibinfo {author} {\bibfnamefont {Shou-Cheng}\ \bibnamefont {Zhang}},\ }\bibfield  {title} {\enquote {\bibinfo {title} {Quantum anomalous {Hall} effect in {Hg$_{1-y}$Mn$_y$Te} quantum wells},}\ }\href {\doibase 10.1103/PhysRevLett.101.146802} {\bibfield  {journal} {\bibinfo  {journal} {Phys. Rev. Lett.}\ }\textbf {\bibinfo {volume} {101}},\ \bibinfo {pages} {146802} (\bibinfo {year} {2008})}\BibitemShut {NoStop}%
\bibitem [{\citenamefont {Dietl}\ \emph {et~al.}(1997)\citenamefont {Dietl}, \citenamefont {Haury},\ and\ \citenamefont {{Merle d'Aubign\'e}}}]{Dietl:1997_PRB}%
  \BibitemOpen
  \bibfield  {author} {\bibinfo {author} {\bibfnamefont {T.}~\bibnamefont {Dietl}}, \bibinfo {author} {\bibfnamefont {A.}~\bibnamefont {Haury}}, \ and\ \bibinfo {author} {\bibfnamefont {Y.}~\bibnamefont {{Merle d'Aubign\'e}}},\ }\bibfield  {title} {\enquote {\bibinfo {title} {Free carrier-induced ferromagnetism in structures of diluted magnetic semiconductors},}\ }\href {\doibase 10.1103/PhysRevB.55.R3347} {\bibfield  {journal} {\bibinfo  {journal} {Phys. Rev. B}\ }\textbf {\bibinfo {volume} {55}},\ \bibinfo {pages} {R3347--R3350} (\bibinfo {year} {1997})}\BibitemShut {NoStop}%
\bibitem [{\citenamefont {Ferrand}\ \emph {et~al.}(2000)\citenamefont {Ferrand}, \citenamefont {Cibert}, \citenamefont {Bourgognon}, \citenamefont {Tatarenko}, \citenamefont {Wasiela}, \citenamefont {Fishman}, \citenamefont {Bonanni}, \citenamefont {Sitter}, \citenamefont {Kole\'snik}, \citenamefont {Jaroszy\'nski}, \citenamefont {Barcz},\ and\ \citenamefont {Dietl}}]{Ferrand:2000_JCG}%
  \BibitemOpen
  \bibfield  {author} {\bibinfo {author} {\bibfnamefont {D.}~\bibnamefont {Ferrand}}, \bibinfo {author} {\bibfnamefont {J.}~\bibnamefont {Cibert}}, \bibinfo {author} {\bibfnamefont {C.}~\bibnamefont {Bourgognon}}, \bibinfo {author} {\bibfnamefont {S.}~\bibnamefont {Tatarenko}}, \bibinfo {author} {\bibfnamefont {A.}~\bibnamefont {Wasiela}}, \bibinfo {author} {\bibfnamefont {G.}~\bibnamefont {Fishman}}, \bibinfo {author} {\bibfnamefont {A.}~\bibnamefont {Bonanni}}, \bibinfo {author} {\bibfnamefont {H.}~\bibnamefont {Sitter}}, \bibinfo {author} {\bibfnamefont {S.}~\bibnamefont {Kole\'snik}}, \bibinfo {author} {\bibfnamefont {J.}~\bibnamefont {Jaroszy\'nski}}, \bibinfo {author} {\bibfnamefont {A.}~\bibnamefont {Barcz}}, \ and\ \bibinfo {author} {\bibfnamefont {T.}~\bibnamefont {Dietl}},\ }\bibfield  {title} {\enquote {\bibinfo {title} {Carrier-induced ferromagnetic interactions in p-doped {Zn$_{1-x}$Mn$_x$Te} epilayers},}\ }\href {\doibase https://doi.org/10.1016/S0022-0248(00)00114-7} {\bibfield  {journal}
  {\bibinfo  {journal} {J. Crys. Growth}\ }\textbf {\bibinfo {volume} {214-215}},\ \bibinfo {pages} {387 -- 390} (\bibinfo {year} {2000})}\BibitemShut {NoStop}%
\bibitem [{\citenamefont {Mycielski}\ \emph {et~al.}(1984)\citenamefont {Mycielski}, \citenamefont {Rigaux}, \citenamefont {Menant}, \citenamefont {Dietl},\ and\ \citenamefont {Otto}}]{Mycielski:1984_SSC}%
  \BibitemOpen
  \bibfield  {author} {\bibinfo {author} {\bibfnamefont {A.}~\bibnamefont {Mycielski}}, \bibinfo {author} {\bibfnamefont {C.}~\bibnamefont {Rigaux}}, \bibinfo {author} {\bibfnamefont {M.}~\bibnamefont {Menant}}, \bibinfo {author} {\bibfnamefont {T.}~\bibnamefont {Dietl}}, \ and\ \bibinfo {author} {\bibfnamefont {M.}~\bibnamefont {Otto}},\ }\bibfield  {title} {\enquote {\bibinfo {title} {Spin glass phase transition in {Hg$_{1-k}$Mn$_k$Te} semimagnetic semiconductors},}\ }\href {\doibase https://doi.org/10.1016/0038-1098(84)90807-X} {\bibfield  {journal} {\bibinfo  {journal} {Solid State Commun.}\ }\textbf {\bibinfo {volume} {50}},\ \bibinfo {pages} {257 -- 260} (\bibinfo {year} {1984})}\BibitemShut {NoStop}%
\bibitem [{\citenamefont {\'Sliwa}\ and\ \citenamefont {Dietl}(2023)}]{Sliwa:2023_arXiv}%
  \BibitemOpen
  \bibfield  {author} {\bibinfo {author} {\bibfnamefont {C.}~\bibnamefont {\'Sliwa}}\ and\ \bibinfo {author} {\bibfnamefont {T.}~\bibnamefont {Dietl}},\ }\bibfield  {title} {\enquote {\bibinfo {title} {Tight-binding theory of spin-spin interactions, {Curie} temperatures, and quantum {Hall} effects in topological {(Hg,Cr)Te} in comparison to non-topological {(Zn,Cr)Te}, and {(Ga,Mn)N}},}\ }\href {\doibase 10.48550/arXiv.2310.19856} {\bibfield  {journal} {\bibinfo  {journal} {arXiv e-prints}\ } (\bibinfo {year} {2023}),\ 10.48550/arXiv.2310.19856},\ \Eprint {http://arxiv.org/abs/2310.19856} {arXiv:2310.19856} \BibitemShut {NoStop}%
\bibitem [{\citenamefont {Dietl}\ \emph {et~al.}(2015)\citenamefont {Dietl}, \citenamefont {Sato}, \citenamefont {Fukushima}, \citenamefont {Bonanni}, \citenamefont {Jamet}, \citenamefont {Barski}, \citenamefont {Kuroda}, \citenamefont {Tanaka}, \citenamefont {Hai},\ and\ \citenamefont {Katayama-Yoshida}}]{Dietl:2015_RMP}%
  \BibitemOpen
  \bibfield  {author} {\bibinfo {author} {\bibfnamefont {T.}~\bibnamefont {Dietl}}, \bibinfo {author} {\bibfnamefont {K.}~\bibnamefont {Sato}}, \bibinfo {author} {\bibfnamefont {T.}~\bibnamefont {Fukushima}}, \bibinfo {author} {\bibfnamefont {A.}~\bibnamefont {Bonanni}}, \bibinfo {author} {\bibfnamefont {M.}~\bibnamefont {Jamet}}, \bibinfo {author} {\bibfnamefont {A.}~\bibnamefont {Barski}}, \bibinfo {author} {\bibfnamefont {S.}~\bibnamefont {Kuroda}}, \bibinfo {author} {\bibfnamefont {M.}~\bibnamefont {Tanaka}}, \bibinfo {author} {\bibfnamefont {Pham~Nam}\ \bibnamefont {Hai}}, \ and\ \bibinfo {author} {\bibfnamefont {H.}~\bibnamefont {Katayama-Yoshida}},\ }\bibfield  {title} {\enquote {\bibinfo {title} {Spinodal nanodecomposition in semiconductors doped with transition metals},}\ }\href {\doibase 10.1103/RevModPhys.87.1311} {\bibfield  {journal} {\bibinfo  {journal} {Rev. Mod. Phys.}\ }\textbf {\bibinfo {volume} {87}},\ \bibinfo {pages} {1311} (\bibinfo {year} {2015})}\BibitemShut {NoStop}%
\bibitem [{\citenamefont {Serre}\ \emph {et~al.}(2006)\citenamefont {Serre}, \citenamefont {Bastard}, \citenamefont {Rigaux}, \citenamefont {Mycielski},\ and\ \citenamefont {Furdyna}}]{Serre:2006_Pr}%
  \BibitemOpen
  \bibfield  {author} {\bibinfo {author} {\bibfnamefont {H.}~\bibnamefont {Serre}}, \bibinfo {author} {\bibfnamefont {G.}~\bibnamefont {Bastard}}, \bibinfo {author} {\bibfnamefont {C.}~\bibnamefont {Rigaux}}, \bibinfo {author} {\bibfnamefont {J.}~\bibnamefont {Mycielski}}, \ and\ \bibinfo {author} {\bibfnamefont {J.}~\bibnamefont {Furdyna}},\ }\enquote {\bibinfo {title} {Infrared magnetoabsorption in zero gap {Hg$_{1-x}$Fe$_x$Te} and {Hg$_{1-x}$Fe$_x$Se} mixed crystals},}\ in\ \href {\doibase 10.1007/3-540-11191-3_56} {\emph {\bibinfo {booktitle} {Physics of Narrow Gap Semiconductors}}},\ \bibinfo {editor} {edited by\ \bibinfo {editor} {\bibfnamefont {E.}~\bibnamefont {Gornik}}, \bibinfo {editor} {\bibfnamefont {H.}~\bibnamefont {Heinrich}}, \ and\ \bibinfo {editor} {\bibfnamefont {L.}~\bibnamefont {Palmetshofer}}}\ (\bibinfo  {publisher} {Springer Berlin Heidelberg},\ \bibinfo {year} {2006})\ pp.\ \bibinfo {pages} {321--325}\BibitemShut {NoStop}%
\bibitem [{\citenamefont {Blinowski}\ \emph {et~al.}(1996)\citenamefont {Blinowski}, \citenamefont {Kacman},\ and\ \citenamefont {Majewski}}]{Blinowski:1996_PRB}%
  \BibitemOpen
  \bibfield  {author} {\bibinfo {author} {\bibfnamefont {J.}~\bibnamefont {Blinowski}}, \bibinfo {author} {\bibfnamefont {P.}~\bibnamefont {Kacman}}, \ and\ \bibinfo {author} {\bibfnamefont {J.~A.}\ \bibnamefont {Majewski}},\ }\bibfield  {title} {\enquote {\bibinfo {title} {Ferromagnetic superexchange in {Cr}-based diluted magnetic semiconductors},}\ }\href {\doibase 10.1103/PhysRevB.53.9524} {\bibfield  {journal} {\bibinfo  {journal} {Phys. Rev. B}\ }\textbf {\bibinfo {volume} {53}},\ \bibinfo {pages} {9524--9527} (\bibinfo {year} {1996})}\BibitemShut {NoStop}%
\bibitem [{\citenamefont {Sawicki}\ \emph {et~al.}(1983)\citenamefont {Sawicki}, \citenamefont {Dietl}, \citenamefont {Plesiewicz}, \citenamefont {S{\c{e}}kowski}, \citenamefont {\'Sniadower}, \citenamefont {Baj},\ and\ \citenamefont {Dmowski}}]{Sawicki:1983_Pr}%
  \BibitemOpen
  \bibfield  {author} {\bibinfo {author} {\bibfnamefont {M.}~\bibnamefont {Sawicki}}, \bibinfo {author} {\bibfnamefont {T.}~\bibnamefont {Dietl}}, \bibinfo {author} {\bibfnamefont {W.}~\bibnamefont {Plesiewicz}}, \bibinfo {author} {\bibfnamefont {P.}~\bibnamefont {S{\c{e}}kowski}}, \bibinfo {author} {\bibfnamefont {L.}~\bibnamefont {\'Sniadower}}, \bibinfo {author} {\bibfnamefont {M.}~\bibnamefont {Baj}}, \ and\ \bibinfo {author} {\bibfnamefont {L.}~\bibnamefont {Dmowski}},\ }\bibfield  {title} {\enquote {\bibinfo {title} {Influence of an acceptor state on transport in zero-gap {Hg$_{1-x}$Mn$_x$Te}},}\ }in\ \href {\doibase 10.1007/3-540-11996-5_55} {\emph {\bibinfo {booktitle} {Application of High Magnetic Fields in \\ Semiconductor Physics}}},\ \bibinfo {editor} {edited by\ \bibinfo {editor} {\bibfnamefont {G.}~\bibnamefont {Landwehr}}}\ (\bibinfo  {publisher} {Springer Berlin Heidelberg},\ \bibinfo {address} {Berlin, Heidelberg},\ \bibinfo {year} {1983})\ pp.\ \bibinfo {pages} {382--385}\BibitemShut
  {NoStop}%
\bibitem [{\citenamefont {Langer}\ \emph {et~al.}(1988)\citenamefont {Langer}, \citenamefont {Delerue}, \citenamefont {Lannoo},\ and\ \citenamefont {Heinrich}}]{Langer:1988_PRB}%
  \BibitemOpen
  \bibfield  {author} {\bibinfo {author} {\bibfnamefont {J.~M.}\ \bibnamefont {Langer}}, \bibinfo {author} {\bibfnamefont {C.}~\bibnamefont {Delerue}}, \bibinfo {author} {\bibfnamefont {M.}~\bibnamefont {Lannoo}}, \ and\ \bibinfo {author} {\bibfnamefont {H.}~\bibnamefont {Heinrich}},\ }\bibfield  {title} {\enquote {\bibinfo {title} {Transition-metal impurities in semiconductors and heterojunction band lineups},}\ }\href {\doibase 10.1103/PhysRevB.38.7723} {\bibfield  {journal} {\bibinfo  {journal} {Phys. Rev. B}\ }\textbf {\bibinfo {volume} {38}},\ \bibinfo {pages} {7723--7739} (\bibinfo {year} {1988})}\BibitemShut {NoStop}%
\bibitem [{\citenamefont {Mathieu}\ \emph {et~al.}(1988)\citenamefont {Mathieu}, \citenamefont {Allegre}, \citenamefont {Chatt}, \citenamefont {Lefebvre},\ and\ \citenamefont {Faurie}}]{Mathieu:1988_PRB}%
  \BibitemOpen
  \bibfield  {author} {\bibinfo {author} {\bibfnamefont {H.}~\bibnamefont {Mathieu}}, \bibinfo {author} {\bibfnamefont {J.}~\bibnamefont {Allegre}}, \bibinfo {author} {\bibfnamefont {A.}~\bibnamefont {Chatt}}, \bibinfo {author} {\bibfnamefont {P.}~\bibnamefont {Lefebvre}}, \ and\ \bibinfo {author} {\bibfnamefont {J.~P.}\ \bibnamefont {Faurie}},\ }\bibfield  {title} {\enquote {\bibinfo {title} {Band offsets and lattice-mismatch effects in strained-layer {CdTe/ZnTe} superlattices},}\ }\href {\doibase 10.1103/PhysRevB.38.7740} {\bibfield  {journal} {\bibinfo  {journal} {Phys. Rev. B}\ }\textbf {\bibinfo {volume} {38}},\ \bibinfo {pages} {7740--7748} (\bibinfo {year} {1988})}\BibitemShut {NoStop}%
\bibitem [{\citenamefont {Dietl}\ and\ \citenamefont {Kossut}(1988)}]{Dietl:1988_PRB}%
  \BibitemOpen
  \bibfield  {author} {\bibinfo {author} {\bibfnamefont {T.}~\bibnamefont {Dietl}}\ and\ \bibinfo {author} {\bibfnamefont {J.}~\bibnamefont {Kossut}},\ }\bibfield  {title} {\enquote {\bibinfo {title} {Band offsets in {Hg}{Te}/{Cd}{Te} and {Hg}{Se}/{Cd}{Se} heterostructures from electron mobility limited by alloy scattering},}\ }\href {\doibase 10.1103/PhysRevB.38.10941} {\bibfield  {journal} {\bibinfo  {journal} {Phys. Rev. B}\ }\textbf {\bibinfo {volume} {38}},\ \bibinfo {pages} {10941--10942} (\bibinfo {year} {1988})}\BibitemShut {NoStop}%
\bibitem [{\citenamefont {Becker}\ \emph {et~al.}(2000)\citenamefont {Becker}, \citenamefont {Latussek}, \citenamefont {Pfeuffer-Jeschke}, \citenamefont {Landwehr},\ and\ \citenamefont {Molenkamp}}]{Becker:2000_PRB}%
  \BibitemOpen
  \bibfield  {author} {\bibinfo {author} {\bibfnamefont {C.~R.}\ \bibnamefont {Becker}}, \bibinfo {author} {\bibfnamefont {V.}~\bibnamefont {Latussek}}, \bibinfo {author} {\bibfnamefont {A.}~\bibnamefont {Pfeuffer-Jeschke}}, \bibinfo {author} {\bibfnamefont {G.}~\bibnamefont {Landwehr}}, \ and\ \bibinfo {author} {\bibfnamefont {L.~W.}\ \bibnamefont {Molenkamp}},\ }\bibfield  {title} {\enquote {\bibinfo {title} {Band structure and its temperature dependence for type-{III ${\mathrm{H}\mathrm{g}\mathrm{T}\mathrm{e}/\mathrm{H}\mathrm{g}}_{1\ensuremath{-}x}{\mathrm{Cd}}_{x}\mathrm{Te}$} superlattices and their semimetal constituent},}\ }\href {\doibase 10.1103/PhysRevB.62.10353} {\bibfield  {journal} {\bibinfo  {journal} {Phys. Rev. B}\ }\textbf {\bibinfo {volume} {62}},\ \bibinfo {pages} {10353--10363} (\bibinfo {year} {2000})}\BibitemShut {NoStop}%
\bibitem [{\citenamefont {Selber}\ \emph {et~al.}(1999)\citenamefont {Selber}, \citenamefont {Peka}, \citenamefont {Biernacki}, \citenamefont {Schulz}, \citenamefont {Schwarz},\ and\ \citenamefont {Benz}}]{Selber:1999_SST}%
  \BibitemOpen
  \bibfield  {author} {\bibinfo {author} {\bibfnamefont {H.~R.}\ \bibnamefont {Selber}}, \bibinfo {author} {\bibfnamefont {P.}~\bibnamefont {Peka}}, \bibinfo {author} {\bibfnamefont {S.~W.}\ \bibnamefont {Biernacki}}, \bibinfo {author} {\bibfnamefont {H.-J.}\ \bibnamefont {Schulz}}, \bibinfo {author} {\bibfnamefont {R.}~\bibnamefont {Schwarz}}, \ and\ \bibinfo {author} {\bibfnamefont {K.~W.}\ \bibnamefont {Benz}},\ }\bibfield  {title} {\enquote {\bibinfo {title} {Analysis of charge-transfer and charge-conserving optical transitions at vanadium ions in {CdTe}},}\ }\href {\doibase 10.1088/0268-1242/14/6/306} {\bibfield  {journal} {\bibinfo  {journal} {Semicond. Sci. Techn.}\ }\textbf {\bibinfo {volume} {14}},\ \bibinfo {pages} {521} (\bibinfo {year} {1999})}\BibitemShut {NoStop}%
\bibitem [{\citenamefont {Cieplak}\ \emph {et~al.}(1975)\citenamefont {Cieplak}, \citenamefont {Godlewski},\ and\ \citenamefont {Baranowski}}]{Cieplak:1975_pssb}%
  \BibitemOpen
  \bibfield  {author} {\bibinfo {author} {\bibfnamefont {M.~Z.}\ \bibnamefont {Cieplak}}, \bibinfo {author} {\bibfnamefont {M.}~\bibnamefont {Godlewski}}, \ and\ \bibinfo {author} {\bibfnamefont {J.~M.}\ \bibnamefont {Baranowski}},\ }\bibfield  {title} {\enquote {\bibinfo {title} {Optical charge transfer spectra and {EPR} spectra of {Cr$^{2+}(d^4)$} and {Cr$^{1+}(d^5)$} in {CdTe}},}\ }\href {\doibase 10.1002/pssb.2220700133} {\bibfield  {journal} {\bibinfo  {journal} {phys. stat. solidi (b)}\ }\textbf {\bibinfo {volume} {70}},\ \bibinfo {pages} {323--331} (\bibinfo {year} {1975})}\BibitemShut {NoStop}%
\bibitem [{\citenamefont {Kuroda}\ \emph {et~al.}(2007)\citenamefont {Kuroda}, \citenamefont {Nishizawa}, \citenamefont {Takita}, \citenamefont {Mitome}, \citenamefont {Bando}, \citenamefont {Osuch},\ and\ \citenamefont {Dietl}}]{Kuroda:2007_NM}%
  \BibitemOpen
  \bibfield  {author} {\bibinfo {author} {\bibfnamefont {S.}~\bibnamefont {Kuroda}}, \bibinfo {author} {\bibfnamefont {N.}~\bibnamefont {Nishizawa}}, \bibinfo {author} {\bibfnamefont {K.}~\bibnamefont {Takita}}, \bibinfo {author} {\bibfnamefont {M.}~\bibnamefont {Mitome}}, \bibinfo {author} {\bibfnamefont {Y.}~\bibnamefont {Bando}}, \bibinfo {author} {\bibfnamefont {K.}~\bibnamefont {Osuch}}, \ and\ \bibinfo {author} {\bibfnamefont {T.}~\bibnamefont {Dietl}},\ }\bibfield  {title} {\enquote {\bibinfo {title} {Origin and control of high-temperature ferromagnetism in semiconductors},}\ }\href {\doibase 10.1038/nmat1910} {\bibfield  {journal} {\bibinfo  {journal} {Nature Mat.}\ }\textbf {\bibinfo {volume} {6}},\ \bibinfo {pages} {440} (\bibinfo {year} {2007})}\BibitemShut {NoStop}%
\bibitem [{\citenamefont {Sato}\ \emph {et~al.}(2010)\citenamefont {Sato}, \citenamefont {Bergqvist}, \citenamefont {Kudrnovsk\'y}, \citenamefont {Dederichs}, \citenamefont {Eriksson}, \citenamefont {Turek}, \citenamefont {Sanyal}, \citenamefont {Bouzerar}, \citenamefont {Katayama-Yoshida}, \citenamefont {Dinh}, \citenamefont {Fukushima}, \citenamefont {Kizaki},\ and\ \citenamefont {Zeller}}]{Sato:2010_RMP}%
  \BibitemOpen
  \bibfield  {author} {\bibinfo {author} {\bibfnamefont {K.}~\bibnamefont {Sato}}, \bibinfo {author} {\bibfnamefont {L.}~\bibnamefont {Bergqvist}}, \bibinfo {author} {\bibfnamefont {J.}~\bibnamefont {Kudrnovsk\'y}}, \bibinfo {author} {\bibfnamefont {P.~H.}\ \bibnamefont {Dederichs}}, \bibinfo {author} {\bibfnamefont {O.}~\bibnamefont {Eriksson}}, \bibinfo {author} {\bibfnamefont {I.}~\bibnamefont {Turek}}, \bibinfo {author} {\bibfnamefont {B.}~\bibnamefont {Sanyal}}, \bibinfo {author} {\bibfnamefont {G.}~\bibnamefont {Bouzerar}}, \bibinfo {author} {\bibfnamefont {H.}~\bibnamefont {Katayama-Yoshida}}, \bibinfo {author} {\bibfnamefont {V.~A.}\ \bibnamefont {Dinh}}, \bibinfo {author} {\bibfnamefont {T.}~\bibnamefont {Fukushima}}, \bibinfo {author} {\bibfnamefont {H.}~\bibnamefont {Kizaki}}, \ and\ \bibinfo {author} {\bibfnamefont {R.}~\bibnamefont {Zeller}},\ }\bibfield  {title} {\enquote {\bibinfo {title} {First-principles theory of dilute magnetic semiconductors},}\ }\href {\doibase 10.1103/RevModPhys.82.1633}
  {\bibfield  {journal} {\bibinfo  {journal} {Rev. Mod. Phys.}\ }\textbf {\bibinfo {volume} {82}},\ \bibinfo {pages} {1633--1690} (\bibinfo {year} {2010})}\BibitemShut {NoStop}%
\bibitem [{\citenamefont {Zhang}\ \emph {et~al.}(2012)\citenamefont {Zhang}, \citenamefont {Zhu}, \citenamefont {Zhang}, \citenamefont {Xiao},\ and\ \citenamefont {Yao}}]{Zhang:2012_PRL}%
  \BibitemOpen
  \bibfield  {author} {\bibinfo {author} {\bibfnamefont {Jian-Min}\ \bibnamefont {Zhang}}, \bibinfo {author} {\bibfnamefont {Wenguang}\ \bibnamefont {Zhu}}, \bibinfo {author} {\bibfnamefont {Ying}\ \bibnamefont {Zhang}}, \bibinfo {author} {\bibfnamefont {Di}~\bibnamefont {Xiao}}, \ and\ \bibinfo {author} {\bibfnamefont {Yugui}\ \bibnamefont {Yao}},\ }\bibfield  {title} {\enquote {\bibinfo {title} {Tailoring magnetic doping in the topological insulator ${\mathrm{bi}}_{2}{\mathrm{se}}_{3}$},}\ }\href {\doibase 10.1103/PhysRevLett.109.266405} {\bibfield  {journal} {\bibinfo  {journal} {Phys. Rev. Lett.}\ }\textbf {\bibinfo {volume} {109}},\ \bibinfo {pages} {266405} (\bibinfo {year} {2012})}\BibitemShut {NoStop}%
\bibitem [{\citenamefont {Vergniory}\ \emph {et~al.}(2014)\citenamefont {Vergniory}, \citenamefont {Otrokov}, \citenamefont {Thonig}, \citenamefont {Hoffmann}, \citenamefont {Maznichenko}, \citenamefont {Geilhufe}, \citenamefont {Zubizarreta}, \citenamefont {Ostanin}, \citenamefont {Marmodoro}, \citenamefont {Henk}, \citenamefont {Hergert}, \citenamefont {Mertig}, \citenamefont {Chulkov},\ and\ \citenamefont {Ernst}}]{Vergniory:2014_PRB}%
  \BibitemOpen
  \bibfield  {author} {\bibinfo {author} {\bibfnamefont {M.~G.}\ \bibnamefont {Vergniory}}, \bibinfo {author} {\bibfnamefont {M.~M.}\ \bibnamefont {Otrokov}}, \bibinfo {author} {\bibfnamefont {D.}~\bibnamefont {Thonig}}, \bibinfo {author} {\bibfnamefont {M.}~\bibnamefont {Hoffmann}}, \bibinfo {author} {\bibfnamefont {I.~V.}\ \bibnamefont {Maznichenko}}, \bibinfo {author} {\bibfnamefont {M.}~\bibnamefont {Geilhufe}}, \bibinfo {author} {\bibfnamefont {X.}~\bibnamefont {Zubizarreta}}, \bibinfo {author} {\bibfnamefont {S.}~\bibnamefont {Ostanin}}, \bibinfo {author} {\bibfnamefont {A.}~\bibnamefont {Marmodoro}}, \bibinfo {author} {\bibfnamefont {J.}~\bibnamefont {Henk}}, \bibinfo {author} {\bibfnamefont {W.}~\bibnamefont {Hergert}}, \bibinfo {author} {\bibfnamefont {I.}~\bibnamefont {Mertig}}, \bibinfo {author} {\bibfnamefont {E.~V.}\ \bibnamefont {Chulkov}}, \ and\ \bibinfo {author} {\bibfnamefont {A.}~\bibnamefont {Ernst}},\ }\bibfield  {title} {\enquote {\bibinfo {title} {Exchange interaction and its tuning in
  magnetic binary chalcogenides},}\ }\href {\doibase 10.1103/PhysRevB.89.165202} {\bibfield  {journal} {\bibinfo  {journal} {Phys. Rev. B}\ }\textbf {\bibinfo {volume} {89}},\ \bibinfo {pages} {165202} (\bibinfo {year} {2014})}\BibitemShut {NoStop}%
\bibitem [{\citenamefont {Bouaziz}\ \emph {et~al.}(2018)\citenamefont {Bouaziz}, \citenamefont {dos Santos~Dias},\ and\ \citenamefont {Lounis}}]{Bouaziz:2018_PRB}%
  \BibitemOpen
  \bibfield  {author} {\bibinfo {author} {\bibfnamefont {J.}~\bibnamefont {Bouaziz}}, \bibinfo {author} {\bibfnamefont {J.}~\bibnamefont {dos Santos~Dias}, \bibfnamefont {M.and Iba\~nez-Azpiroz}}, \ and\ \bibinfo {author} {\bibfnamefont {S.}~\bibnamefont {Lounis}},\ }\bibfield  {title} {\enquote {\bibinfo {title} {Ab initio investigation of impurity-induced in-gap states in ${\mathrm{bi}}_{2}{\mathrm{te}}_{3}$ and ${\mathrm{bi}}_{2}{\mathrm{se}}_{3}$},}\ }\href {\doibase 10.1103/PhysRevB.98.035119} {\bibfield  {journal} {\bibinfo  {journal} {Phys. Rev. B}\ }\textbf {\bibinfo {volume} {98}},\ \bibinfo {pages} {035119} (\bibinfo {year} {2018})}\BibitemShut {NoStop}%
\bibitem [{\citenamefont {Cuono}\ \emph {et~al.}(2023)\citenamefont {Cuono}, \citenamefont {Sattigeri}, \citenamefont {Autieri},\ and\ \citenamefont {Dietl}}]{Cuono23Eu}%
  \BibitemOpen
  \bibfield  {author} {\bibinfo {author} {\bibfnamefont {Giuseppe}\ \bibnamefont {Cuono}}, \bibinfo {author} {\bibfnamefont {Raghottam~M.}\ \bibnamefont {Sattigeri}}, \bibinfo {author} {\bibfnamefont {Carmine}\ \bibnamefont {Autieri}}, \ and\ \bibinfo {author} {\bibfnamefont {Tomasz}\ \bibnamefont {Dietl}},\ }\bibfield  {title} {\enquote {\bibinfo {title} {Ab initio overestimation of the topological region in eu-based compounds},}\ }\href {\doibase 10.1103/PhysRevB.108.075150} {\bibfield  {journal} {\bibinfo  {journal} {Phys. Rev. B}\ }\textbf {\bibinfo {volume} {108}},\ \bibinfo {pages} {075150} (\bibinfo {year} {2023})}\BibitemShut {NoStop}%
\bibitem [{\citenamefont {Islam}\ \emph {et~al.}(2022)\citenamefont {Islam}, \citenamefont {Ghosh}, \citenamefont {Cuono}, \citenamefont {Lau}, \citenamefont {Brzezicki}, \citenamefont {Bansil}, \citenamefont {Agarwal}, \citenamefont {Singh}, \citenamefont {Dietl},\ and\ \citenamefont {Autieri}}]{Islam22}%
  \BibitemOpen
  \bibfield  {author} {\bibinfo {author} {\bibfnamefont {Rajibul}\ \bibnamefont {Islam}}, \bibinfo {author} {\bibfnamefont {Barun}\ \bibnamefont {Ghosh}}, \bibinfo {author} {\bibfnamefont {Giuseppe}\ \bibnamefont {Cuono}}, \bibinfo {author} {\bibfnamefont {Alexander}\ \bibnamefont {Lau}}, \bibinfo {author} {\bibfnamefont {Wojciech}\ \bibnamefont {Brzezicki}}, \bibinfo {author} {\bibfnamefont {Arun}\ \bibnamefont {Bansil}}, \bibinfo {author} {\bibfnamefont {Amit}\ \bibnamefont {Agarwal}}, \bibinfo {author} {\bibfnamefont {Bahadur}\ \bibnamefont {Singh}}, \bibinfo {author} {\bibfnamefont {Tomasz}\ \bibnamefont {Dietl}}, \ and\ \bibinfo {author} {\bibfnamefont {Carmine}\ \bibnamefont {Autieri}},\ }\bibfield  {title} {\enquote {\bibinfo {title} {Topological states in superlattices of hgte class of materials for engineering three-dimensional flat bands},}\ }\href {\doibase 10.1103/PhysRevResearch.4.023114} {\bibfield  {journal} {\bibinfo  {journal} {Phys. Rev. Res.}\ }\textbf {\bibinfo {volume} {4}},\ \bibinfo
  {pages} {023114} (\bibinfo {year} {2022})}\BibitemShut {NoStop}%
\bibitem [{\citenamefont {Islam}\ \emph {et~al.}(2023)\citenamefont {Islam}, \citenamefont {Mardanya}, \citenamefont {Lau}, \citenamefont {Cuono}, \citenamefont {Chang}, \citenamefont {Singh}, \citenamefont {Canali}, \citenamefont {Dietl},\ and\ \citenamefont {Autieri}}]{PhysRevB.107.125102}%
  \BibitemOpen
  \bibfield  {author} {\bibinfo {author} {\bibfnamefont {Rajibul}\ \bibnamefont {Islam}}, \bibinfo {author} {\bibfnamefont {Sougata}\ \bibnamefont {Mardanya}}, \bibinfo {author} {\bibfnamefont {Alexander}\ \bibnamefont {Lau}}, \bibinfo {author} {\bibfnamefont {Giuseppe}\ \bibnamefont {Cuono}}, \bibinfo {author} {\bibfnamefont {Tay-Rong}\ \bibnamefont {Chang}}, \bibinfo {author} {\bibfnamefont {Bahadur}\ \bibnamefont {Singh}}, \bibinfo {author} {\bibfnamefont {Carlo~M.}\ \bibnamefont {Canali}}, \bibinfo {author} {\bibfnamefont {Tomasz}\ \bibnamefont {Dietl}}, \ and\ \bibinfo {author} {\bibfnamefont {Carmine}\ \bibnamefont {Autieri}},\ }\bibfield  {title} {\enquote {\bibinfo {title} {Engineering axion insulator and other topological phases in superlattices without inversion symmetry},}\ }\href {\doibase 10.1103/PhysRevB.107.125102} {\bibfield  {journal} {\bibinfo  {journal} {Phys. Rev. B}\ }\textbf {\bibinfo {volume} {107}},\ \bibinfo {pages} {125102} (\bibinfo {year} {2023})}\BibitemShut {NoStop}%
\bibitem [{\citenamefont {Paier}\ \emph {et~al.}(2006)\citenamefont {Paier}, \citenamefont {Marsman}, \citenamefont {Hummer}, \citenamefont {Kresse}, \citenamefont {Gerber},\ and\ \citenamefont {\'Angy\'an}}]{Paier06}%
  \BibitemOpen
  \bibfield  {author} {\bibinfo {author} {\bibfnamefont {J.}~\bibnamefont {Paier}}, \bibinfo {author} {\bibfnamefont {M.}~\bibnamefont {Marsman}}, \bibinfo {author} {\bibfnamefont {K.}~\bibnamefont {Hummer}}, \bibinfo {author} {\bibfnamefont {G.}~\bibnamefont {Kresse}}, \bibinfo {author} {\bibfnamefont {I.~C.}\ \bibnamefont {Gerber}}, \ and\ \bibinfo {author} {\bibfnamefont {J.~G.}\ \bibnamefont {\'Angy\'an}},\ }\bibfield  {title} {\enquote {\bibinfo {title} {Screened hybrid density functionals applied to solids},}\ }\href {\doibase 10.1063/1.2187006} {\bibfield  {journal} {\bibinfo  {journal} {J. Chem. Phys.}\ }\textbf {\bibinfo {volume} {124}},\ \bibinfo {pages} {154709} (\bibinfo {year} {2006})},\ \Eprint {http://arxiv.org/abs/https://doi.org/10.1063/1.2187006} {https://doi.org/10.1063/1.2187006} \BibitemShut {NoStop}%
\bibitem [{\citenamefont {Dou}\ \emph {et~al.}(2021)\citenamefont {Dou}, \citenamefont {Sun},\ and\ \citenamefont {Wei}}]{Dou21}%
  \BibitemOpen
  \bibfield  {author} {\bibinfo {author} {\bibfnamefont {Baoying}\ \bibnamefont {Dou}}, \bibinfo {author} {\bibfnamefont {Qingde}\ \bibnamefont {Sun}}, \ and\ \bibinfo {author} {\bibfnamefont {Su-Huai}\ \bibnamefont {Wei}},\ }\bibfield  {title} {\enquote {\bibinfo {title} {Effects of co-doping in semiconductors: {CdTe}},}\ }\href {\doibase 10.1103/PhysRevB.104.245202} {\bibfield  {journal} {\bibinfo  {journal} {Phys. Rev. B}\ }\textbf {\bibinfo {volume} {104}},\ \bibinfo {pages} {245202} (\bibinfo {year} {2021})}\BibitemShut {NoStop}%
\bibitem [{\citenamefont {Lyons}\ \emph {et~al.}(2017)\citenamefont {Lyons}, \citenamefont {Alkauskas}, \citenamefont {Janotti},\ and\ \citenamefont {de~Walle}}]{Lyons2017}%
  \BibitemOpen
  \bibfield  {author} {\bibinfo {author} {\bibfnamefont {J.~L.}\ \bibnamefont {Lyons}}, \bibinfo {author} {\bibfnamefont {A.}~\bibnamefont {Alkauskas}}, \bibinfo {author} {\bibfnamefont {A.}~\bibnamefont {Janotti}}, \ and\ \bibinfo {author} {\bibfnamefont {C.~G.~Van}\ \bibnamefont {de~Walle}},\ }\bibfield  {title} {\enquote {\bibinfo {title} {Deep donor state of the copper acceptor as a source of green luminescence in {ZnO}},}\ }\href {\doibase 10.1063/1.4995404} {\bibfield  {journal} {\bibinfo  {journal} {Appl. Phys. Lett.}\ }\textbf {\bibinfo {volume} {111}},\ \bibinfo {pages} {042101} (\bibinfo {year} {2017})}\BibitemShut {NoStop}%
\bibitem [{\citenamefont {Kavanagh}\ \emph {et~al.}(2021)\citenamefont {Kavanagh}, \citenamefont {Walsh},\ and\ \citenamefont {Scanlon}}]{Kavanagh2021}%
  \BibitemOpen
  \bibfield  {author} {\bibinfo {author} {\bibfnamefont {Se{\'{a}}n~R.}\ \bibnamefont {Kavanagh}}, \bibinfo {author} {\bibfnamefont {Aron}\ \bibnamefont {Walsh}}, \ and\ \bibinfo {author} {\bibfnamefont {David~O.}\ \bibnamefont {Scanlon}},\ }\bibfield  {title} {\enquote {\bibinfo {title} {Rapid recombination by {Cadmium} vacancies in {CdTe}},}\ }\href {\doibase 10.1021/acsenergylett.1c00380} {\bibfield  {journal} {\bibinfo  {journal} {{ACS} Energy Lett.}\ }\textbf {\bibinfo {volume} {6}},\ \bibinfo {pages} {1392--1398} (\bibinfo {year} {2021})}\BibitemShut {NoStop}%
\bibitem [{\citenamefont {Skauli}\ and\ \citenamefont {Colin}(2001)}]{Skauli01}%
  \BibitemOpen
  \bibfield  {author} {\bibinfo {author} {\bibfnamefont {T.}~\bibnamefont {Skauli}}\ and\ \bibinfo {author} {\bibfnamefont {T.}~\bibnamefont {Colin}},\ }\bibfield  {title} {\enquote {\bibinfo {title} {Accurate determination of the lattice constant of molecular beam epitaxial {Cd}{Hg}{Te}},}\ }\href {\doibase https://doi.org/10.1016/S0022-0248(00)01005-8} {\bibfield  {journal} {\bibinfo  {journal} {Journal of Crystal Growth}\ }\textbf {\bibinfo {volume} {222}},\ \bibinfo {pages} {719--725} (\bibinfo {year} {2001})}\BibitemShut {NoStop}%
\bibitem [{\citenamefont {Autieri}\ \emph {et~al.}(2021)\citenamefont {Autieri}, \citenamefont {\'{S}liwa}, \citenamefont {Islam}, \citenamefont {Cuono},\ and\ \citenamefont {Dietl}}]{Autieri21}%
  \BibitemOpen
  \bibfield  {author} {\bibinfo {author} {\bibfnamefont {C.}~\bibnamefont {Autieri}}, \bibinfo {author} {\bibfnamefont {Cezary}\ \bibnamefont {\'{S}liwa}}, \bibinfo {author} {\bibfnamefont {R.}~\bibnamefont {Islam}}, \bibinfo {author} {\bibfnamefont {G.}~\bibnamefont {Cuono}}, \ and\ \bibinfo {author} {\bibfnamefont {T.}~\bibnamefont {Dietl}},\ }\bibfield  {title} {\enquote {\bibinfo {title} {Momentum-resolved spin splitting in {Mn}-doped trivial {CdTe} and topological {HgTe} semiconductors},}\ }\href {\doibase 10.1103/PhysRevB.103.115209} {\bibfield  {journal} {\bibinfo  {journal} {Phys. Rev. B}\ }\textbf {\bibinfo {volume} {103}},\ \bibinfo {pages} {115209} (\bibinfo {year} {2021})}\BibitemShut {NoStop}%
\bibitem [{\citenamefont {Furdyna}(1988)}]{Furdyna:1988_JAP}%
  \BibitemOpen
  \bibfield  {author} {\bibinfo {author} {\bibfnamefont {J.~K.}\ \bibnamefont {Furdyna}},\ }\bibfield  {title} {\enquote {\bibinfo {title} {{Diluted magnetic semiconductors}},}\ }\href {\doibase 10.1063/1.341700} {\bibfield  {journal} {\bibinfo  {journal} {J. Appl. Phys.}\ }\textbf {\bibinfo {volume} {64}},\ \bibinfo {pages} {R29--R64} (\bibinfo {year} {1988})}\BibitemShut {NoStop}%
\bibitem [{\citenamefont {Bonanni}\ \emph {et~al.}(2021)\citenamefont {Bonanni}, \citenamefont {Dietl},\ and\ \citenamefont {Ohno}}]{Bonanni:2021_HB}%
  \BibitemOpen
  \bibfield  {author} {\bibinfo {author} {\bibfnamefont {A.}~\bibnamefont {Bonanni}}, \bibinfo {author} {\bibfnamefont {T.}~\bibnamefont {Dietl}}, \ and\ \bibinfo {author} {\bibfnamefont {H.}~\bibnamefont {Ohno}},\ }\bibfield  {title} {\enquote {\bibinfo {title} {Dilute magnetic materials},}\ }in\ \href@noop {} {\emph {\bibinfo {booktitle} {Handbook of Magnetism and Magnetic Materials}}},\ \bibinfo {editor} {edited by\ \bibinfo {editor} {\bibfnamefont {M.}~\bibnamefont {Coey}}\ and\ \bibinfo {editor} {\bibfnamefont {S.}~\bibnamefont {Parkin}}}\ (\bibinfo  {publisher} {Springer, Berlin},\ \bibinfo {year} {2021})\BibitemShut {NoStop}%
\bibitem [{\citenamefont {Mycielski}(1988)}]{Mycielski:1988_JAP}%
  \BibitemOpen
  \bibfield  {author} {\bibinfo {author} {\bibfnamefont {A.}~\bibnamefont {Mycielski}},\ }\bibfield  {title} {\enquote {\bibinfo {title} {{Fe-based semimagnetic semiconductors}},}\ }\href {\doibase 10.1063/1.340813} {\bibfield  {journal} {\bibinfo  {journal} {J. Appl. Phys.}\ }\textbf {\bibinfo {volume} {63}},\ \bibinfo {pages} {3279--3284} (\bibinfo {year} {1988})}\BibitemShut {NoStop}%
\bibitem [{\citenamefont {G\l\'od}\ \emph {et~al.}(1994)\citenamefont {G\l\'od}, \citenamefont {Dietl}, \citenamefont {Fromherz}, \citenamefont {Bauer},\ and\ \citenamefont {Miotkowski}}]{Glod:1994_PRB}%
  \BibitemOpen
  \bibfield  {author} {\bibinfo {author} {\bibfnamefont {P.}~\bibnamefont {G\l\'od}}, \bibinfo {author} {\bibfnamefont {T.}~\bibnamefont {Dietl}}, \bibinfo {author} {\bibfnamefont {T.}~\bibnamefont {Fromherz}}, \bibinfo {author} {\bibfnamefont {G.}~\bibnamefont {Bauer}}, \ and\ \bibinfo {author} {\bibfnamefont {I.}~\bibnamefont {Miotkowski}},\ }\bibfield  {title} {\enquote {\bibinfo {title} {Resonant donors in semiconductors: {Sc} impurity in {Cd}{Se} and {Cd$_{1-x}$}{Mn$_x$}{Se}},}\ }\href {\doibase 10.1103/PhysRevB.49.7797} {\bibfield  {journal} {\bibinfo  {journal} {Phys. Rev. B}\ }\textbf {\bibinfo {volume} {49}},\ \bibinfo {pages} {7797--7800} (\bibinfo {year} {1994})}\BibitemShut {NoStop}%
\bibitem [{\citenamefont {Cruz}\ \emph {et~al.}(2024)\citenamefont {Cruz}, \citenamefont {Tang}, \citenamefont {Jia}, \citenamefont {Watson},\ and\ \citenamefont {Zhang}}]{APSMarch}%
  \BibitemOpen
  \bibfield  {author} {\bibinfo {author} {\bibfnamefont {G.-J.}\ \bibnamefont {Cruz}}, \bibinfo {author} {\bibfnamefont {Z.}~\bibnamefont {Tang}}, \bibinfo {author} {\bibfnamefont {F.}~\bibnamefont {Jia}}, \bibinfo {author} {\bibfnamefont {D.~F.}\ \bibnamefont {Watson}}, \ and\ \bibinfo {author} {\bibfnamefont {P.}~\bibnamefont {Zhang}},\ }\href@noop {} {\enquote {\bibinfo {title} {{A}{P}{S} -{A}{P}{S} {M}arch {M}eeting 2024 - {E}vent - {J}ahn-{T}eller {D}efects in {S}olids: {I}ntegrated {I}nfinite-mode {C}oupling {T}heory and {A}pplications --- meetings.aps.org},}\ }\bibinfo {howpublished} {\url{https://meetings.aps.org/Meeting/MAR24/Session/J00.132}} (\bibinfo {year} {2024})\BibitemShut {NoStop}%
\bibitem [{\citenamefont {Sarkar}\ \emph {et~al.}(2022)\citenamefont {Sarkar}, \citenamefont {Eriksson}, \citenamefont {Sarma},\ and\ \citenamefont {Marco}}]{PhysRevB.105.184201}%
  \BibitemOpen
  \bibfield  {author} {\bibinfo {author} {\bibfnamefont {S.}~\bibnamefont {Sarkar}}, \bibinfo {author} {\bibfnamefont {O.}~\bibnamefont {Eriksson}}, \bibinfo {author} {\bibfnamefont {D.~D.}\ \bibnamefont {Sarma}}, \ and\ \bibinfo {author} {\bibfnamefont {I.~Di}\ \bibnamefont {Marco}},\ }\bibfield  {title} {\enquote {\bibinfo {title} {Structural and electronic properties of the random alloy {ZnSe$_{x}$S$_{1-x}$}},}\ }\href {\doibase 10.1103/PhysRevB.105.184201} {\bibfield  {journal} {\bibinfo  {journal} {Phys. Rev. B}\ }\textbf {\bibinfo {volume} {105}},\ \bibinfo {pages} {184201} (\bibinfo {year} {2022})}\BibitemShut {NoStop}%
\bibitem [{\citenamefont {{Forte}}\ \emph {et~al.}(2018)\citenamefont {{Forte}}, \citenamefont {{Guerra}}, \citenamefont {{Avella}}, \citenamefont {{Autieri}}, \citenamefont {{Romano}},\ and\ \citenamefont {{Noce}}}]{Forte18}%
  \BibitemOpen
  \bibfield  {author} {\bibinfo {author} {\bibfnamefont {F.}~\bibnamefont {{Forte}}}, \bibinfo {author} {\bibfnamefont {D.}~\bibnamefont {{Guerra}}}, \bibinfo {author} {\bibfnamefont {A.}~\bibnamefont {{Avella}}}, \bibinfo {author} {\bibfnamefont {C.}~\bibnamefont {{Autieri}}}, \bibinfo {author} {\bibfnamefont {A.}~\bibnamefont {{Romano}}}, \ and\ \bibinfo {author} {\bibfnamefont {C.}~\bibnamefont {{Noce}}},\ }\bibfield  {title} {\enquote {\bibinfo {title} {{Interplay Between Spin-Orbit Coupling and Structural Deformations in Heavy Transition-Metal Oxides with Tetrahedral Coordination}},}\ }\href {\doibase 10.12693/APhysPolA.133.394} {\bibfield  {journal} {\bibinfo  {journal} {Acta Phys. Polon. A}\ }\textbf {\bibinfo {volume} {133}},\ \bibinfo {pages} {394--397} (\bibinfo {year} {2018})}\BibitemShut {NoStop}%
\bibitem [{\citenamefont {Ni}\ \emph {et~al.}(2020)\citenamefont {Ni}, \citenamefont {Liu}, \citenamefont {Wu}, \citenamefont {Lei}, \citenamefont {Xu}, \citenamefont {Jun},\ and\ \citenamefont {Ouyang}}]{Ni20}%
  \BibitemOpen
  \bibfield  {author} {\bibinfo {author} {\bibfnamefont {Dixing}\ \bibnamefont {Ni}}, \bibinfo {author} {\bibfnamefont {Shude}\ \bibnamefont {Liu}}, \bibinfo {author} {\bibfnamefont {Musheng}\ \bibnamefont {Wu}}, \bibinfo {author} {\bibfnamefont {Xueling}\ \bibnamefont {Lei}}, \bibinfo {author} {\bibfnamefont {Bo}~\bibnamefont {Xu}}, \bibinfo {author} {\bibfnamefont {Seong-Chan}\ \bibnamefont {Jun}}, \ and\ \bibinfo {author} {\bibfnamefont {Chuying}\ \bibnamefont {Ouyang}},\ }\bibfield  {title} {\enquote {\bibinfo {title} {Strong jahn-teller effect at nio4 tetrahedron in {Ni}{Co$_2$}{O$_4$} spinel},}\ }\href {\doibase https://doi.org/10.1016/j.physleta.2019.126114} {\bibfield  {journal} {\bibinfo  {journal} {Phys. Lett. A}\ }\textbf {\bibinfo {volume} {384}},\ \bibinfo {pages} {126114} (\bibinfo {year} {2020})}\BibitemShut {NoStop}%
\bibitem [{\citenamefont {Kawamura}\ \emph {et~al.}(2017)\citenamefont {Kawamura}, \citenamefont {Yoshimi}, \citenamefont {Tsukazaki}, \citenamefont {Takahashi}, \citenamefont {Kawasaki},\ and\ \citenamefont {Tokura}}]{Kawamura:2017_PRL}%
  \BibitemOpen
  \bibfield  {author} {\bibinfo {author} {\bibfnamefont {M.}~\bibnamefont {Kawamura}}, \bibinfo {author} {\bibfnamefont {R.}~\bibnamefont {Yoshimi}}, \bibinfo {author} {\bibfnamefont {A.}~\bibnamefont {Tsukazaki}}, \bibinfo {author} {\bibfnamefont {K.~S.}\ \bibnamefont {Takahashi}}, \bibinfo {author} {\bibfnamefont {M.}~\bibnamefont {Kawasaki}}, \ and\ \bibinfo {author} {\bibfnamefont {Y.}~\bibnamefont {Tokura}},\ }\bibfield  {title} {\enquote {\bibinfo {title} {{Current-Driven Instability of the Quantum Anomalous Hall Effect in Ferromagnetic Topological Insulators}},}\ }\href {\doibase 10.1103/PhysRevLett.119.016803} {\bibfield  {journal} {\bibinfo  {journal} {Phys. Rev. Lett.}\ }\textbf {\bibinfo {volume} {119}},\ \bibinfo {pages} {016803} (\bibinfo {year} {2017})}\BibitemShut {NoStop}%
\bibitem [{\citenamefont {\ifmmode~\acute{S}\else \'{S}\fi{}liwa}\ \emph {et~al.}(2021)\citenamefont {\ifmmode~\acute{S}\else \'{S}\fi{}liwa}, \citenamefont {Autieri}, \citenamefont {Majewski},\ and\ \citenamefont {Dietl}}]{Sliwa:2021_PRB}%
  \BibitemOpen
  \bibfield  {author} {\bibinfo {author} {\bibfnamefont {Cezary}\ \bibnamefont {\ifmmode~\acute{S}\else \'{S}\fi{}liwa}}, \bibinfo {author} {\bibfnamefont {Carmine}\ \bibnamefont {Autieri}}, \bibinfo {author} {\bibfnamefont {Jacek~A.}\ \bibnamefont {Majewski}}, \ and\ \bibinfo {author} {\bibfnamefont {Tomasz}\ \bibnamefont {Dietl}},\ }\bibfield  {title} {\enquote {\bibinfo {title} {Superexchange dominates in magnetic topological insulators},}\ }\href {\doibase 10.1103/PhysRevB.104.L220404} {\bibfield  {journal} {\bibinfo  {journal} {Phys. Rev. B}\ }\textbf {\bibinfo {volume} {104}},\ \bibinfo {pages} {L220404} (\bibinfo {year} {2021})}\BibitemShut {NoStop}%
\bibitem [{\citenamefont {Novik}\ \emph {et~al.}(2005)\citenamefont {Novik}, \citenamefont {Pfeuffer-Jeschke}, \citenamefont {Jungwirth}, \citenamefont {Latussek}, \citenamefont {Becker}, \citenamefont {Landwehr}, \citenamefont {Buhmann},\ and\ \citenamefont {Molenkamp}}]{Novik:2005_PRB}%
  \BibitemOpen
  \bibfield  {author} {\bibinfo {author} {\bibfnamefont {E.~G.}\ \bibnamefont {Novik}}, \bibinfo {author} {\bibfnamefont {A.}~\bibnamefont {Pfeuffer-Jeschke}}, \bibinfo {author} {\bibfnamefont {T.}~\bibnamefont {Jungwirth}}, \bibinfo {author} {\bibfnamefont {V.}~\bibnamefont {Latussek}}, \bibinfo {author} {\bibfnamefont {C.~R.}\ \bibnamefont {Becker}}, \bibinfo {author} {\bibfnamefont {G.}~\bibnamefont {Landwehr}}, \bibinfo {author} {\bibfnamefont {H.}~\bibnamefont {Buhmann}}, \ and\ \bibinfo {author} {\bibfnamefont {L.~W.}\ \bibnamefont {Molenkamp}},\ }\bibfield  {title} {\enquote {\bibinfo {title} {Band structure of semimagnetic {Hg$_{1-y}$Mn$_y$Te} quantum wells},}\ }\href {\doibase 10.1103/PhysRevB.72.035321} {\bibfield  {journal} {\bibinfo  {journal} {Phys. Rev. B}\ }\textbf {\bibinfo {volume} {72}},\ \bibinfo {pages} {035321} (\bibinfo {year} {2005})}\BibitemShut {NoStop}%
\bibitem [{\citenamefont {Dietl}\ and\ \citenamefont {Ohno}(2014)}]{Dietl:2014_RMP}%
  \BibitemOpen
  \bibfield  {author} {\bibinfo {author} {\bibfnamefont {T.}~\bibnamefont {Dietl}}\ and\ \bibinfo {author} {\bibfnamefont {H.}~\bibnamefont {Ohno}},\ }\bibfield  {title} {\enquote {\bibinfo {title} {Dilute ferromagnetic semiconductors: Physics and spintronic structures},}\ }\href {\doibase 10.1103/RevModPhys.86.187} {\bibfield  {journal} {\bibinfo  {journal} {Rev. Mod. Phys.}\ }\textbf {\bibinfo {volume} {86}},\ \bibinfo {pages} {187--251} (\bibinfo {year} {2014})}\BibitemShut {NoStop}%
\end{thebibliography}%
\end{document}